\documentclass[aps,pra,amsmath,twocolumn,amssymb,superscriptaddress]{revtex4-1}

\usepackage{amssymb,amsfonts,amsmath}
\usepackage{color}
\usepackage[apple mac]{inputenc}
\usepackage[english]{babel}
\usepackage{times}
\usepackage{latexsym}
\usepackage{fancyhdr}
\usepackage{verbatim}
\usepackage{graphicx}
\usepackage{cancel}
\usepackage{epsf}
\usepackage{subfigure}
\usepackage{bm}
\usepackage{epstopdf}

\usepackage[colorlinks=true,
            linkcolor=magenta,
            urlcolor=blue,
            citecolor=blue]{hyperref}

\definecolor{Nathanblue}{rgb}{0.,0.24,0.51}

\newcommand{\blue}{\color{Nathanblue}}

\definecolor{orange}{rgb}{0.96,0.24,0.00}

\usepackage{mathtools}

\def\be{\begin{equation}}
\def\ee{\end{equation}}

\begin{document}

\title{{\blue Floquet-engineered nonlinearities and controllable pair-hopping processes: \\ From optical Kerr cavities to correlated quantum matter}}



\author{N. Goldman}
\email[]{ngoldman@ulb.ac.be}
\affiliation{CENOLI,
Universit\'e Libre de Bruxelles, CP 231, Campus Plaine, B-1050 Brussels, Belgium}
\author{O. K. Diessel}
\affiliation{Max-Planck-Institute of Quantum Optics, 85748 Garching, Germany}
\author{L. Barbiero}
\affiliation{Institute for Condensed Matter Physics and Complex Systems, DISAT, Politecnico di Torino, I-10129 Torino, Italy}
\author{M. Pr\"ufer}
\affiliation{Vienna Center for Quantum Science and Technology,
Atominstitut, TU Wien, Stadionallee 2, 1020 Vienna, Austria}
\author{M. Di Liberto}
\affiliation{Dipartimento di Fisica e Astronomia ``G. Galilei" \& Padua Quantum Technologies
Research Center, Università degli Studi di Padova, I-35131, Padova, Italy}
\affiliation{INFN Istituto Nazionale di Fisica Nucleare, Sezione di Padova, I-35131, Padova, Italy}
\author{L. Peralta Gavensky}
\affiliation{CENOLI,
Universit\'e Libre de Bruxelles, CP 231, Campus Plaine, B-1050 Brussels, Belgium}

\begin{abstract}

This work explores the possibility of creating and controlling unconventional nonlinearities by periodic driving, in a broad class of systems described by the nonlinear Schr\"odinger equation (NLSE). By means of a parent quantum many-body description, we demonstrate that such driven systems are well captured by an effective NLSE with emergent nonlinearities, which can be finely controlled by tuning the driving sequence. We first consider a general class of two-mode nonlinear systems -- relevant to optical Kerr cavities, waveguides and Bose-Einstein condensates -- where we find an emergent four-wave mixing nonlinearity, which originates from pair-hopping processes in the parent quantum picture. Tuning this drive-induced nonlinearity is shown to modify the phase-space topology, which can be detected through relative population and phase measurements. We then couple individual (two-mode) dimers in view of designing extended lattice models with unconventional nonlinearities and controllable pair-hopping processes. Following this general dimerization construction, we obtain an effective lattice model with drive-induced interactions, whose ground-state exhibits orbital order, chiral currents and emergent magnetic fluxes through the spontaneous breaking of time-reversal symmetry. We analyze these intriguing properties both in the weakly-interacting (mean-field) regime, captured by the effective NLSE, and in the strongly-correlated quantum regime. Our general approach opens a route for the engineering of unconventional optical nonlinearities in photonic devices and controllable drive-induced interactions in ultracold quantum matter.

\end{abstract}

\date{\today}

\maketitle

\section{Introduction}
Physical systems can be controlled and enriched by subjecting them to a time-periodic drive. Widely explored in the context of atomic physics since the 70's~\cite{haroche1970modified,avan1976effect,raab2000motional,lignier2007dynamical,eckardt2017colloquium}, this general approach recently became the leitmotif of a vaste and pluridisciplinary program known as Floquet engineering~\cite{eckardt2017colloquium,oka2019floquet,rudner2020band,weitenberg2021tailoring}. Today, it concerns a wide range of physical platforms, including ultracold quantum gases~\cite{eckardt2017colloquium,weitenberg2021tailoring}, solid-state materials~\cite{oka2019floquet,rudner2020band}, universal quantum simulators and computers~\cite{georgescu2014quantum,altman2021quantum}, mechanical~\cite{salerno2016floquet} and acoustical~\cite{fleury2016floquet} systems, and photonic devices~\cite{rechtsman2013photonic,schine2016synthetic,roushan2017chiral,ozawa2019topological}. 

More specifically, Floquet engineering can be applied to modify the band structure of lattice systems~\cite{eckardt2017colloquium,rudner2020band}, generate artificial gauge fields~\cite{goldman2014periodically,aidelsburger2018artificial} and design complex interaction processes~\cite{berman2001manipulating,saito2003dynamically,ma2011photon,rapp2012ultracold,ajoy2013quantum,di2014quantum,daley2014effective,bukov2015universal,eckardt2015high,anisimovas2015role,meinert2016floquet,hung2016quantum,pieplow2018generation,lee2018floquet,choi2020robust,barbiero2020bose,dehghani2021light,geier2021floquet,zahn2022formation,sanz2022interaction}. These remarkable possibilities open new avenues for the experimental exploration of a broad range of intriguing physical phenomena, such as light-induced high-temperature superconductivity~\cite{fausti2011light,mitrano2016possible}, magnetism~\cite{struck2011quantum,struck2013engineering,gorg2018enhancement}, topological physics~\cite{ozawa2019topological,rudner2020band,weitenberg2021tailoring}, many-body localization~\cite{ponte2015many,abanin2019colloquium}, chaos-assisted tunneling~\cite{hensinger2001dynamical,arnal2020chaos}, and lattice gauge theories~\cite{barbiero2019coupling,schweizer2019floquet}.

Floquet engineering has recently entered the realm of photonics, where various settings and periodic-driving scenarios have been proposed and experimentally realized. In laser-written optical waveguide arrays~\cite{szameit2010discrete}, where waveguides can be finely modulated along the propagation direction, Floquet schemes were implemented in view of generating topological band structures~\cite{rechtsman2013photonic,mukherjee2017experimental,maczewsky2017observation,mukherjee2018state,mukherjee2020observation,mukherjee2021observation}, synthetic dimensions~\cite{lustig2019photonic} and artificial magnetic fields~\cite{mukherjee2018experimental} for light.
In the context of optical resonators, electro-optical modulators were used to resonantly couple different cavity modes and realize synthetic dimensions~\cite{yuan2018synthetic,dutt2019experimental,dutt2020single,balvcytis2021synthetic,englebert2021bloch}, while non-planar geometries were designed to create stroboscopic dynamics reflecting an effective magnetic field for photons~\cite{schine2016synthetic}. In circuit-QED, time-modulated couplers connecting superconducting qubits were exploited to create artificial magnetic fields for strongly-interacting photons hopping on a lattice~\cite{roushan2017chiral}. Finally, drive-induced optical nonlinearities recently emerged as an exciting avenue in the context of polaritons~\cite{clark2019interacting,johansen2020multimode},  insulating materials~\cite{shan2021giant}, and high-Q microwave cavities coupled to transmon qubits~\cite{zhang2022drive}.

The scope of this work is two-fold. First, we introduce a general and practical theoretical method to treat a broad class of periodically-driven nonlinear systems described by the nonlinear Schr\"odinger equation (NLSE). By exploiting a parent quantum many-body description, we show that such driven nonlinear systems are well captured by an effective NLSE with emergent nonlinearities, which can be finely controlled by tuning the driving sequence [Fig.~\ref{fig_approach}]. This approach is first analyzed for a generic two-mode nonlinear system subjected to a repeated pulse sequence that mixes the two modes periodically in time. In this case, an emergent nonlinearity known as four-wave mixing~\cite{agrawal2001applications,delque2007polarization,agrawal2012nonlinear} is shown to originate from drive-induced pair-hopping processes in the parent quantum picture. This framework captures a broad range of nonlinear optical settings, including two-mode optical Kerr cavities~\cite{cao2017experimental,hill2020effects,garbin2020asymmetric,coen2023nonlinear}, optical waveguide couplers~\cite{szameit2010discrete,szameit2009inhibition} and coupled superconducting microwave cavities~\cite{roushan2017chiral}, but also ultracold atomic gases trapped in double-well potentials~\cite{andrews1997observation,smerzi1997quantum,albiez2005direct,schumm2005bose,hofferberth_radio-frequency_2006} and two-component Bose-Einstein condensates (BEC)~\cite{zibold2010classical,gross2010nonlinear,recati2022coherently}.

Building on these results, we then couple individual (two-mode) dimers in view of designing extended lattice models with unconventional nonlinearities and controllable pair-hopping processes. Following this general dimerization construction, we obtain an effective lattice model with drive-induced interactions, whose ground-state exhibits orbital order, chiral currents and emergent magnetic fluxes through the spontaneous breaking of time-reversal symmetry (TRS). This rich model is analyzed both in the weakly-interacting (mean-field) regime, captured by the NLSE, and in the strongly-interacting (quantum) regime, through various analytical and numerical methods. We discuss how the exotic properties and phase transitions of this peculiar lattice model could be detected in practice, through static and dynamical probes, in realistic settings. Our general construction leads to controllable \mbox{Hubbard-type} models and quantum spin models, well suited for the exploration of exotic quantum phases of matter emerging from unconventional interactions.

\subsection*{Theoretical approach and outline of the article}

\begin{figure}[h!]
\includegraphics[width = \linewidth]{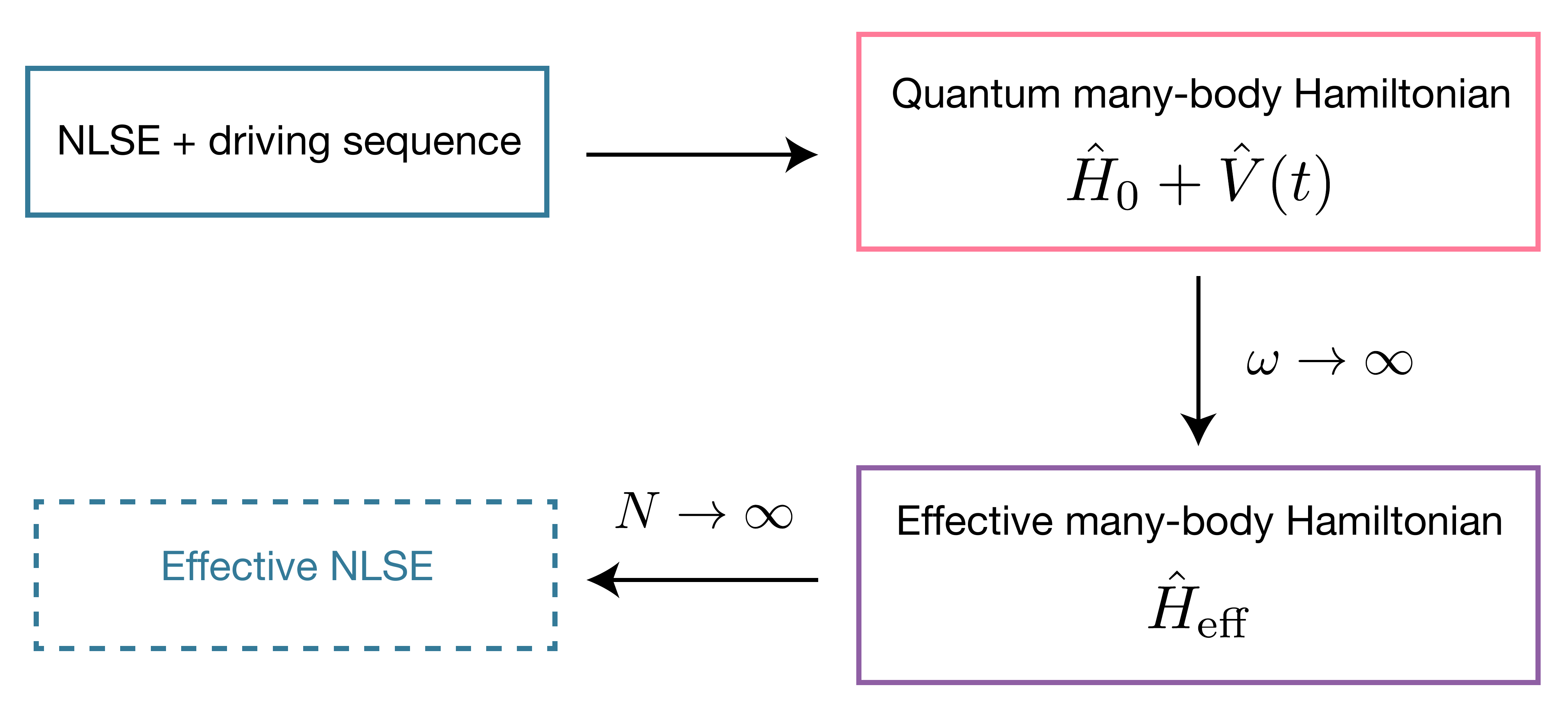}
\caption{Schematic of the approach. We consider a general class of nonlinear systems, driven by a periodic driving sequence and described by the nonlinear Schr\"odinger equation (NLSE). To analyse these settings, we introduce a parent quantum many-body Hamiltonian, $\hat H_0 + \hat V(t)$, which describes interacting bosons subjected to a periodic drive. From this, we derive an effective quantum Hamiltonian $\hat H_{\text{eff}}$ in the high-frequency limit of the drive ($\omega\!\rightarrow\infty$). We then derive the effective NLSE upon taking the classical limit, $N\!\rightarrow\infty$, where $N$ is the number of bosons, hence revealing the effective nonlinearities generated by the driving sequence; see also Fig.~\ref{fig_approach_numerics} in Section~\ref{section_numerics} regarding the numerical validation of this approach.
}
\label{fig_approach}
\end{figure}

The first Sections~\ref{section_two_modes}-\ref{section_quantum} explore how unconventional nonlinearities can emerge in driven nonlinear systems described by the two-mode NLSE. Our theoretical approach uses a parent quantum many-body Hamiltonian, which describes interacting bosons subjected to a periodic drive. Within this Hamiltonian framework, we derive an effective (time-independent) quantum Hamiltonian that well describes the stroboscopic dynamics in the high-frequency regime of the periodic drive~\cite{goldman2014periodically,goldman2015periodically,bukov2015universal,eckardt2015high,mikami2016brillouin}. We then take the classical limit of this effective quantum description~\cite{pitaevskii2016bose,carusotto2013quantum,holthaus2001towards,cao2020reconfigurable} to finally obtain an effective NLSE. This approach, which explicitly reveals the emergent nonlinearities generated and controlled by the drive, is summarized in Fig.~\ref{fig_approach}. 

Effective nonlinearities can be tuned by simply adjusting the driving sequence. Section~\ref{section_pendulum} analyzes how this control over nonlinearities can lead to modifications of the classical phase-space topology. We describe these transitions through a ``fixed-point phase diagram",  which we explain using a simple pendulum analogy. Interestingly, the control over drive-induced nonlinearities is directly reflected in the phase-space topology, which can be detected through the dynamics of the relative population and phase in the two modes.

Section~\ref{section_numerics} explores the validity of our two central approximations:~the high-frequency approximation related to the drive and the mean-field approximation associated with the classical limit. Here we perform numerical simulations of the quantum and classical dynamics, comparing the full time dynamics generated by the drive to the effective descriptions. As a by-product, this numerical analysis further illustrates how effective nonlinearities can be unambiguously detected through the dynamics of the relative population and phase in the two modes. 

We then design lattice systems with controllable drive-induced interactions in Section~\ref{section_lattice}. Using a dimerization construction, by which we couple individual (two-mode) dimers, we derive two classes of lattice models with effective pair-hopping processes. In Section~\ref{section_GS}, we set the focus on the ground-state properties of a specific dimerized lattice model with pair hopping, which gives rise to orbital order, chiral currents and emergent magnetic fluxes through the spontaneous breaking of TRS. These intriguing properties are analyzed using various analytical and numerical methods, both in the weakly-interacting (mean-field) regime captured by the NLSE and in the strongly-correlated quantum regime. As a by-product, we derive effective spin models, deep in the strongly-interacting regime, which are shown to feature   peculiar (Dzyaloshinskii-Moriya-type) interactions.

We conclude this work in Section~\ref{section_exp}, by proposing possible experimental implementations and detection schemes in optics and cold atoms.

\section{Two-mode nonlinear systems and drive-induced nonlinearities}\label{section_two_modes}

We start by considering a broad class of two-mode nonlinear systems, described by the nonlinear Schr\"odinger equation (NLSE)
\begin{align}
&i \frac{\partial \psi_1}{\partial t} = \left ( - \gamma \frac{\partial^2}{\partial x^2} + \vert \psi_1 \vert^2 + \beta \vert \psi_2 \vert^2   \right ) \psi_1 - \frac{\Omega_0}{2} \psi_2, \notag \\
&i \frac{\partial \psi_2}{\partial t} = \left ( -\gamma \frac{\partial^2}{\partial x^2} + \vert \psi_2 \vert^2 + \beta \vert \psi_1 \vert^2   \right ) \psi_2 - \frac{\Omega_0}{2} \psi_1. \label{NLS}
\end{align}
Here, $\psi_{1,2}(x,t)$ denote the complex amplitude of the fields corresponding to the two modes $s=1,2$; they depend on the evolution ``time" $t$ and the ``spatial" coordinate $x$. The focus of this work is set on the ``internal" dynamics associated with the two modes, such that the ``spatial" coordinate $x$ [and the related kinetic-energy term~$\sim \gamma$ in Eq.~\eqref{NLS}] does not play any role in the following. For the sake of generality, the equations of motion~\eqref{NLS} contain two types of nonlinearities, which are generically present in optical cavities~\cite{cao2017experimental,hill2020effects,garbin2020asymmetric,coen2023nonlinear}:~the so-called self-phase modulation and the cross-phase modulation, whose respective strengths are set by the parameter $\beta$; we have also included a static linear coupling of strength $\Omega_0/2$. We point out that the nonlinear equations~\eqref{NLS} are decoupled in the limit $\Omega_0\!=\!\beta\!=\!0$, i.e.~in the absence of linear coupling and cross-phase modulation. 

\begin{figure}[h!]
\includegraphics[width = \linewidth]{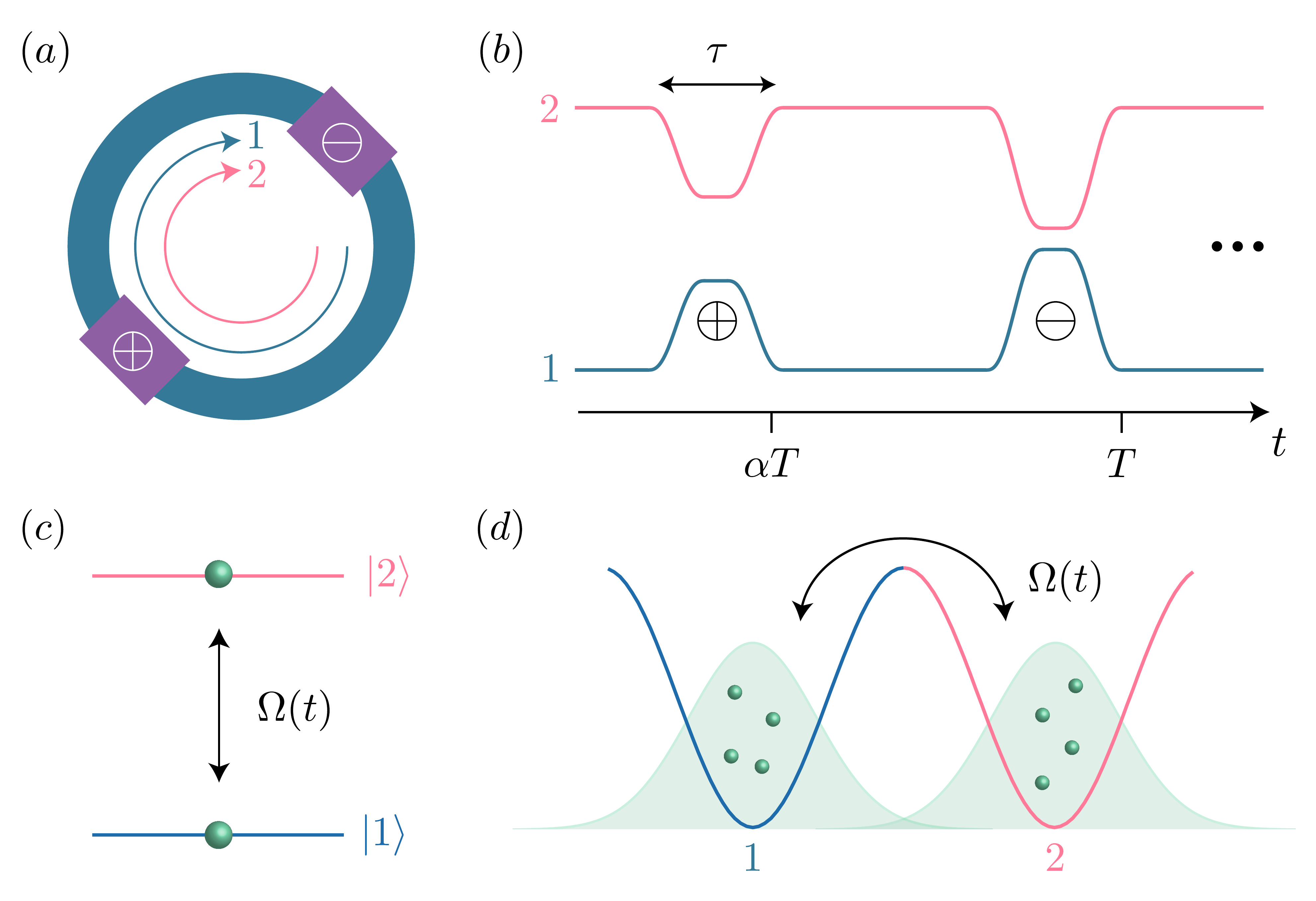}
\caption{Possible realizations in optics and cold atomic gases: (a) Two modes in an optical ring cavity ($1$ and $2$), repeatedly undergoing  mixing operations ($\oplus$ and $\ominus$) along the ring. These operations correspond to a coupling between the two polarization eigenmodes of the cavity, as realized by means of quarter-wave plates; see Eqs.~\eqref{eq_exchange}-\eqref{eq_exchange_2}. (b) Two optical waveguides ($1$ and $2$) with modulated inter-waveguide separation, realizing a ``time-periodic" linear coupling $\Omega (t)$ between the two optical modes [Eq.~\eqref{NLS_time}]. In both cases (a)-(b), the ``time" coordinate corresponds to the propagation direction~\cite{szameit2010discrete,fatome2021self}. (c) Two-component BEC involving two atomic internal states and a time-dependent (microwave) coupling $\Omega (t)$. (d) Bosonic gas in a double-well potential, with a time-modulated tunneling strength $\Omega (t)$.
}
\label{fig_sketch}
\end{figure}

While Eq.~\eqref{NLS} naturally describes the two polarization modes $\psi_{1,2}$ of a light field propagating in a lossless cavity~\cite{cao2017experimental,hill2020effects,garbin2020asymmetric,coen2023nonlinear}, or light propagating in a pair of adjacent waveguides~\cite{szameit2010discrete,szameit2009inhibition}, it should be noted that Eq.~\eqref{NLS}  equally captures the physics of bosonic atomic gases trapped in a double well potential, as well as two-component Bose-Einstein condensates~\cite{smerzi1997quantum,zibold2010classical,recati2022coherently}; see Fig.~\ref{fig_sketch} for an illustration of these four possible realizations. A more detailed discussion on experiment aspects is provided in Section~\ref{section_exp}. 

\begin{figure}[h!]
\includegraphics[width = \linewidth]{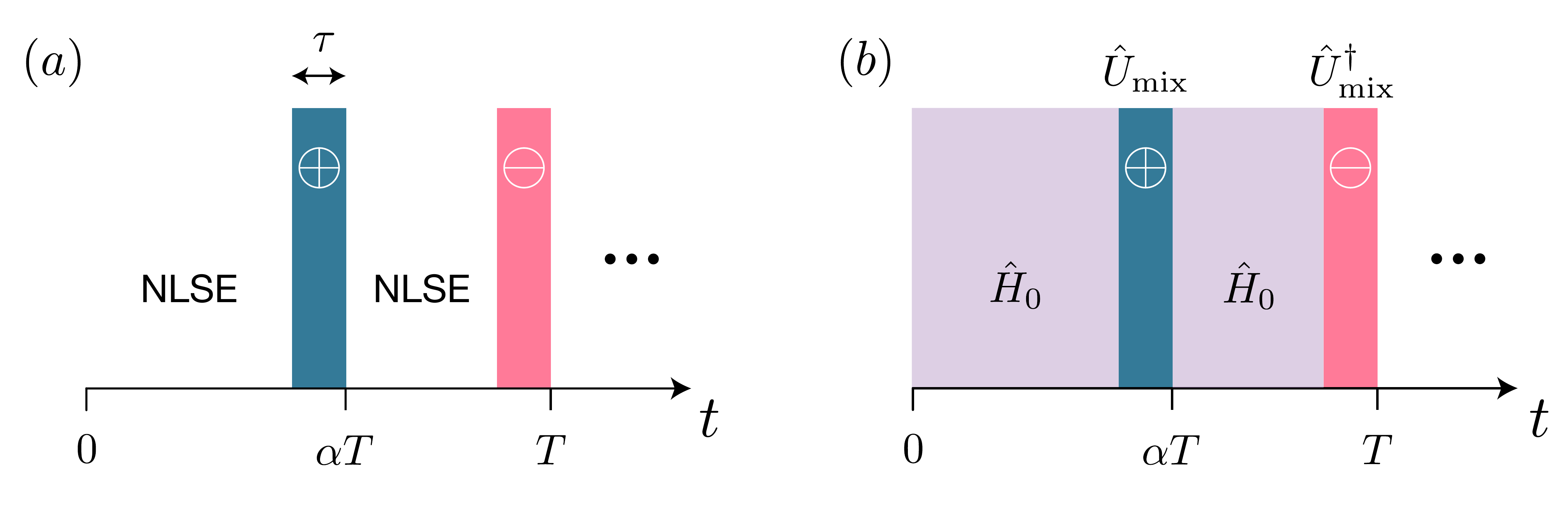}
\caption{(a) The pulse sequence involves free evolution, described by the NLSE in Eq.~\eqref{NLS}, interrupted by two pulses $(\oplus,\ominus)$ described by Eqs.~\eqref{eq_exchange}-\eqref{eq_exchange_2}. (b) The same pulse sequence, now expressed in terms of the time-evolution operator in Eq.~\eqref{U_sequence}, which involves the mixing operations $\hat U_{\text{mix}}^{(\dagger)}$ and the ``free" time evolution operator $\hat H_0$.
}
\label{fig_sequence}
\end{figure}

In order to modify the nonlinearities of a system described by Eq.~\eqref{NLS}, we now introduce a time-periodic pulse sequence of period $T$, which mixes the two modes in a fast and stroboscopic manner. As illustrated in Fig.~\ref{fig_sequence}(a), the sequence is characterized by four successive steps (within each period $T$): 
\begin{itemize}
\item Step 1:~Free evolution according to the NLSE~\eqref{NLS} for a duration $t=\alpha T$, where $\alpha$ is a tunable parameter defined between $[0,1]$.
\item Step 2:~the two components suddenly undergo the mixing operation (Pulse $\oplus$)
\begin{align}\label{eq_exchange}
&\psi_1 \rightarrow (1/\sqrt{2}) \left (\psi_1 +i \psi_2 \right ),  \\
&\psi_2 \rightarrow (1/\sqrt{2}) \left (i \psi_1 + \psi_2 \right).\notag
\end{align}
\item Step 3:~Free evolution according to the NLSE~\eqref{NLS} for a duration $t=(1-\alpha) T$.
\item Step 4:~the two components undergo the reverse mixing operation (Pulse $\ominus$)
\begin{align}\label{eq_exchange_2}
&\psi_1 \rightarrow (1/\sqrt{2}) \left (\psi_1 -i \psi_2 \right ),  \\
& \psi_2 \rightarrow (1/\sqrt{2}) \left (\psi_2 -i \psi_1\right). \notag
\end{align}
\end{itemize}

For certain devices, the mixing operations Eqs.~\eqref{eq_exchange}-\eqref{eq_exchange_2} can be performed readily, on arbitrarily short time scales. For instance, in a two-mode optical cavity~\cite{cao2017experimental,hill2020effects,garbin2020asymmetric}, these operations would correspond to a coupling between the two polarization eigenmodes of the cavity, as directly realized by means of quarter-wave plates~\cite{kockaert2006fast,kozyreff2006fast}; see Fig.~\ref{fig_sketch}(a) and Section~\ref{section_exp}.

More generally, when the mixing processes in Eqs.~\eqref{eq_exchange}-\eqref{eq_exchange_2} cannot be directly performed by a device, they can be realized by activating a linear coupling between the two modes, during a short pulse duration $\tau \ll T$, such that the equations of motion of the driven system can be written in the form
\begin{align}
&i \frac{\partial \psi_1}{\partial t} = \left ( - \gamma \frac{\partial^2}{\partial x^2} + \vert \psi_1 \vert^2 + \beta \vert \psi_2 \vert^2   \right ) \psi_1 - \frac{\Omega (t)}{2} \psi_2, \label{NLS_time}\\
&i \frac{\partial \psi_2}{\partial t} = \left ( -\gamma \frac{\partial^2}{\partial x^2} + \vert \psi_2 \vert^2 + \beta \vert \psi_1 \vert^2   \right ) \psi_2 - \frac{\Omega (t)}{2} \psi_1. \notag
\end{align}
Here, the function $\Omega(t)\!=\!\Omega_0 + f_{\text{pulse}}(t)$ includes the pulse sequence defined by the function 
\begin{align}
f_{\text{pulse}}(t) &= (+\pi/2 + 2 \pi \mathfrak{p})/\tau \qquad  t_{\frak{n}}^{\oplus}-\tau \le t \le t_{\frak{n}}^{\oplus}, \notag\\
&= (-\pi/2 + 2 \pi \mathfrak{p})/\tau \hspace{0.7cm}  t_{\frak{n}}^{\ominus}-\tau \le t \le t_{\frak{n}}^{\ominus}, \notag\\
&=0 \hspace{3.cm} \text{otherwise} , \label{pulse}
\end{align}
where $t_{\frak{n}}^{\oplus}=(\mathfrak{n}+\alpha)T$ and $t_{\frak{n}}^{\ominus}=(\mathfrak{n}+1)T$ denote the successive pulse activation times, with $\mathfrak{n}=0,1,2 \dots $; see Fig.~\ref{fig_sketch}(b) and Fig.~\ref{fig_sequence}(a). The pulse function in Eq.~\eqref{pulse} also includes an arbitrary integer, $\mathfrak{p}\in \mathbb{Z}$, which can be chosen based on practical constraints; for instance, it can be set such that the linear coupling $\Omega(t)$ never changes sign over time, which can be convenient for certain physical realizations; see Figs.~\ref{fig_sketch}(b)-(d) and Section~\ref{section_exp}.

To verify that the drive in Eqs.~\eqref{NLS_time}-\eqref{pulse} indeed realizes the mixing operations in Eqs.~\eqref{eq_exchange}-\eqref{eq_exchange_2}, we restrict ourselves to the (linear) driving terms in the coupled Schrödinger equations~\eqref{NLS_time} and we obtain the time-evolution operators corresponding to the first and second pulses, respectively:
\begin{align}
&\hat U (t_{\frak{n}}^{\oplus}; t_{\frak{n}}^{\oplus} - \tau)=e^{i \frac{\pi}{4} \hat \sigma_x}\equiv \hat U_{\text{mix}},\label{pi_over_two}\\
&\hat U (t_{\frak{n}}^{\ominus}; t_{\frak{n}}^{\ominus} - \tau)=e^{-i \frac{\pi}{4} \hat \sigma_x} = \hat U_{\text{mix}}^{\dagger},\notag
\end{align}
where $\hat \sigma_x$ is the standard Pauli matrix. The operators $\hat U_{\text{mix}}$ and $\hat U_{\text{mix}}^{\dagger}$ in Eq.~\eqref{pi_over_two} indeed realize the mixing operations in Eqs.~\eqref{eq_exchange}-\eqref{eq_exchange_2}, respectively. We note that these mixing operations are known as $\pi/2$ pulses in quantum optics~\cite{gardiner2014quantum,kitagawa1993squeezed,gross2010nonlinear,liu2011spin}. 


In the limit of a fast pulse sequence, namely, when the period of the drive $T\ll T_{\text{eff}}$ is much smaller than the effective ``time" scale of the system (to be discussed below), we find that the stroboscopic time-evolution of the nonlinear system is well described by an \emph{effective} NLSE with modified nonlinearities. Following the method detailed in Section~\ref{section_quantum}, this effective NLSE reads
\begin{align}
&i \frac{\partial \psi_1}{\partial t} = \left ( - \gamma \frac{\partial^2}{\partial x^2} + U_1 \vert \psi_1 \vert^2 + U_2 \vert \psi_2 \vert^2    \right ) \psi_1 \notag \\
&\hspace{1.2cm} +U_3 \psi_1^* \psi_2^2 - \frac{\Omega_0}{2} \psi_2 , \notag\\
&i \frac{\partial \psi_2}{\partial t} = \left ( -\gamma \frac{\partial^2}{\partial x^2} + U_1 \vert \psi_2 \vert^2  + U_2 \vert \psi_1 \vert^2  \right ) \psi_2 \notag \\
&\hspace{1.2cm}+U_3 \psi_2^* \psi_1^2 - \frac{\Omega_0}{2} \psi_1 ,\label{NLS_effective_generalized}
\end{align}
where the three types of nonlinearities are controlled by the parameters
\begin{align}
&U_1= (3\alpha -1)/2, \notag \\
&U_2= \beta (3\alpha -1)/2 , \notag \\
&U_3= (\alpha-1) (1-\beta)/2 . \label{g_def}
\end{align}
In this framework, the system is assumed to be measured stroboscopically at times $t\!=\! T\!\times\!\frak{n}$, with $\frak{n}\!\in\!\mathbb{N}$. 

Comparing Eqs.~\eqref{NLS_effective_generalized}-\eqref{g_def} with the original Eq.~\eqref{NLS}, we find that the repeated mixing processes in Eqs.~\eqref{eq_exchange}-\eqref{eq_exchange_2} effectively produce a new form of nonlinearity, commonly known in optics as four-wave mixing~\cite{agrawal2001applications,delque2007polarization,agrawal2012nonlinear}. The drive also renormalizes the initial nonlinearities (self-phase and cross-phase modulations) by a same factor $(3\alpha-1)/2$. We point out that the effective four-wave mixing is induced even in the limit of two initially decoupled modes ($\beta\!=\!\Omega_0\!=\!0$). We also remark that the NLSE in Eq.~\eqref{NLS} is recovered in the limit $\alpha\!=\!1$, corresponding to a non-driven system.

We stress that the nonlinear system described by the NLSE in Eq.~\eqref{NLS_effective_generalized} is assumed to be lossless, such that $\vert \psi_1 \vert^2 + \vert \psi_2 \vert^2\!=\!N$ is a constant. Under this contraint, one can add any arbitrary constant $\mathfrak{c}$ to the self-phase and cross-phase modulations, $(U_1,U_2) \longrightarrow (U_1 + \mathfrak{c}, U_2 + \mathfrak{c})$, without affecting the physics. In particular, this implies that the pathological case $\beta\!=\!1$ always trivializes to a linear problem.

As another technical note, we point out that the mixing processes in Eqs.~\eqref{eq_exchange}-\eqref{eq_exchange_2} do not modify the kinetic-energy terms in Eq.~\eqref{NLS}. For the sake of presentation, we henceforth set $\gamma\!=\!0$ (except otherwise stated), but we do keep in mind that these terms can be readily added in the description without affecting the results~\cite{footnote}.

It is the aim of the following Sections~\ref{section_quantum}-\ref{section_numerics} to demonstrate the effective description displayed in Eq.~\eqref{NLS_effective_generalized} and to explore its regimes of validity, using analytical and numerical methods. Section~\ref{section_pendulum} analyzes how tuning the relative strengths of effective nonlinearities [Eq.~\eqref{g_def}] can induce topological changes in phase space, hence leading to strong modifications of the dynamics. We then generalize our approach to lattice systems in Section~\ref{section_lattice}.

\section{Quantum many-body approach to drive-induced nonlinearities}\label{section_quantum}

Our approach consists in three successive steps [Fig.~\ref{fig_approach}]:
\begin{itemize}
\item We introduce a parent quantum many-body Hamiltonian, whose semiclassical dynamics reproduces the time evolution of the driven nonlinear system in Eq.~\eqref{NLS_time}; 
\item Within this quantum framework, we derive the effective (Floquet) Hamiltonian that well captures the long time dynamics in the high-frequency limit ($2\pi/T\rightarrow\infty$); \item We then obtain the effective classical equations of motion (i.e.~the effective NLSE) from the effective quantum Hamiltonian. 
\end{itemize}

The validity of this approach will then be verified in Section~\ref{section_numerics}, through numerical studies of both quantum and classical dynamics. 

We point out that the periodically-driven NLSE has been widely explored in optics~\cite{yang1997analysis,towers2002stable,szameit2009inhibition} and in cold atoms~\cite{holthaus2001towards,abdullaev2003controlling,saito2003dynamically,matuszewski2005fully,kramer2005parametric,susanto2008effects,lellouch2017parametric,higashikawa2018floquet} using other theoretical methods.

\subsection{The parent quantum many-body system}

\begin{figure}[h!]
\includegraphics[width = \linewidth]{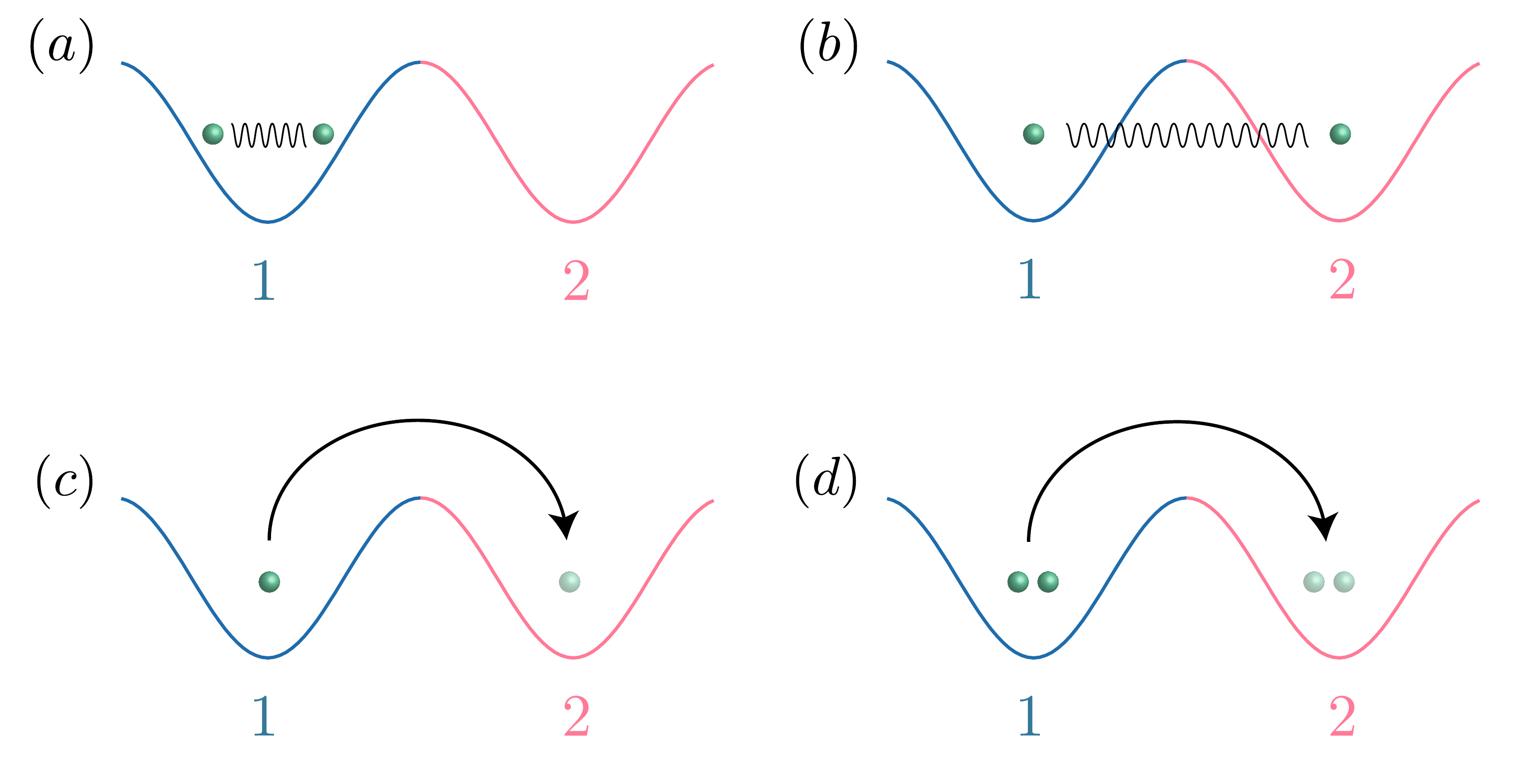}
\caption{Processes in the Hamiltonian in Eq.~\eqref{eq_parent_static}:~(a) Intra-mode (Hubbard) interactions; (b) inter-mode (cross) interactions; and (c) single-particle hopping processes.~(d) The effective Hamiltonian in Eq.~\eqref{eff_bosonic} includes pair-hopping processes, by which two interacting particles in the same mode simultaneously change mode. In this illustration, the two modes $1$ and $2$ correspond to the low-energy orbitals of a double-well potential, and the bosons are represented by green spheres.
}
\label{fig_process}
\end{figure}

Our starting point is the quantum many-body Hamiltonian
\begin{align}
\hat H_0 =& \frac{1}{2} \left ( \hat a_1^{\dagger} \hat a_1^{\dagger} \hat a_1  \hat a_1 + \hat a_2^{\dagger} \hat a_2^{\dagger} \hat a_2 \hat a_2 \right ) \notag \\
&+ \beta \hat a_1^{\dagger} \hat a_2^{\dagger} \hat a_1  \hat a_2 - \frac{\Omega_0}{2} \left (\hat a_1^{\dagger} \hat a_2 + \hat a_2^{\dagger} \hat a_1  \right),\label{eq_parent_static}
\end{align}
where $\hat a_{s}^{\dagger}$ (resp.~$\hat a_{s}$) creates (resp. annihilates) a boson in the mode $s\!=\!1,2$. These operators satisfy the bosonic commutation relations, $[\hat{a}_{s},\hat{a}_{s'}^{\dagger}]=\delta_{s,s'}$. The first line in Eq.~\eqref{eq_parent_static} describes intra-mode (Hubbard) interactions, while the second line describes inter-mode (cross) interactions of strength $\beta$; the Hamiltonian also includes single-particle hopping processes of amplitude $\Omega_0/2$; see Fig.~\ref{fig_process}(a)-(c) for a sketch of the processes and Refs.~\cite{bloch2008many,dutta2015non}. Henceforth, the bare Hubbard interaction strength sets our unit of energy and time.

First of all, we note that the classical equations of motion (NLSE) in Eq.~\eqref{NLS} are readily obtained from Heisenberg's equations, $d \hat a_{s}/dt = i [\hat H_0 , \hat a_{s}]$, upon taking the classical limit $\hat a_{1,2} \rightarrow \psi_{1,2}$; see Refs.~\cite{pitaevskii2016bose,holthaus2001towards,carusotto2013quantum,cao2020reconfigurable}. Specifically, the self-phase modulation in Eq.~\eqref{NLS} stems from the intra-mode (Hubbard) interaction terms in Eq.~\eqref{eq_parent_static}, while the cross-phase modulation stems from the inter-mode (cross) interaction term. Hence, this justifies the choice of Eq.~\eqref{eq_parent_static} as a proper parent quantum Hamiltonian for our initial (non-driven) system. Note that we set $\hbar\!=\!1$ throughout this work.

In fact, for the sake of later convenience, it is instructive to derive the NLSE in Eq.~\eqref{NLS} using a different approach. Indeed, this will allow us to introduce central notions and quantities, which will be used throughout this work. Let us introduce a set of angular momentum (Schwinger) operators~\cite{auerbach2012interacting}, defined as
\begin{align}
&\hat J_x = \frac{1}{2} \left (\hat a_1^{\dagger} \hat a_2 + \hat a_2^{\dagger} \hat a_1 \right ) , \quad \hat J_y = \frac{1}{2i} \left (\hat a_2^{\dagger} \hat a_1 - \hat a_1^{\dagger} \hat a_2 \right ), \notag \\
&\hat J_z = \frac{1}{2} \left (\hat a_2^{\dagger} \hat a_2 - \hat a_1^{\dagger} \hat a_1 \right ) , \quad \hat N = \hat a_1^{\dagger} \hat a_1 + \hat a_2^{\dagger} \hat a_2. \label{Schwinger}
\end{align}
These angular-momentum operators satisfy the spin commutation relations $[\hat J_{\mu}, \hat J_{\nu}]=i\varepsilon_{\mu \nu \lambda} \hat J_{\lambda}$, and the operator $\hat N$ simply counts the total number of bosons in the system (assumed to be constant). We note that these operators also satisfy the sum rule
\be
\hat J_x^2 + \hat J_y^2 + \hat J_z^2= \frac{\hat N}{2} \left (\frac{\hat N}{2} + 1\right ), \label{sum_rule}
\ee
which is a conserved quantity. For a single boson ($N\!=\!1$), we have $\hat J_{\mu}\!=\!\hat \sigma_{\mu}/2$, where $\hat \sigma_{x,y,z}$ denote the Pauli matrices.

Using the operators in Eq.~\eqref{Schwinger}, the parent Hamiltonian in Eq.~\eqref{eq_parent_static} simply reads
\begin{equation}
\hat H_0 = \chi \hat J_z^2 - \Omega_0 \hat J_x + \text{constant}, \qquad \chi\!=\!1 - \beta .\label{Josephson}
\end{equation}
We henceforth neglect the constant terms, which are proportional to $\hat N$ and $\hat N^2$; see Appendix~\ref{sect_app}. We note that the Hamiltonian in Eq.~\eqref{Josephson} has been extensively studied in the context of the bosonic Josephson effect~\cite{smerzi1997quantum,pereverzev1997quantum,raghavan1999coherent,albiez2005direct,levy2007ac,zibold2010classical,zibold2012classical} and  nuclear physics~\cite{lipkin1965validity}. From Eq.~\eqref{Josephson}, we also recover that the pathological case $\beta\!=\!1$ trivializes to a non-interacting problem ($\chi\!=\!0$), as previously noted in Section~\ref{section_two_modes}.


The equations of motion associated with Eq.~\eqref{Josephson} are readily obtained from Heisenberg's equations
\begin{align}
&\frac{d \hat J_z (t)}{dt} = i [ \hat H_0, \hat J_z (t) ] = - \Omega_0 \hat J_y (t), \label{eq_spin} \\
&\frac{d \hat J_y (t)}{dt} = i [ \hat H_0, \hat J_y (t) ] = \Omega_0 \hat J_z (t) \notag \\
&\hspace{3.5cm}+ \chi \left (\hat J_z (t)\hat J_x (t) + \hat J_x (t)\hat J_z (t) \right ). \notag
\end{align}

In order to connect Eq.~\eqref{eq_spin} to the classical NLSE in Eq.~\eqref{NLS}, we take the classical limit and introduce the Bloch-Poincar\'e sphere representation $(\theta,\varphi)$ through the mapping~\cite{zibold2012classical}
\begin{align}
&\hat J_x \rightarrow \frac{N}{2}\sqrt{1-z^2} \cos \varphi , \quad \hat J_y \rightarrow -\frac{N}{2}\sqrt{1-z^2} \sin \varphi ,\notag\\
&\hat J_z \rightarrow - \frac{N}{2} z, \hspace{2cm} z=\cos \theta . \label{mapping}
\end{align}
We note that this Bloch-sphere representation relies on Eq.~\eqref{sum_rule} and particle conservation. Injecting this Eq.~\eqref{mapping} into Eq.~\eqref{eq_spin}, one obtains the classical equations of motion
\begin{align}
&\dot z=- \Omega_0 \sqrt{1-z^2} \sin \varphi , \notag \\
&\dot \varphi= N \chi z + \Omega_0 \frac{z}{\sqrt{1-z^2}} \cos \varphi , \label{pendulum}
\end{align}
for the two canonical conjugate variables $z(t)$ and $\varphi(t)$~\cite{zibold2010classical,smerzi1997quantum,raghavan1999coherent}. 

We point out that Eq.~\eqref{pendulum} is equivalent to the NLSE in Eq.~\eqref{NLS} upon representing the complex amplitudes $\psi_{1,2}$ on the Bloch-Poincar\'e sphere~\cite{paraoanu2001josephson,cao2020reconfigurable}
\begin{align}
& \psi_1=\sqrt{N}\cos (\theta/2) = \sqrt{\frac{N}{2}+n}, \notag \\
& \psi_2=\sqrt{N}\sin (\theta/2)\, e^{i \varphi} =\sqrt{\frac{N}{2}-n} \, e^{i \varphi},\label{psi_z_phi}
\end{align}
where we introduced the relative phase $\varphi$ between the two modes, the relative population (or relative light intensity)
\begin{equation}
z=\cos \theta = \frac{2n}{N}=\left ( \vert \psi_1 \vert^2 - \vert \psi_2 \vert^2  \right)/N,\label{z_eq}
\end{equation}
and the total population (or total light intensity) 
\begin{equation}
N=\vert \psi_1 \vert^2 + \vert \psi_2 \vert^2. 
\end{equation}
We emphasize that the dynamics in phase space, i.e.~the trajectories ($z(t),\varphi(t))$ resulting from Eq.~\eqref{pendulum}, can be simply monitored in experiments by measuring the relative population (intensity) and relative phase of the two modes; see also Section~\ref{section_exp} for a more detailed discussion.

For the sake of completeness, we note that the equations of motion in Eq.~\eqref{pendulum} can be derived from Hamilton's equation, using the classical Hamiltonian~\cite{zibold2010classical,smerzi1997quantum,di2019nonlinear}
\begin{equation}
\mathcal{H}_0 (z,\varphi)= \frac{\chi N}{2} z^2 - \Omega_0 \sqrt{1-z^2} \cos \varphi .\label{classical_ham}
\end{equation}
The classical dynamics of the non-driven system hence relies on a competition between the ``mean-field" interaction parameter $g\!=\!\chi N$ and the linear coupling $\Omega_0$. This competition is at the core of bifurcations and symmetry breaking in bosonic Josephson junctions~\cite{smerzi1997quantum,raghavan1999coherent,zibold2010classical,cao2017experimental,cao2020reconfigurable}. These concepts will be further discussed in Section~\ref{section_pendulum}.


\subsection{The pulse sequence and the effective Floquet Hamiltonian}\label{sect_effective}

We now introduce the quantum-many-body analogue of the pulse sequence introduced in Section~\ref{section_two_modes}; see Fig.~\ref{fig_sequence}. We write the time-evolution operator over one period $T$ in the form [Fig.~\ref{fig_sequence}(b)]
\begin{equation}
\hat U (T;0) = \hat U_{\text{mix}}^{\dagger} \, 
e^{-i (1- \alpha) T \hat H_0} \hat U_{\text{mix}} \, e^{-i \alpha T \hat H_0},\label{U_sequence}
\end{equation}
where the mixing operator is defined as
\begin{equation}
\hat U_{\text{mix}}=e^{i \frac{\pi}{2} \hat J_x},\label{mix_operator}
\end{equation}
and we remind that $\alpha\!\in\![0,1]$ is a tunable parameter.

We note that the operator $\hat U_{\text{mix}}$ in Eq.~\eqref{mix_operator} indeed corresponds to the $\pi/2$-pulse operator in Eq.~\eqref{pi_over_two} for a single boson ($N\!=\!1$), which is consistent with the fact that the mixing operation is a single-particle process. When writing Eq.~\eqref{U_sequence}, we explicitly took the limit $\tau\!\rightarrow\!0$, where $\tau$ is the pulse duration; see Eq.~\eqref{pulse}.

The state of the quantum many-body system at time $t_{\frak{n}}\!=\!T\!\times\!\frak{n} $ is then obtained as
\begin{equation}
\vert \psi (t_{\frak{n}}) \rangle = \hat U (t_{\frak{n}}; 0) \vert \psi (0) \rangle = \left (\hat U (T;0) \right )^{\frak{n}}\vert \psi (0) \rangle ,\label{time_evolution}
\end{equation}
where $\vert \psi (0) \rangle$ denotes the initial state of the system.

We now derive the effective (Floquet) Hamiltonian~\cite{goldman2014periodically,goldman2015periodically,bukov2015universal}, which captures the stroboscopic dynamics of the driven system, and hence, its time evolution over long time scales $t_{\frak{n}}\!\gg\!T$. The effective Hamiltonian is defined through the time-evolution operator over one period~\cite{kitagawa2010topological,goldman2014periodically}
\begin{equation}
\hat U (T;0) = e^{-i T \hat H_{\text{eff}}},\label{effective_definition}
\end{equation}
and it can be evaluated explicitly through a $1/\omega$-expansion, where $\omega\!=\!2\pi/T$ denotes the drive frequency; see Refs.~\cite{goldman2014periodically,goldman2015periodically,bukov2015universal,eckardt2015high,mikami2016brillouin}. In order to reach convergence of this infinite series expansion, we partially resum the series~\cite{goldman2014periodically} by splitting the time-evolution operator in Eq.~\eqref{U_sequence} into two parts
\begin{equation}
\hat U (T;0) = e^{-i (1-\alpha)T \hat H_1} e^{-i  \alpha T \hat H_0},\label{split_eq}
\end{equation}
where we introduced the operator $\hat H_1$ defined as
\begin{equation}
e^{-i t \hat H_1}  \equiv  e^{-i \frac{\pi}{2} \hat J_x} 
e^{-i t \hat H_0} e^{i \frac{\pi}{2} \hat J_x}.\label{H1_def}
\end{equation}
The time-evolution operator in Eq.~\eqref{H1_def} has a simple interpretation:~it corresponds to free time-evolution in a rotated basis.

Then, assuming that $T \omega_{\text{eff}}\!\ll\!1$, where $\omega_{\text{eff}}$ is the characteristic frequency associated with the processes included in the Hamiltonians $\hat H_{0}$ and $\hat H_{1}$, we apply the Trotter approximation to Eq.~\eqref{split_eq},
\begin{equation}
\hat U (T;0) \approx e^{-i T \left ( \alpha \hat H_0+ (1 - \alpha) \hat H_1 \right )},\label{Trotter}
\end{equation}
from which we directly obtain the effective Hamiltonian [Eq.~\eqref{effective_definition}]
\begin{equation}
\hat H_{\text{eff}}= \alpha \hat H_0+ (1-\alpha) \hat H_1 + \mathcal{O} (T).\label{eff_interm}
\end{equation}

Our problem of finding the effective Hamiltonian thus reduces to the calculation of $\hat H_1$ defined in Eq.~\eqref{H1_def}. This step can be performed exactly, by noting that
\begin{equation}
\hat H_1 = e^{-i \frac{\pi}{2} \hat J_x} \hat H_0 \, e^{i \frac{\pi}{2} \hat J_x}= \chi \left (e^{-i \frac{\pi}{2} \hat J_x} \hat J_z^2 \, e^{i \frac{\pi}{2} \hat J_x} \right ) - \Omega_0 \hat J_x ,\notag
\end{equation}
where we used the definition of $\hat H_0$ in Eq.~\eqref{Josephson}.

Using the Baker-Campbell-Hausdorff formula, one obtains~\cite{kidd2019quantum}
\begin{equation}
e^{-i \frac{\pi}{2} \hat J_x} \hat J_z^2 \, e^{i \frac{\pi}{2} \hat J_x} = \hat J_y^2,\label{BCH}
\end{equation}
such that 
\begin{equation}
\hat H_1 = \chi \hat J_y^2 - \Omega_0 \hat J_x.\label{H1}
\end{equation}
The effective Hamiltonian in Eq.~\eqref{eff_interm} finally reads
\begin{equation}
\hat H_{\text{eff}}=\chi \left ( \alpha \hat J_z^2 + (1-\alpha) \hat J_y^2 \right ) - \Omega_0 \hat J_x + \mathcal{O} (T).\label{eff_final}
\end{equation}
From this result, we find that the Trotter approximation [Eq.~\eqref{Trotter}] is valid for a sufficiently short driving period satisfying $T\!\ll\!1/\chi$ and $T\!\ll\!1/\Omega_0$. 

The Hamiltonian displayed in Eq.~\eqref{eff_final} involves unconventional interactions, which are known as two-axis twisting interactions in the context of quantum optics~\cite{kitagawa1993squeezed,ma2011quantum}. They have been proposed in view of creating squeezed spin states with optimal squeezing in various physical settings~\cite{kitagawa1993squeezed,liu2011spin,ma2011quantum,huang2015two,kajtoch2015quantum,kajtoch2016spin,he2019engineering,groszkowski2020heisenberg,roscilde2021spin,yanes2022one,kruckenhauser2022high}, as we further discuss in Section~\ref{section_husimi}.

\subsubsection{A few limiting cases}

At this stage, it is insightful to analyze the effective Hamiltonian in Eq.~\eqref{eff_final} for a few limiting cases:

\begin{itemize}

\item When $\alpha\!=\!1$, one finds $\hat H_{\text{eff}}\!=\!\hat H_0$, which reflects the triviality of the sequence in Eq.~\eqref{U_sequence} in this case. 

\item When $\alpha\!=\!0$, one finds the effective Hamiltonian
\begin{align}
\hat H_{\text{eff}} = \chi \hat J_y^2 - \Omega_0 \hat J_x = e^{-i \frac{\pi}{2} \hat J_x} (\hat H_0) \, e^{i \frac{\pi}{2} \hat J_x} ,\label{different_frame}
\end{align}
which is thus strictly equivalent to the non-driven Hamiltonian $\hat H_0$ up to a unitary transformation [Eq.~\eqref{BCH}]:~the Hamiltonians $\hat H_0$ and $\hat H_{\text{eff}}$ share the same spectrum. In this case, the driving sequence simply generates an initial and final kick~\cite{goldman2014periodically}, as can be deduced by explicitly writing the time-evolving state at some arbitrary stroboscopic time $t\!=\!t_{\frak{n}}$ [Eq.~\eqref{time_evolution}]
\begin{align}
\vert \psi (t_{\frak{n}}) \rangle &= \left (\hat U_{\alpha=0} (T;0) \right )^{\frak{n}}\vert \psi (0) \rangle ,\notag \\
&=e^{-i \frac{\pi}{2} \hat J_x} e^{-i t_{\frak{n}} \hat H_0} e^{i \frac{\pi}{2} \hat J_x} \vert \psi (0) \rangle.\label{time_evolution_kick}
\end{align}
The long-time dynamics in Eq.~\eqref{time_evolution_kick} is indeed dictated by the static Hamiltonian $\hat H_0$, but it is also affected by the initial and final kicks, $e^{\pm i \frac{\pi}{2} \hat J_x}$, associated with the change of basis (rotation on the Bloch-Poincar\'e sphere)~\footnote{In optics, this change of basis corresponds to a change of representation, from linear polarization modes to circular-polarization modes.}. 
In Section~\ref{section_lattice}, we will see that this situation can nonetheless lead to intriguing phenomena upon coupling individual dimers in a time-periodic manner [Fig.~\ref{fig_lattice_effective}(b)].

\item When $\alpha=1/2$, the effective Hamiltonian reads
\begin{align}
\hat H_{\text{eff}} &= \frac{\chi}{2} \left (\hat J_y^2 + \hat J_z^2 \right ) - \Omega_0 \hat J_x + \mathcal{O} (T),\notag\\
&=-\frac{\chi}{2} \hat J_x^2 - \Omega_0 \hat J_x + \mathcal{O} (T),\label{special_symmetry}
\end{align}
where we used the sum rule~\eqref{sum_rule} and omitted constant terms. In this case, the system displays the special symmetry $[\hat H_{\text{eff}}, \hat J_x]\!=\!0$, such that the many-body eigenstates and energies can be written exactly. 

\end{itemize}

\subsubsection{The effective Hamiltonian in the bosonic representation}

It is instructive to rewrite the effective Hamiltonian in Eq.~\eqref{eff_final} using the original bosonic operators [Appendix~\ref{sect_app}],
\begin{align}
\hat H_{\text{eff}} &= \frac{U_1}{2} \left ( \hat a_1^{\dagger} \hat a_1^{\dagger} \hat a_1  \hat a_1 + \hat a_2^{\dagger} \hat a_2^{\dagger} \hat a_2 \hat a_2 \right ) \notag \\
&+U_2 \left ( \hat a_1^{\dagger} \hat a_2^{\dagger} \hat a_1  \hat a_2 \right )\notag \\
&+\frac{U_3}{2} \left ( \hat a_1^{\dagger} \hat a_1^{\dagger} \hat a_2  \hat a_2 + \hat a_2^{\dagger} \hat a_2^{\dagger} \hat a_1 \hat a_1 \right )\notag \\
& -\frac{\Omega_0}{2} \left (\hat a_1^{\dagger} \hat a_2 + \hat a_2^{\dagger} \hat a_1  \right) + \mathcal{O} (T) , \label{eff_bosonic}
\end{align}
where the interaction strengths are given in Eq.~\eqref{g_def}.

A comparison with the initial Hamiltonian $\hat H_0$ in Eq.~\eqref{eq_parent_static} indicates that the driving pulse sequence has effectively generated novel interaction terms; see the third line of Eq.~\eqref{eff_bosonic}. These pair-hopping terms~\cite{liang2009atom,dutta2015non,anisimovas2015role,jurgensen2015twisted,luhmann2016twisted,hemmerich2019bosons,lin2020novel}, which stem from the $\hat J_y^2$ interactions in Eq.~\eqref{eff_final}, describe processes by which two particles in mode $s$ collide and end up in the other mode $s'\!\ne\!s$; see Fig.~\ref{fig_process}(d). As we now discuss below in Section~\ref{sect_classical_eff}, these pair-hopping terms are at the origin of the four-wave mixing nonlinearity announced in Eq.~\eqref{NLS_effective_generalized}.

As a technical remark, we remind the reader that adding a constant shift to the intra-mode (Hubbard) and inter-mode interactions, $(U_1,U_2) \longrightarrow (U_1 + \mathfrak{c}, U_2 + \mathfrak{c})$, does not modify the physics, due to the number-conserving nature of the system. 

\subsection{Effective classical equations of motion}\label{sect_classical_eff}

First of all, we find that the effective NLSE in Eq.~\eqref{NLS_effective_generalized} is directly obtained from the effective Hamiltonian $\hat H_{\text{eff}}$ in Eq.~\eqref{eff_bosonic}, using Heisenberg’s equations $d \hat a_{s}/dt = i [\hat H_{\text{eff}} , \hat a_{s}]$, and upon taking the classical limit $\hat a_{1,2} \rightarrow \psi_{1,2}$. In particular, the effective four-wave mixing in Eq.~\eqref{NLS_effective_generalized} originates from the effective pair-hopping terms in Eq.~\eqref{eff_bosonic}.

In analogy with Eqs.~\eqref{eq_spin}-\eqref{pendulum}, we now explicitly derive the classical equations of motion for the two canonically conjugate variables $z(t)$ and $\varphi(t)$, describing the relative population and phase of the two modes. Using the effective Hamiltonian in Eq.~\eqref{eff_final} and Heisenberg’s equations, we find
\begin{align}
&\frac{d \hat J_z (t)}{dt} = i [ \hat H_{\text{eff}}, \hat J_z (t) ] = - \Omega_0 \hat J_y (t) \notag \\
&\hspace{2.8cm}-(1-\alpha)\chi  \left (\hat J_y (t)\hat J_x (t) + \hat J_x (t)\hat J_y (t) \right ), \notag \\
&\frac{d \hat J_y (t)}{dt} = i [ \hat H_{\text{eff}}, \hat J_y (t) ] = \Omega_0 \hat J_z (t) \notag \\
&\hspace{2.8cm}+\alpha\chi \left (\hat J_z (t)\hat J_x (t) + \hat J_x (t)\hat J_z (t) \right ). \notag
\end{align}
Applying the Bloch-Poincar\'e-sphere mapping [Eq.~\eqref{mapping}], we obtain the classical equations of motion
\begin{align}
&\dot z=- \chi N (1-\alpha) (1-z^2) \cos \varphi \sin \varphi - \Omega_0 \sqrt{1-z^2} \sin \varphi, \notag \\
&\dot \varphi= \chi N z \left (\alpha - (1-\alpha)  \sin^2 \varphi \right ) + \Omega_0 \frac{z}{\sqrt{1-z^2}} \cos \varphi.
\label{pendulum_eff_imbalanced}
\end{align}
The classical equations of motion in Eq.~\eqref{pendulum_eff_imbalanced} are physically equivalent to the effective NLSE announced in Eq.~\eqref{NLS_effective_generalized}, through the mapping provided by Eq.~\eqref{psi_z_phi}. One verifies that Eq.~\eqref{pendulum_eff_imbalanced} reduces to the equations of motion of the undriven system [Eq.~\eqref{pendulum}] in the limit $\alpha\!=\!1$.

Drive-induced nonlinearities, which are controlled by the parameter $\alpha$, strongly affect the dynamics of the two-mode system, as we now analyze in the following Section~\ref{section_pendulum}.

\section{Drive-induced nonlinear dynamics \\ and the pendulum analogy}\label{section_pendulum}
%
%

\subsection{Symmetries, phase-space topology and transitions}

First of all, we find that the equations of motion in Eq.~\eqref{pendulum_eff_imbalanced} can be derived from Hamilton's equation, using the classical Hamiltonian
\begin{align}
&\mathcal{H}_{\text{eff}}(z,\varphi;\alpha)=- \Omega_0 \sqrt{1-z^2} \cos \varphi  \label{classical_eff_imbalanced} \\
&\hspace{1.2cm}+\frac{\chi N}{2} \alpha z^2 + \frac{\chi N}{2} (1-\alpha) (1- z^2)\, \sin^2 \varphi  .\notag
\end{align}
It is useful to note that the Hamiltonian in Eq.~\eqref{classical_eff_imbalanced}, and the resulting equations of motions in Eq.~\eqref{pendulum_eff_imbalanced} satisfy, for any value of the parameters, the following discrete symmetries:
\begin{align}
S_1: & \, z \to -z,\notag\\
S_2: & \, \varphi \to -\varphi  .\label{eq_symmetries}
\end{align}
We remind that $\varphi$ is defined modulo $2\pi$, given its angular nature, and that $z\!=\!\cos \theta$ is defined in the interval $[-1, 1]$. We also note that additional symmetries exist for specific values of the parameters~\footnote{For $\Omega_0\!=\!0$, there is the additional discrete symmetry:~$\varphi \to \varphi + \pi$. For $\alpha\!=\!1/2$, there is the additional continuous symmetry:~$\sqrt{1-z^2} \to c \sqrt{1-z^2},\, \cos(\varphi) \to \cos(\varphi)/c$ for any real $c>0$.}.
\begin{figure}[t!]
\includegraphics[width = \linewidth]{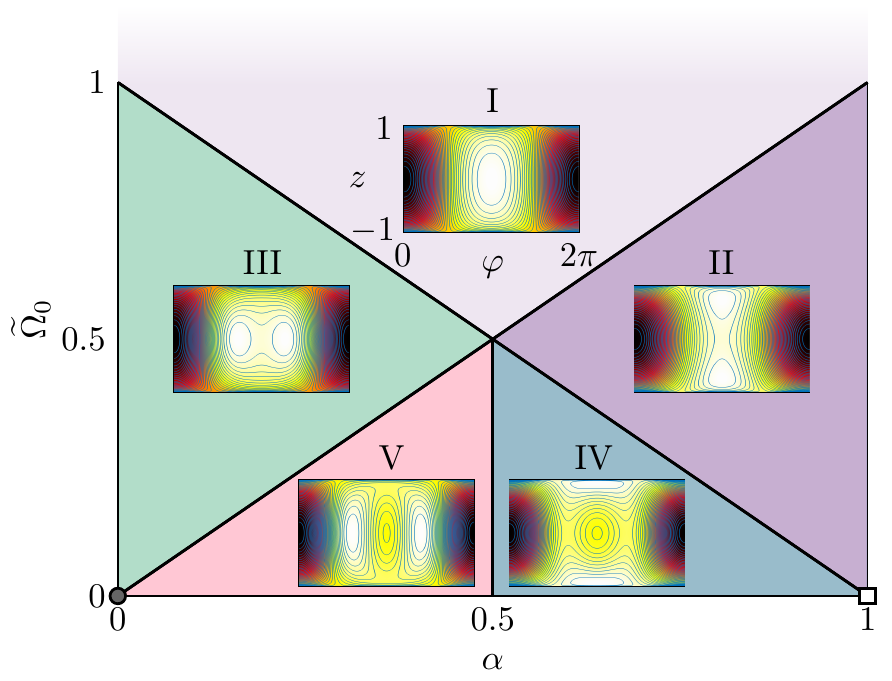}
\caption{Phase diagram associated with the effective classical Hamiltonian in Eq.~\eqref{classical_eff_imbalanced}, as a function of the drive-induced nonlinearity parameter $\alpha$ and the dimensionless linear coupling $\tilde\Omega_0\!=\!\Omega_0/(\chi N)$. Each phase is characterized by a distinctive phase-space topology (fixed-point configuration); see Eq.~\eqref{phases_def}. A few trajectories are indicated as thin blue curves (equipotential lines of the energy landscape) for each representative case. The axis used in the phase-space diagrams is shown in Phase I; the colormap is the same as in Fig.~\ref{fig_landscape_imbalanced}, where it is explicitly defined. The black circle (resp.~white square) at $(\alpha,\tilde\Omega_0)\!=\!(0, 0)$ [resp. $(\alpha,\tilde\Omega_0)\!=\!(1, 0)$] indicates the singular point at which only $\text{FP}_*$ (resp.~$\text{FP}_{\pm}$) are stable fixed points.}
\label{Fig:PhaseDiagram}
\end{figure}

The classical Hamiltonian $\mathcal{H}_{\text{eff}}(z,\varphi;\alpha)$ describes an energy landscape over the Bloch sphere, from which one can readily deduce all possible trajectories and related fixed points ($\dot z\!=\!\dot\varphi\!=\!0$)~\cite{zibold2010classical}. In order to reveal the impact of drive-induced nonlinearities on the dynamics, we determine a ``fixed-point phase diagram", as a function of the dimensionless parameters $\alpha$ and $\tilde\Omega_0\!=\!\Omega_0/(\chi N)$. Here, we identify a ``phase" as a region in parameter space that is characterized by a distinctive phase-space topology (fixed-point configuration); see Fig.~\ref{Fig:PhaseDiagram}.

By performing a stability analysis on the classical Hamiltonian~\eqref{classical_eff_imbalanced}, we obtain the following fixed points:
%
\begin{align}
    \text{FP}_0: \quad  & z=0, \varphi=0, \notag \\
    \text{FP}_\pi: \quad  & z=0, \varphi=\pi, \notag \\
    \text{FP}_*: \quad  & z=0, \varphi = \pm \arccos{[\widetilde{\Omega}_0/(1-\alpha)]}, \notag \\
    \text{FP}_\pm: \quad  & z=\pm \sqrt{1-\widetilde{\Omega}_0^2/ \alpha^2}, \varphi=\pi . \label{FP_eq}
\end{align}    
These fixed points can be stable or unstable, depending on the values of $\alpha$ and $\widetilde{\Omega}_0$. In particular, the emergence of certain fixed points can be associated with a spontaneous breaking of the aforementioned symmetries [Eq.~\eqref{eq_symmetries}]:~the fixed points FP$_\pm$ break $S_1$, while the fixed points FP$_*$ break $S_2$. We note that neither FP$_0$ nor FP$_\pi$ break a symmetry. 

From the stability analysis, we identify five distinct phases:
%
\begin{align}
    \text{Phase I:} &\quad \text{FP}_0, \text{FP}_\pi \,\text{stable} \notag \\   
    \text{Phase II:} &\quad \text{FP}_0, \text{FP}_\pm \,\text{stable} \qquad \qquad (\cancel{S_1}) \notag \\   
    \text{Phase III:} &\quad \text{FP}_0, \text{FP}_* \,\text{stable} \qquad \qquad \hspace{0.08cm}(\cancel{S_2}) \notag \\   
    \text{Phase IV:} &\quad \text{FP}_0, \text{FP}_{\pi}, \text{FP}_\pm  \,\text{stable} \hspace{0.67cm} (\cancel{S_1}) \notag \\
    \text{Phase V:} &\quad \text{FP}_0, \text{FP}_\pi, \text{FP}_* \,\text{stable}\hspace{0.75cm} (\cancel{S_2})\label{phases_def}
\end{align}
%
We note that a spontaneous symmetry breaking (involving either $S_1$ or $S_2$) occurs for every phase, except for Phase I. The complete phase diagram is displayed in Fig.~\ref{Fig:PhaseDiagram}. \\

At this stage, it is worth considering three limiting cases:

\begin{itemize}

\item When $\alpha\!=\!1$, the classical Hamiltonian $\mathcal{H}_{\text{eff}}(z,\varphi;\alpha)$ reduces to the bosonic Josephson Junction (BJJ) Hamiltonian $\mathcal{H}_{0}(z,\varphi)$ in Eq.~\eqref{classical_ham}. The BJJ model features a bifurcation point at $\widetilde{\Omega}_0=1$:~the fixed point $\text{FP}_{\pi}$, which is stable for $\widetilde{\Omega}_0>1$ (Phase I), becomes unstable for $\widetilde{\Omega}_0<1$, giving rise to two new stable fixed points $\text{FP}_{\pm}$ (Phase II); see Fig.~\ref{Fig:PhaseDiagram}. This transition, which is associated with the spontaneous breaking of the $S_1$ symmetry, was observed in cold atoms~\cite{zibold2010classical} and  microresonators~\cite{cao2017experimental}.

\item When $\alpha\!=\!0$, the system corresponds to the BJJ model in a rotated basis; see Eq.~\eqref{different_frame}. The aforementioned bifurcation then corresponds to a transition from Phase I ($\widetilde{\Omega}_0>1$) to Phase III ($\widetilde{\Omega}_0<1$), characterized by the two stable fixed points $\text{FP}_{*}$ and the spontaneous breaking of the $S_2$ symmetry; see Fig.~\ref{Fig:PhaseDiagram}.

\item When $\alpha\!=\!1/2$, the effective Hamiltonian satisfies the symmetry $[\hat H_{\text{eff}}, \hat J_x]\!=\!0$; see Eq.~\eqref{special_symmetry}. Classically, this corresponds to the following constant of motion 
\begin{equation}
C=\sqrt{1-z(t)^2} \cos \varphi(t).
\end{equation}
One verifies that this special constant of motion implies that the system remains in Phase I for any arbitrary value of $\widetilde{\Omega}_0$.
\end{itemize}

For any other values of $\alpha$, the system is allowed to enter two new phases (Phases IV and V), which are both characterized by four stable fixed points but are associated with different symmetry breaking; see Fig.~\ref{Fig:PhaseDiagram}. 

\begin{figure}[h!]
\includegraphics[width = \linewidth]{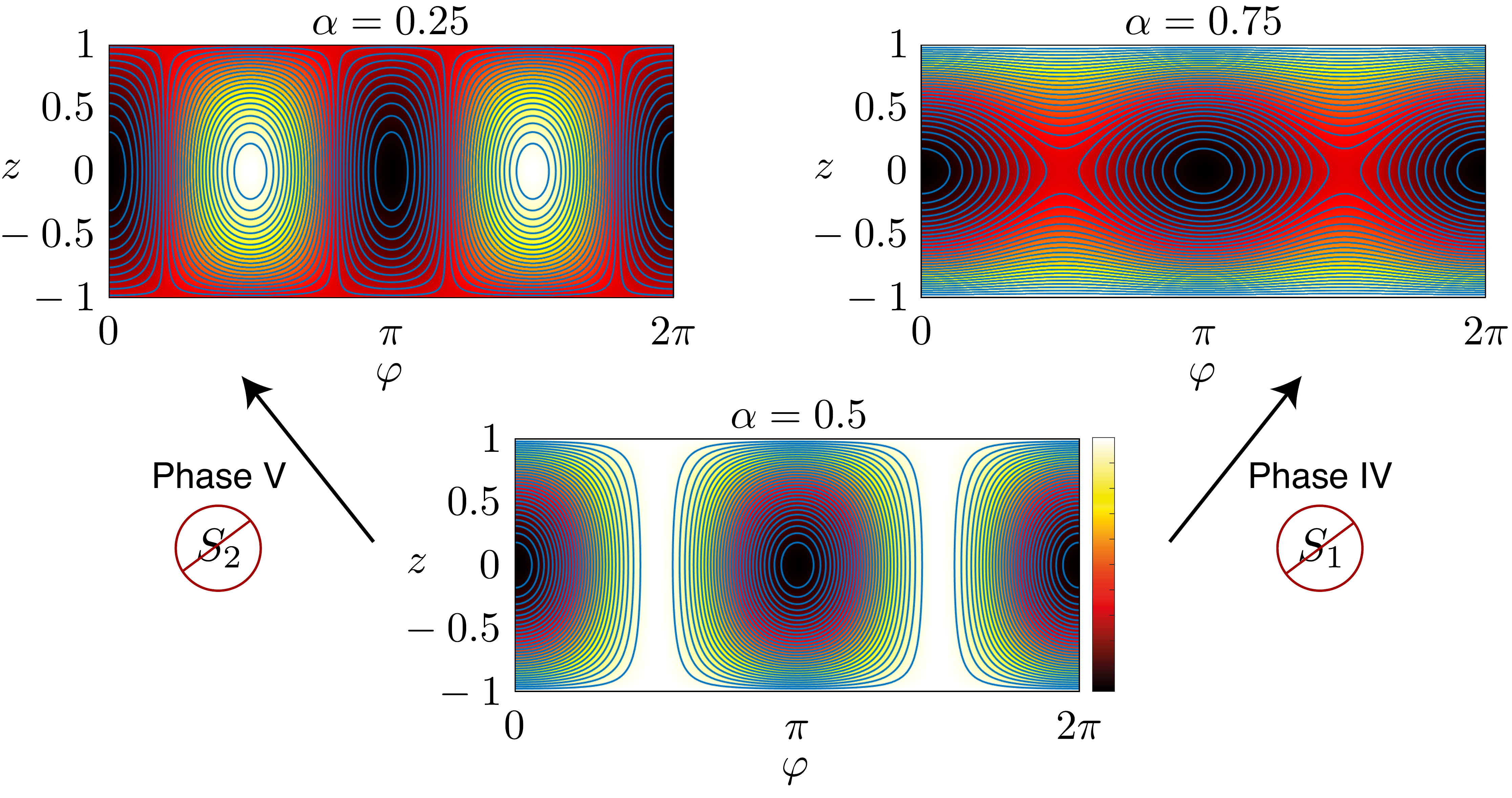}
\caption{Energy landscape associated with the effective classical Hamiltonian $\mathcal{H}_{\text{eff}}(z,\varphi;\alpha)$ displayed in Eq.~\eqref{classical_eff_imbalanced}, for vanishing linear coupling $\Omega_0\!=\!0$, and three different values of the drive-induced nonlinearity parameter: $\alpha\!=\!0.25$, $\alpha\!=\!0.5$ and $\alpha\!=\!0.75$. A few trajectories are indicated as thin blue curves (equipotential lines of the energy landscape) for each case. Note the emergence and disappearance of stable fixed points on the Bloch-Poincar\'e sphere, as the nonlinearity parameter $\alpha$ is varied:~Phase space undergoes a transition from Phase V ($\alpha\!=\!0.25$) to Phase I ($\alpha\!=\!0.5$) to Phase IV ($\alpha\!=\!0.75$); see Eq.~\eqref{phases_def}.
}
\label{fig_landscape_imbalanced}
\end{figure}

Importantly, one can induce transitions between these various phases by simply tuning the drive-induced nonlinearity parameter $\alpha$. This is illustrated in Fig.~\ref{fig_landscape_imbalanced}, which shows two successive transitions in the absence of linear coupling ($\Omega_0\!=\!0$):~When $\alpha\!=\!1/2$, the system is in Phase I, with two stable fixed points ($\text{FP}_{0,\pi}$) satisfying both symmetries $S_{1,2}$. Reducing the nonlinearity parameter ($\alpha\!<\!1/2$) stabilizes two additional fixed points $\text{FP}_{*}$, which break $S_2$ symmetry (Phase V). Instead, setting $\alpha\!>\!1/2$ stabilizes the two  fixed points $\text{FP}_{\pm}$, associated with the breaking of $S_1$ symmetry (Phase IV).

From the quantum effective Hamiltonian in Eq.~\eqref{eff_final}, we observe that the phase-space transitions (and related symmetry breaking) illustrated in Fig.~\ref{fig_landscape_imbalanced} stem from a competition between the two types of interaction terms, $\hat J_z^2$ and $\hat J_y^2$. From the microscopic point of view [Eq.~\eqref{eff_bosonic}], this corresponds to a competition between the intra-mode (Hubbard) interaction and the drive-induced pair-hopping processes. This is different from the transition discussed in the context of the BJJ model~\cite{raghavan1999coherent,zibold2012classical} -- from Phase I to Phase II -- which involves a competition between the Hubbard interaction and the single-particle hopping (or linear coupling) $\Omega_0$, and which is associated with the breaking of a single symmetry ($S_1$).

Last but not least, we note that the transition from Phase III to Phase I, which involves a competition between the drive-induced interaction term $\hat J_y^2$ and the linear coupling $\hat J_x$ in Eq.~\eqref{eff_final}, will be at the origin of the chiral superfluid to trivial superfluid phase transition discussed in Section~\ref{section_GS}, in the context of dimerized lattice systems.

\subsection{The pendulum analogy}

Tuning drive-induced nonlinearities can change the topology of phase-space, hence radically altering nonlinear dynamics. Interestingly, this phenomenon can be captured by a simple pendulum analogy~\cite{raghavan1999coherent,zibold2012classical,pigneur2018analytical}, as we now explain.

\begin{figure}[h!]
\includegraphics[width = \linewidth]{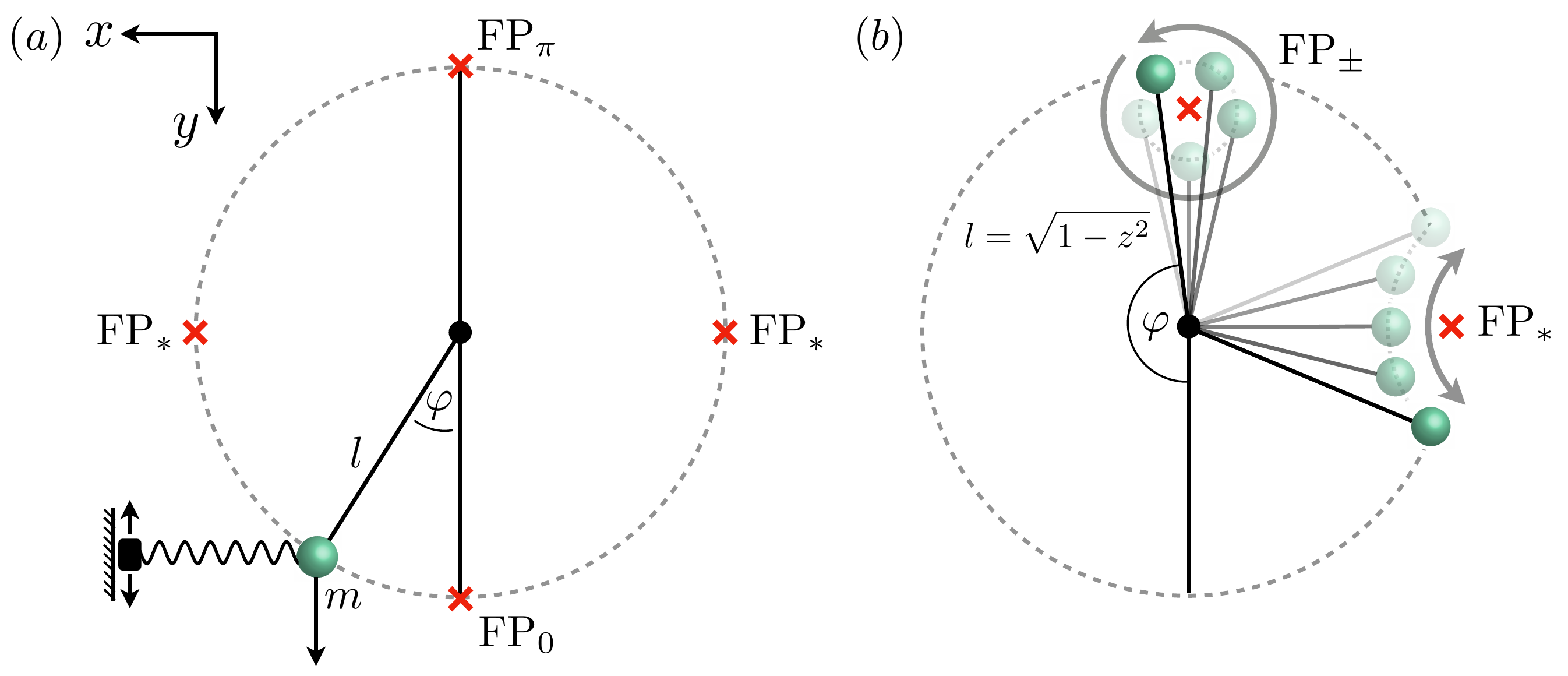}
\caption{Pendulum analogy for the effective Hamiltonian $\mathcal{H}_{\text{eff}}(z,\varphi;\alpha)$ in Eq.~\eqref{classical_eff_imbalanced}: (a) Schematics of a classical pendulum of mass $m$ and length $l$, subjected to gravity, and horizontally attached to a spring (allowed to slide along the y axis); see Eqs.~\eqref{eq_pendulum_simple}-\eqref{V_spring}. The red crosses indicate the positions associated with several fixed points; see Eq.~\eqref{FP_eq}. (b) Due to the momentum-dependent length, $l\!=\!\sqrt{1 - z^2}$, two types of fixed points can be stabilized depending on the strength of the spring:~$\text{FP}_{\pm}$  (weak spring, $\alpha\!>\!1/2$) and $\text{FP}_{*}$ (strong spring, $\alpha\!<\!1/2$). The sketch shows typical trajectories around the $\text{FP}_{\pm}$ and $\text{FP}_{*}$ stable fixed points. Figure inspired from~\cite{zibold2012classical}.}
\label{Fig:PendulumSketch}
\end{figure}

Let us consider a standard pendulum of mass $m$ and length $l$, subjected to gravity. Defining the angular-displacement variable $\varphi$ through [Fig.~\ref{Fig:PendulumSketch}(a)]
\begin{align}
x= l\, \text{sin}(\varphi)  , \quad y=l \,\text{cos}(\varphi), 
\label{Eq:PendulumDerivation}
\end{align}
we write the classical Hamiltonian of this simple pendulum as
\begin{align}
\mathcal{H}_{\text{pendulum}}(z,\varphi)=\frac{z^2}{2 I} - m g l  \cos \varphi .\label{eq_pendulum_simple}
\end{align}
Here, $z$ denotes the angular-momentum variable and $I$ is the moment of inertia of the simple pendulum. 

As pointed out in Ref.~\cite{raghavan1999coherent}, the Hamiltonian $\mathcal{H}_0(z,\varphi)$ describing the BJJ in Eq.~\eqref{classical_ham} is precisely of the form~\eqref{eq_pendulum_simple}, upon establishing the following dictionary:
\begin{align}
I \rightarrow (\chi N)^{-1} , \quad  m g \rightarrow \Omega_0 , \quad l \rightarrow \sqrt{1 - z^2}.\label{dictionary}
\end{align}
In this sense, the BJJ model can be mapped onto a non-rigid pendulum, with momentum-dependent length~\cite{raghavan1999coherent,zibold2012classical}. While the stable fixed point FP$_0$ of the BJJ model is naturally associated with the position at rest of a rigid pendulum, the other stable fixed points FP$_{\pi}$ ($\widetilde{\Omega}_0>1$) and FP$_{\pm}$ ($\widetilde{\Omega}_0<1$) stem from the non-rigid nature of the pendulum; see Fig.~\ref{Fig:PendulumSketch}(b) for a sketch of a typical trajectory around FP$_{\pm}$. In the BJJ model, the angular-momentum variable $z$ is restricted to take values in the interval $z\!\in\![-1 , 1]$; see Eq.~\eqref{z_eq}.

Having reviewed the pendulum analogy for the BJJ model~\cite{raghavan1999coherent,zibold2012classical}, we now apply this analogy to the effective Hamiltonian $\mathcal{H}_{\text{eff}}(z,\varphi;\alpha)$ in Eq.~\eqref{classical_eff_imbalanced}. Compared to the BJJ Hamiltonian $\mathcal{H}_{0}(z,\varphi)$, the effective Hamiltonian in Eq.~\eqref{classical_eff_imbalanced} features a new term $\sim \sin^2 \varphi$. Using the coordinates defined in Eq.~\eqref{Eq:PendulumDerivation}, we find that this term can be interpreted as an additional contribution to the potential energy of the non-rigid pendulum, given by 
\begin{equation}
V_{\text{spring}}=\frac{\chi N}{2} (1- \alpha) \, x^2. \label{V_spring}
\end{equation}
Consequently, the driven nonlinear system described by the effective Hamiltonian $\mathcal{H}_{\text{eff}}(z,\varphi;\alpha)$ in Eq.~\eqref{classical_eff_imbalanced} can be mapped onto a non-rigid pendulum that is horizontally attached to a spring (allowed to slide along the y axis); see Fig~\ref{Fig:PendulumSketch} (a) for a sketch. We note that the strength of the spring scales as $(1- \alpha)$, such that it vanishes in the limit $\alpha\!=\!1$ (BJJ model).

The spring has an intuitive effect on the trajectories of the pendulum:~activating the spring ($\alpha\!\lesssim1$) naturally reduces the amplitude of oscillations around the equilibrium point ($z\!=\!\varphi\!=\!0$). Besides, for a sufficiently strong spring ($\alpha\!<\!1/2$), full rotations around the pendulum's pivot (i.e.~trajectories associated with a full scan of the $\varphi$ axis and a well-defined chirality, sign$(z)$) become strictly forbidden. These intuitive effects are visible in the phase-space diagrams illustrated in Fig.~\ref{fig_landscape_imbalanced}.  

It is also interesting to note that the addition of the spring leads to two new fixed points $\text{FP}_*$, which become stable for a sufficiently strong spring ($\alpha\!<\!1/2$); see Figs.~\ref{fig_landscape_imbalanced} and~\ref{Fig:PendulumSketch}. These new equilibrium points correspond to a finite angle $\varphi$, set by the strength of the spring [Eq.~\eqref{FP_eq}]. 

To gain further insight, let us focus on the case of vanishing linear coupling $\Omega_0\!=\!0$, which is displayed in Fig.~\ref{fig_landscape_imbalanced}. In terms of the pendulum analogy, this corresponds to a vanishing force of gravity [Eq.~\eqref{dictionary}], implying that the pendulum is only subjected to the elastic force of the spring. For a weak spring ($\alpha\!\lesssim1$) and small angular momentum ($\vert z \vert\!\ll\!1$), one can assume that the length of the pendulum is constant, and one obtains a set of intuitive fixed points:~the two stable fixed points ($\text{FP}_0$ and $\text{FP}_{\pi}$) simply correspond to the two rest positions of the spring, while the two unstable fixed points $\text{FP}_{*}$ correspond to the positions where the spring is maximally stretched and its force is exactly balanced by the constraint reaction of the rigid pendulum. As previously noted, the fixed points $\text{FP}_{*}$ become stable for a sufficiently strong spring ($\alpha\!<\!1/2$), a peculiar effect which finds its origin in the non-rigid (momentum-dependent) length of the pendulum; see Fig.~\ref{Fig:PendulumSketch}(b) for a sketch of a typical trajectory around FP$_{*}$.


\section{Numerical validation of the effective-Hamiltonian approach}\label{section_numerics}

\begin{figure}[h!]
\includegraphics[width = 1.\linewidth]{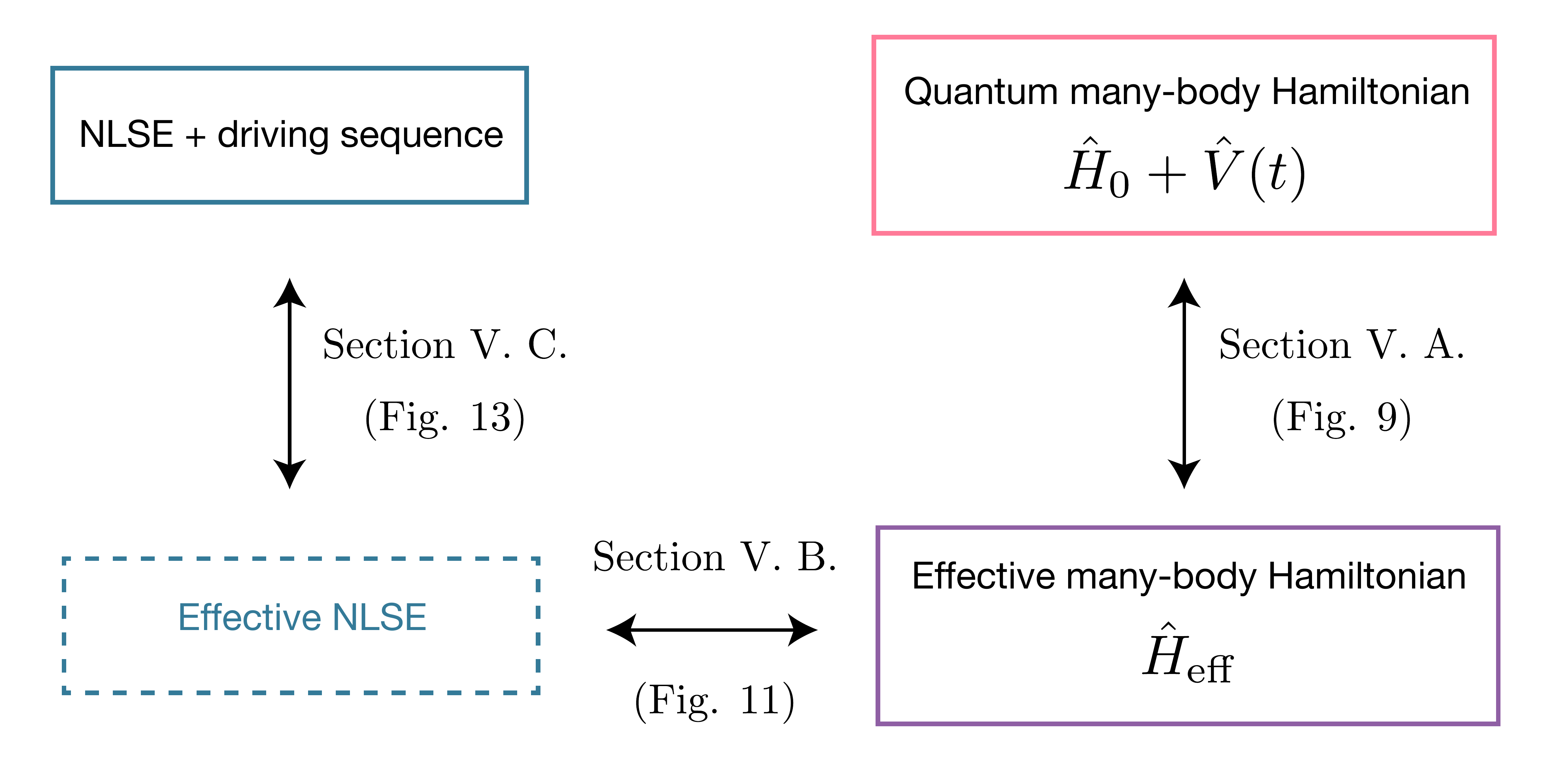}
\caption{Outline of the numerical study, which validates the approach originally displayed in Fig.~\ref{fig_approach}.
}
\label{fig_approach_numerics}
\end{figure}

This Section aims at exploring the validity of the effective-Hamiltonian analysis developed in Section~\ref{sect_effective} and its classical limit presented in Section~\ref{sect_classical_eff}. In particular, this Section demonstrates that the stroboscopic dynamics of the driven NLSE in Eq.~\eqref{NLS_time} is well described by the effective NLSE with emergent nonlinearities in Eq.~\eqref{NLS_effective_generalized}, as announced in Section~\ref{section_two_modes}. The outline of our numerical study is displayed in Fig.~\ref{fig_approach_numerics}. We hereby set the drive-induced interaction parameter to the value $\alpha\!=\!1/2$; we verified that our conclusions hold in general.

\subsection{Validating the effective quantum Hamiltonian}

First, we demonstrate that the dynamics associated with the effective Hamiltonian in Eq.~\eqref{eff_bosonic} [or equally Eq.~\eqref{eff_final}] reproduces the stroboscopic dynamics of the driven system described by Eqs.~\eqref{U_sequence}-\eqref{time_evolution}. To this end, we choose a coherent spin state as an initial state~\cite{kitagawa1993squeezed,zibold2012classical}
\begin{equation}
\vert \psi (0) \rangle = \vert N , \theta, \varphi \rangle = \left (\hat a_{\theta,\varphi}^{\dagger} \right )^{N} \vert \emptyset \rangle,\label{CSS}
\end{equation}
which corresponds to a macroscopic occupation of the single-particle state,
\begin{equation}
\vert \theta, \varphi \rangle = \cos (\theta/2) \vert 1 \rangle + \sin(\theta/2)e^{i \varphi} \vert 2 \rangle ,
\end{equation}
defined on the Bloch sphere. Here we introduced the single-particle states $\vert 1 \rangle\!=\!\hat a_{1}^{\dagger} \vert \emptyset \rangle$ and $\vert 2 \rangle\!=\!\hat a_{2}^{\dagger} \vert \emptyset \rangle$, associated with the two modes, as well as the creation operator $\hat a_{\theta,\varphi}^{\dagger}\vert \emptyset \rangle\!=\!\vert \theta, \varphi \rangle$. We note that the chosen initial state in Eq.~\eqref{CSS} behaves classically in the limit $N\!\rightarrow\!\infty$~\cite{zibold2012classical}, which will be convenient for later purposes (i.e.~when comparing quantum and classical dynamics). 

\begin{figure}[h!]
\includegraphics[width = \linewidth]{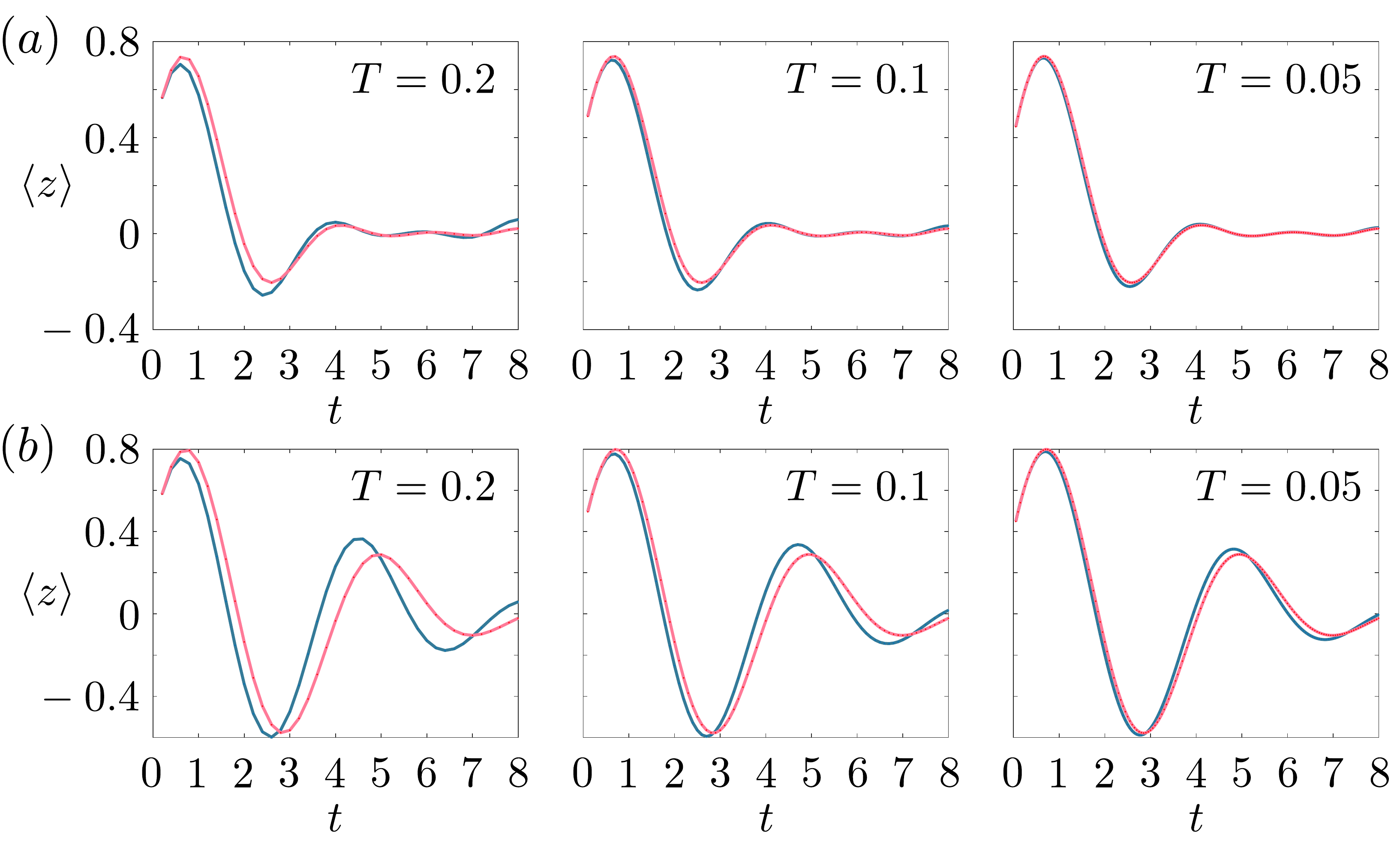}
\caption{Population imbalance $\langle z \rangle$ as a function of time, as obtained from the quantum dynamics of the driven system (blue curve) and the effective-Hamiltonian quantum dynamics (red curve) for: (a) $N\!=\!10$ bosons and (b) $N\!=\!50$ bosons. For each case, the full time dynamics of the driven system is generated using the sequence in Eq.~\eqref{U_sequence} with a period $T\!=\!0.2$, $T\!=\!0.1$ and $T\!=\!0.05$. Here the interaction parameter is set to $g\!=\!\chi N\!=\!5$, the static linear coupling is set to $\Omega_0\!=\!0$ and $\alpha\!=\!1/2$; the initial coherent spin state $\vert N , \theta, \varphi \rangle$ corresponds to $z\!=\!\cos \theta\!=\!0.4$ and $\varphi\!=\!2.25$. In all plots, the time-evolved state is evaluated at stroboscopic times $t_{\frak{n}}\!=\!T\!\times\!\frak{n} $.
}
\label{fig_effective_full_quantum}
\end{figure}

We analyze the quantum dynamics through the evaluation of the population imbalance 
\begin{equation}
\langle z (t_{\frak{n}}) \rangle = (2/N) \langle \psi (t_{\frak{n}}) \vert \hat J_z \vert \psi (t_{\frak{n}}) \rangle , \quad t_{\frak{n}}\!=\!T\!\times\!\frak{n} , \notag
\end{equation}
where the time-evolved state $\vert \psi (t_{\frak{n}}) \rangle$ is obtained from:~(i) the full time dynamics of the driven system [Eqs.~\eqref{U_sequence}-\eqref{time_evolution}], and (ii) the effective Hamiltonian [Eq.~\eqref{eff_bosonic}]. Figure~\ref{fig_effective_full_quantum} compares these two results for both $N\!=\!10$ and $N\!=\!50$ bosons, and the same ``mean-field" interaction parameter $g\!=\!\chi N\!=\!5$. In both cases, one obtains that the effective description well captures the stroboscopic dynamics when the driving period is sufficiently small, $T\!\lesssim\!0.1$ in the current units [see Eq.~\eqref{eq_parent_static}]. This analysis validates the effective Hamiltonian in Eq.~\eqref{eff_bosonic} in the high-frequency regime.

\subsection{Effective semiclassical dynamics and quantum effects}\label{section_husimi}

As a next step, we now show that the effective Hamiltonian $\hat H_{\text{eff}}$ in Eq.~\eqref{eff_bosonic} well captures the classical dynamics generated by the equations of motion in Eq.~\eqref{pendulum_eff_imbalanced}. We remind that the latter classical description is associated with the  Hamiltonian function $\mathcal{H}_{\text{eff}}(z,\varphi; \alpha)$ displayed in Eq.~\eqref{classical_eff_imbalanced}, where $z$ and $\varphi$ describe the relative population and phase of the two modes; see Eqs.~\eqref{psi_z_phi}-\eqref{z_eq}. The agreement between the quantum and classical descriptions is expected to be reached in the limit $N\!\rightarrow\infty$, where quantum fluctuations become negligible~\cite{smerzi1997quantum,gajda1997fluctuations,zibold2010classical,zibold2012classical,carusotto2013quantum,larre2015propagation}. We also remind the reader that the classical equations of motion in Eq.~\eqref{pendulum_eff_imbalanced}, which are analyzed in this Section, are equivalent to the effective NLSE in Eq.~\eqref{NLS_effective_generalized}, through the mapping defined in Eq.~\eqref{psi_z_phi}.

\begin{figure}[h!]
\includegraphics[width = 0.8\linewidth]{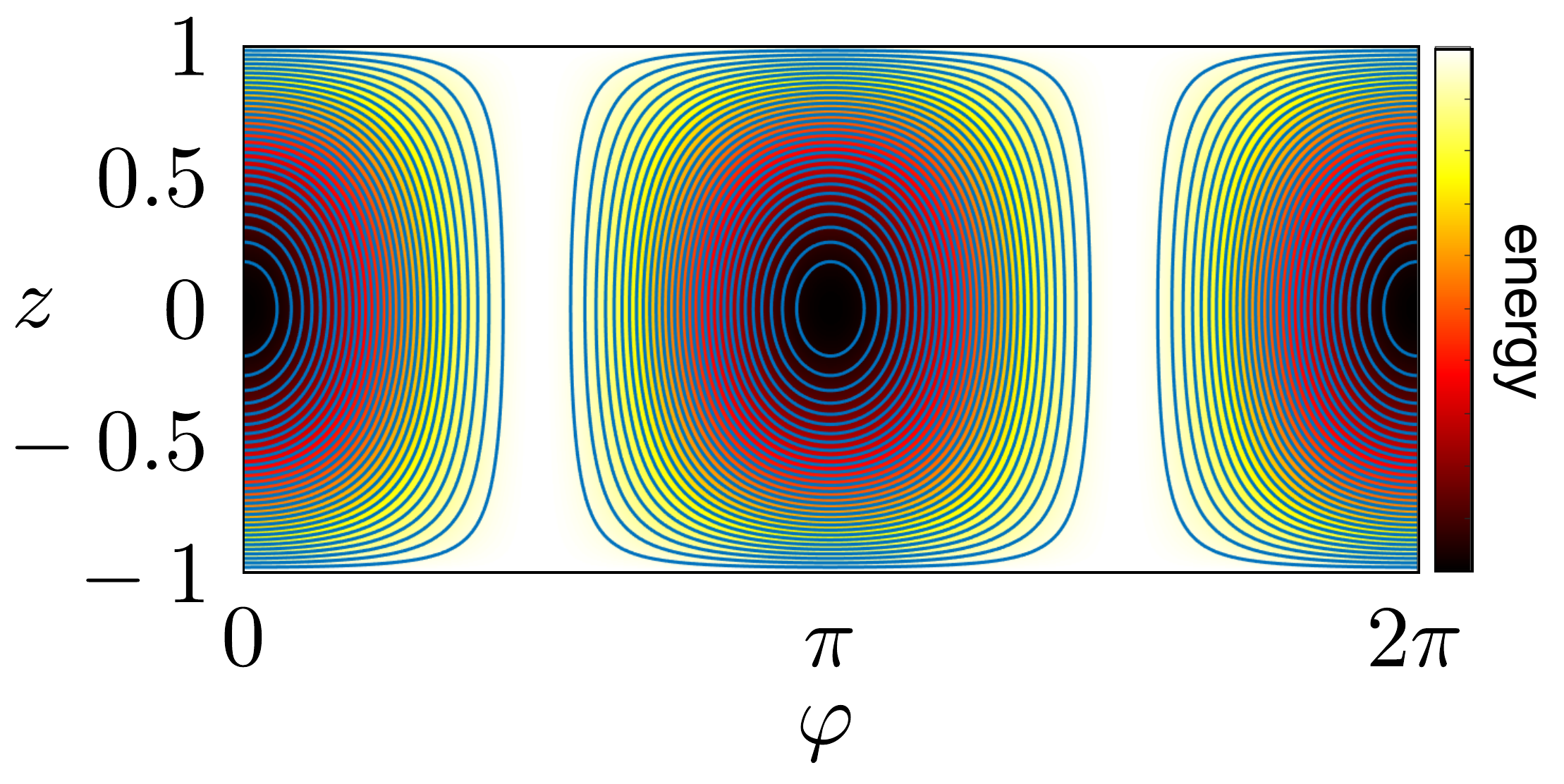}
\caption{Energy landscape associated with the classical Hamiltonian $\mathcal{H}_{\text{eff}}(z,\varphi; \alpha)$ displayed in Eq.~\eqref{classical_eff_imbalanced}, for $\Omega_0\!=\!0$ and $\alpha\!=\!1/2$. A few trajectories are indicated as thin blue curves (equipotential lines of the energy landscape).
}
\label{fig_eff_landscape}
\end{figure}

First of all, let us analyze the dynamics generated by the effective classical equations of motion in Eq.~\eqref{pendulum_eff_imbalanced}. In order to highlight the role of nonlinearities, we hereby set the static linear coupling to $\Omega_0\!=\!0$ and we remind that $\alpha\!=\!1/2$. In Fig.~\ref{fig_eff_landscape}, we display a few representative trajectories over the energy landscape $\mathcal{H}_{\text{eff}}(z,\varphi; \alpha)$ defined in Eq.~\eqref{classical_eff_imbalanced}. These trajectories reflect the presence of two stable fixed points at $\text{FP}_0\!=\!(z\!=\!0,\varphi\!=\!0)$ and $\text{FP}_\pi\!=\!(z\!=\!0,\varphi\!=\!\pi)$. We stress that this configuration of fixed points radically differs from that associated with the non-driven system [see $\mathcal{H}_0(z,\varphi)$ in Eq.~\eqref{classical_ham}], which are located at $z\!=\!\pm1$ for the same choice of $\Omega_0\!=\!0$; see Section~\ref{section_pendulum}.

\begin{figure}[h!]
\includegraphics[width = \linewidth]{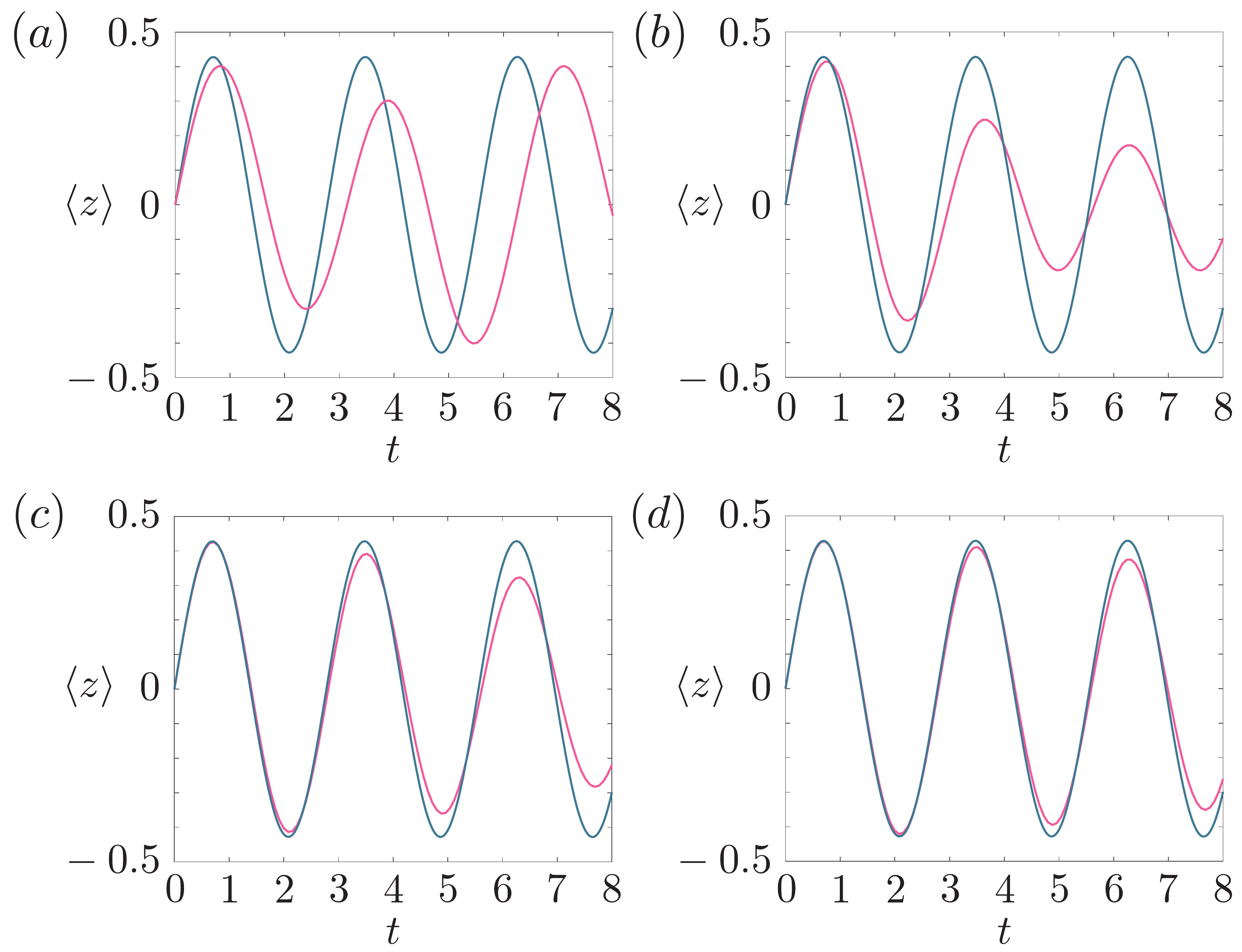}
\caption{Population imbalance $\langle z \rangle$ as a function of time, as obtained from the effective-Hamiltonian quantum dynamics (red curve) and the effective classical equations of motion (blue curve). The number of bosons is: (a) $N\!=\!5$; (b) $N\!=\!10$; (c) $N\!=\!80$; (d) $N\!=\!170$. Here the interaction parameter is set to $g\!=\!\chi N\!=\!5$, while the static linear coupling is set to $\Omega_0\!=\!0$ and $\alpha\!=\!1/2$. The initial coherent spin state $\vert N , \theta, \varphi \rangle$ corresponds to $z\!=\!\cos \theta\!=\!0$ and $\varphi\!=\!2.7$; the same initial condition is chosen for the effective classical dynamics. The dynamics $z(t)$ should be compared with the trajectories depicted in Fig.~\ref{fig_eff_landscape}, close to the stable fixed point $\text{FP}_\pi\!=\!(z\!=\!0,\varphi\!=\!\pi)$.
}
\label{fig_eff_classical}
\end{figure}

We now compare these classical predictions to the quantum dynamics associated with the effective Hamiltonian $\hat H_{\text{eff}}$ in Eq.~\eqref{eff_bosonic}, using a coherent spin state $\vert N , \theta, \varphi \rangle$ as an initial condition; see Eq.~\eqref{CSS}. Figure~\ref{fig_eff_classical} shows the trajectories $\langle z(t) \rangle$ for $N\!=\!5, 10, 80, 170$ bosons, while keeping the ``mean-field" interaction parameter $\chi N\!=\!5$ constant. From these results, we confirm that a good agreement between the effective classical and quantum descriptions is indeed obtained in the large $N$ limit. 

\begin{figure}
\includegraphics[width=\linewidth]{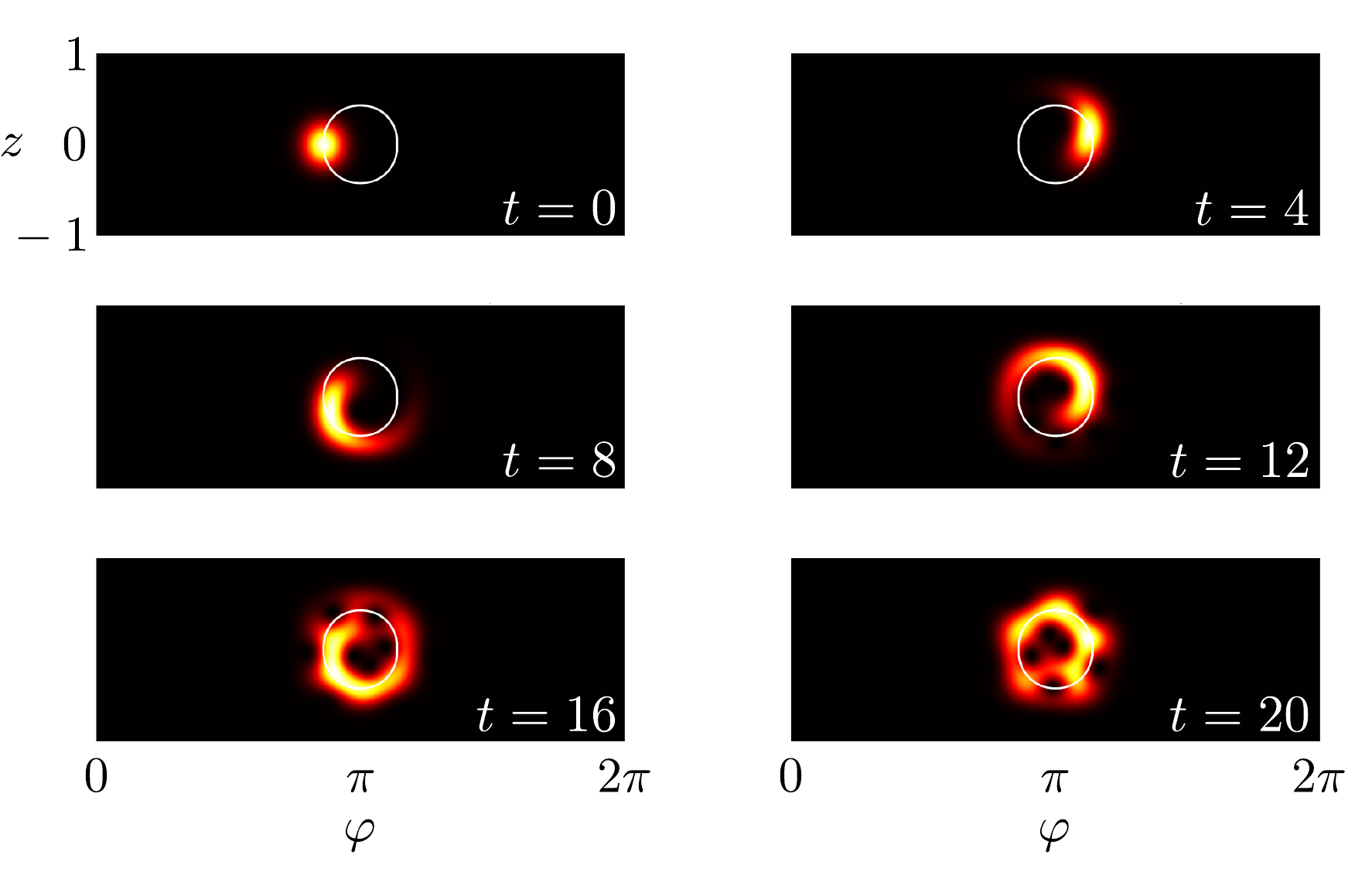}
\caption{Time-evolving Husimi function $Q(z,\varphi; t)$ for a state $\vert \psi (t) \rangle$ that evolves according to the effective Hamiltonian $\hat H_{\text{eff}}$ in Eq.~\eqref{eff_bosonic}. Here, the number of bosons is $N\!=\!80$, and the other parameters are the same as in Fig.~\ref{fig_eff_classical}(c). The initial coherent spin state $\vert \psi (0) \rangle\!=\!\vert N , \theta, \varphi \rangle$, at $z\!=\!\cos \theta\!=\!0$ and $\varphi\!=\!2.7$, becomes substantially squeezed around time $t\!\approx\!3$, hence signaling the breakdown of its classical description. An oversqueezed state, exhibiting Majorana stars, appears around $t\!\approx\!12$. The trajectory predicted by the effective classical equations of motion [Eq.~\eqref{pendulum_eff_imbalanced}] is depicted in white.
}
\label{fig_husimi}
\end{figure}

In order to further appreciate the residual deviations between the quantum and classical dynamics in the small $N$ regime, we depict the time-evolving Husimi function $Q(z,\varphi; t)$ in Fig.~\ref{fig_husimi} for the case $N\!=\!80$. The Husimi function~\cite{zibold2010classical,zibold2012classical,julia2012dynamic,bruno2012quantum,strobel2014fisher,evrard2019enhanced,nascimbene2020quantum} is obtained by evaluating the squared overlap of the time-evolving state $\vert \psi (t) \rangle$ with the coherent spin states defined over the Bloch sphere (with same particle number $N$),
\begin{equation}
Q(z,\varphi; t) = \vert \langle N , \theta, \varphi \vert \psi (t) \rangle \vert^2 , \quad z=\cos \theta .
\end{equation}
Here the state $\vert \psi (t) \rangle$ is evolved according to the effective Hamiltonian $\hat H_{\text{eff}}$ in Eq.~\eqref{eff_bosonic}, so that the evolution of the Husimi function in Fig.~\ref{fig_husimi} is to be compared with the quantum dynamics displayed in Fig.~\ref{fig_eff_classical}(c) for $N\!=\!80$ bosons. The time-evolution of the Husimi function $Q(z,\varphi; t)$ shown in Fig.~\ref{fig_husimi} indicates that the initial coherent spin state $\vert \psi (0) \rangle\!=\!\vert N , \theta, \varphi \rangle$ becomes substantially squeezed~\cite{kitagawa1993squeezed} around time $t\!\approx3$, which also corresponds to the time around which the classical trajectory starts deviating from the effective-Hamiltonian quantum dynamics in Fig.~\ref{fig_eff_classical}(c). At later times, $t\!\approx\!12$, the state becomes oversqueezed and it exhibits Majorana stars in the Husimi distribution~\cite{bruno2012quantum,evrard2019enhanced,nascimbene2020quantum}. We find that these non-classical features are postponed to later evolution times upon increasing the number of bosons $N$ while keeping the interaction parameter $g\!=\!\chi N$ fixed. Despite these non-classical features, the center of mass of the Husimi function is found to approximately follow a classical orbit around the stable fixed point $\text{FP}_\pi\!=\!(z\!=\!0,\varphi\!=\!\pi)$, as depicted in Fig.~\ref{fig_eff_landscape}.

\begin{figure}
\includegraphics[width=0.8\linewidth]{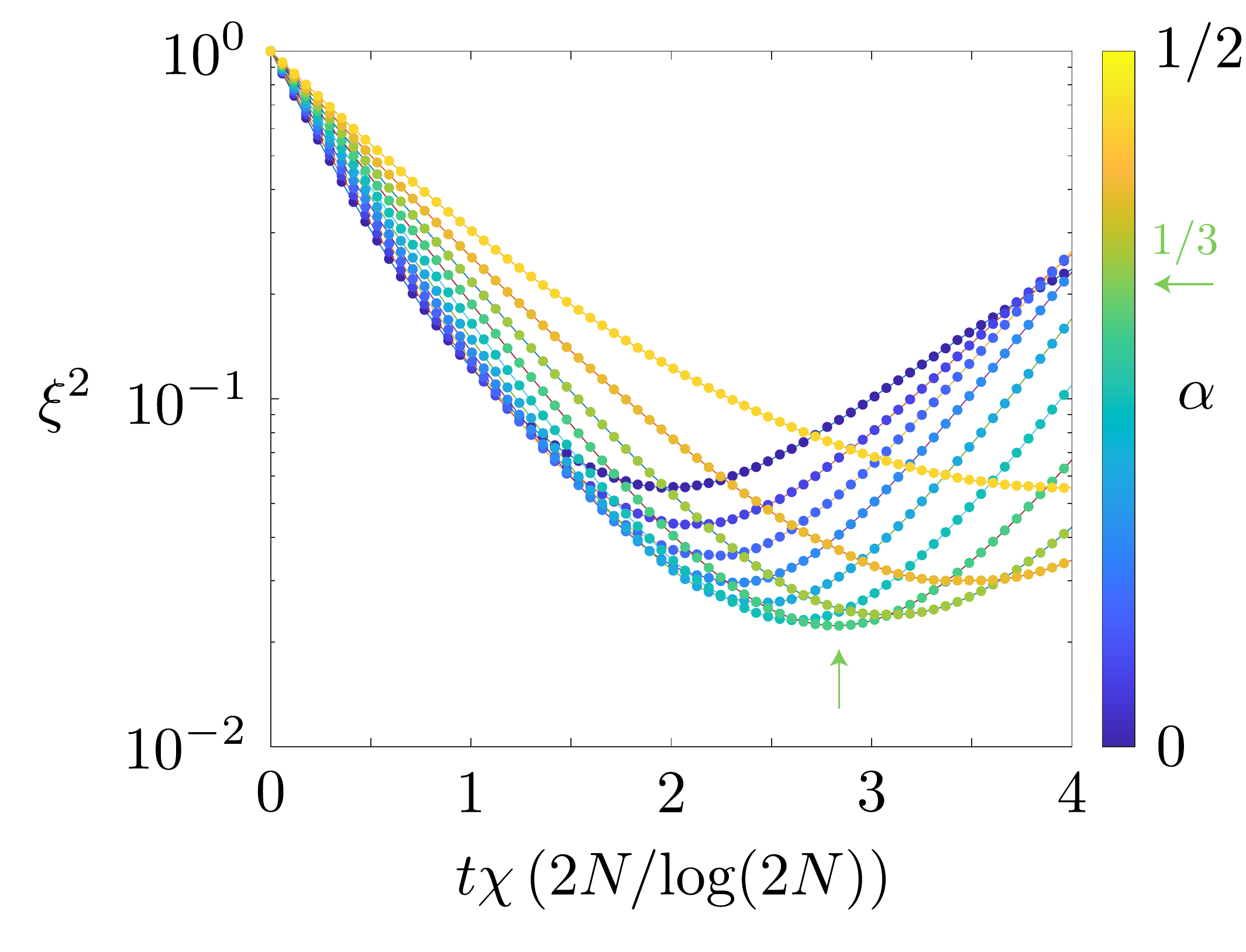}
\caption{Squeezing parameter $\xi^2$ as a function of time $t$, for various values of the drive parameter $\alpha$. The green arrow indicates optimal squeezing, which is reached in the limit $\alpha\!\rightarrow\!1/3$, after a duration $t_{\text{opt}}\!\approx\! 2.8\, \text{log}(2N)/2N\chi$. Here, the initial state is a coherent spin state of $N\!=\!80$ bosons located at the North pole ($z\!=\!1$), and the system is evolved using the effective Hamiltonian $\hat H_{\text{eff}}$ in Eq.~\eqref{eff_bosonic}. The interaction parameter is set to $g\!=\!\chi N\!=\!5$ as in Fig.~\ref{fig_husimi}.
}
\label{fig_squeeze}
\end{figure}

Within this framework, quantum effects such as squeezing and entanglement can be further enhanced by properly adjusting the initial state and the system parameters. In particular, setting the system at an unstable fixed point of a two-axis twisting (TAT) Hamiltonian is expected to produce optimal squeezing~\cite{kitagawa1993squeezed,ma2011quantum,kajtoch2015quantum}. In our setting, the effective Hamiltonian in Eq.~\eqref{eff_final} reduces to the TAT Hamiltonian in the limit $\alpha\!=\!1/3$, in which case
\begin{align}
\hat H_{\text{eff}}(\alpha=1/3)
&=\frac{\chi}{3} \left (\hat J_y^2 - \hat J_x^2 \right ) - \Omega_0 \hat J_x + \mathcal{O} (T),\label{eff_TAT}
\end{align}
where we used Eq.~\eqref{sum_rule} and omitted an irrelevant constant~\cite{liu2011spin}. Considering the unstable fixed point located at the North pole ($z\!=\!1$), we study how the squeezing parameter $\xi^2$ behaves as the parameter $\alpha$ is tuned within the interval $\alpha\!\in\![0, 0.5]$. Here, we evaluate the squeezing parameter according to the definition~\cite{ma2011quantum}
\begin{equation}
\xi^2=\frac{4 (\Delta \hat J_{\perp}^2)_{\text{min}}}{N},
\end{equation}
where $(\Delta \hat J_{\perp}^2)_{\text{min}}$ denotes the minimal variance normal to the mean spin vector $\langle \hat{\boldsymbol{J}} \rangle$. Using this definition, one has $\xi^2\!=\!1$ for a coherent spin state [Eq.~\eqref{CSS}] and $\xi^2\!<\!1$ for a squeezed state~\cite{kitagawa1993squeezed,ma2011quantum}. The time-evolving squeezing parameter $\xi^2(t)$ is represented in Fig.~\ref{fig_squeeze}, for various values of the parameter $\alpha$. These results demonstrate that an optimal squeezing is indeed reached for $\alpha\!=\!1/3$, after a duration $t_{\text{opt}}\!\approx\! 2.8\, \text{log}(2N)/2N\chi$. We verified that the optimal squeezing value scales as $\xi^2_{\text{opt}}\!\sim\!N^{-1}$ and that the corresponding optimal squeezing time scales as $t_{\text{opt}}\sim\text{log}(2N)/2N\chi$ in the vicinity of $\alpha\!=\!1/3$~\cite{ma2011quantum,kajtoch2015quantum}.

\subsection{The driven nonlinear Schr\"odinger equation \\ and its effective description}

In this Section, we finally analyze the agreement between the classical dynamics associated with the driven NLSE [Eqs.~\eqref{NLS_time}-\eqref{pulse}] and the dynamics generated by the \emph{effective} classical equations of motion [Eq.~\eqref{pendulum_eff_imbalanced}], which derive from the Hamiltonian $\mathcal{H}_{\text{eff}}(z,\varphi; \alpha)$ in Eq.~\eqref{classical_eff_imbalanced}. We remind that these effective equations of motion are equivalent to the \emph{effective} NLSE announced in Eq.~\eqref{NLS_effective_generalized}.

In practice, we numerically solve the following classical equations of motion [Eq.~\eqref{pendulum}]
\begin{align}
&\dot z=f_{\text{pulse}}(t) \sqrt{1-z^2} \sin \varphi , \notag \\
&\dot \varphi= N \chi z - f_{\text{pulse}}(t) \frac{z}{\sqrt{1-z^2}} \cos \varphi , \label{pendulum_driven}
\end{align}
where the pulse function $f_{\text{pulse}}(t)$ is defined in Eq.~\eqref{pulse}; here we again set the static coupling $\Omega_0\!=\!0$. The equations of motion in Eq.~\eqref{pendulum_driven} are equivalent to the driven NLSE in Eqs.~\eqref{NLS_time}-\eqref{pulse} through the mapping provided by Eq.~\eqref{psi_z_phi}.

\begin{figure}[h!]
\includegraphics[width = \linewidth]{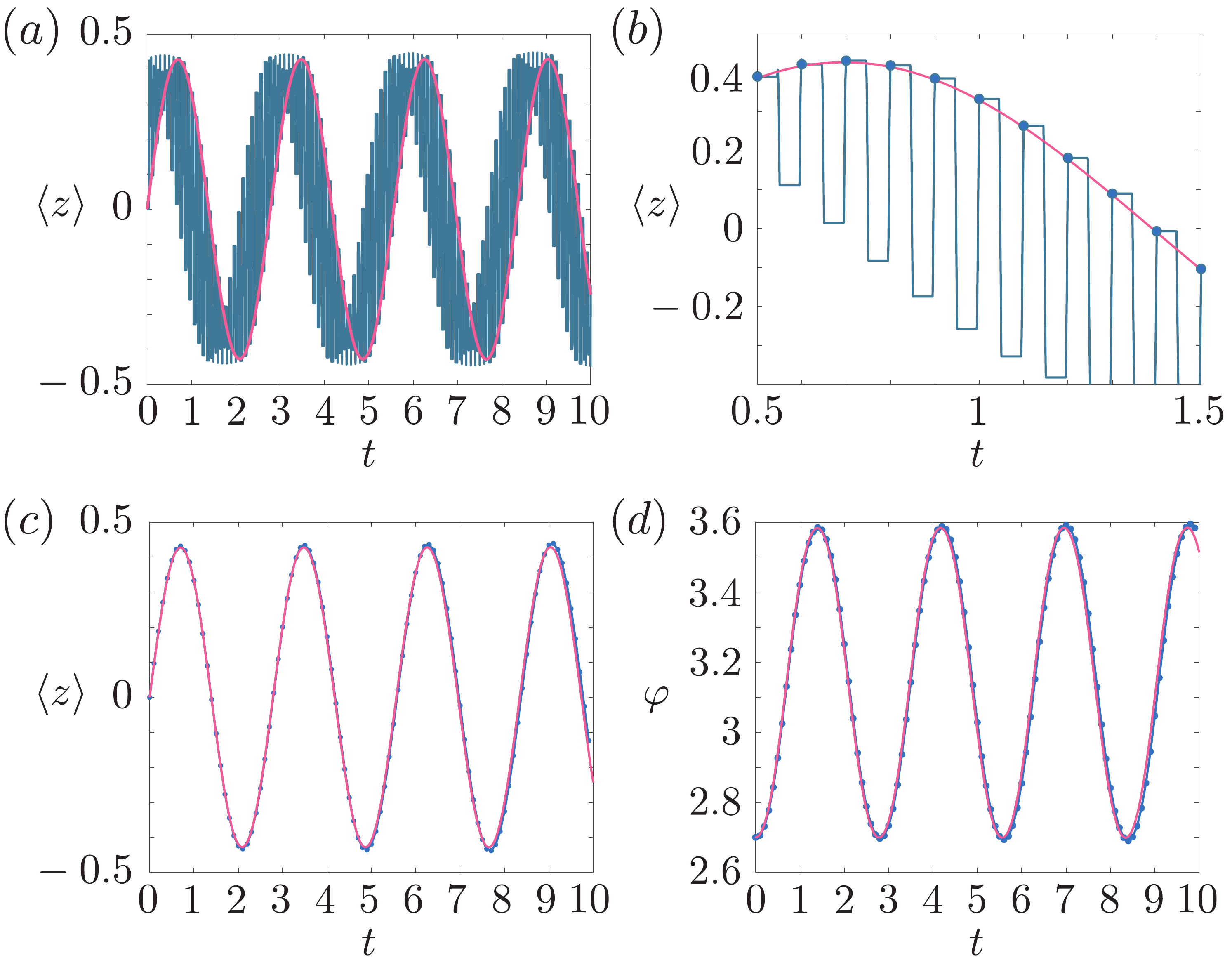}
\caption{The driven NLSE versus the effective NLSE descriptions:~(a) Population imbalance $z(t)$ as a function of time, as obtained from the driven NLSE in Eq.~\eqref{pendulum_driven} (blue curve) and from the effective classical equations of motion in Eq.~\eqref{pendulum_eff_imbalanced} (red curve). (b) Zoom in the panel (a):~the blue dots highlight the stroboscopic dynamics at times $t_{\frak{n}}$; note the micromotion at arbitrary times $t\!\ne\!t_{\frak{n}}$. (c) Stroboscopic dynamics $z(t_{\frak{n}})$ obtained from the driven NLSE (blue curve and dots), compared with the effective classical description (red curve). (d) Same as in panel (c) but for the other canonical variable $\varphi$. In all panels, the period of the drive is set to $T\!=\!0.1$ and the pulse duration to $\tau\!=\!T/20$; the interaction parameter is set to $g\!=\!\chi N\!=\!5$, the static linear coupling is set to $\Omega_0\!=\!0$ and $\alpha\!=\!1/2$; the initial condition corresponds to $z\!=\!\cos \theta\!=\!0$ and $\varphi\!=\!2.7$ as in Fig.~\ref{fig_eff_classical}. We note that the stroboscopic dynamics $(z(t_{\frak{n}}),\varphi(t_{\frak{n}}))$ reflects the \emph{effective} energy landscape depicted in Fig.~\ref{fig_eff_landscape}, close to the stable fixed point $\text{FP}_\pi\!=\!(z\!=\!0,\varphi\!=\!\pi)$.
}
\label{fig_classical_complete}
\end{figure}

The resulting dynamics is displayed in Fig.~\ref{fig_classical_complete}, together with the dynamics generated from the effective classical Hamiltonian $\mathcal{H}_{\text{eff}}(z,\varphi; \alpha)$ in Eq.~\eqref{classical_eff_imbalanced}. The results in Fig.~\ref{fig_classical_complete} confirm that the effective classical description very well captures the  dynamics of the driven nonlinear system at stroboscopic times $t\!=\!t_{\frak{n}}$, while a finite micromotion is observed at intermediate times $t\!\ne\!t_{\frak{n}}$.

We also emphasize that the trajectories $(z(t),\varphi(t))$ generated by the effective equations of motion [Fig.~\ref{fig_classical_complete}] reflect the presence of a stable fixed point at $\text{FP}_\pi\!=\!(z\!=\!0,\varphi\!=\!\pi)$; see Fig.~\ref{fig_eff_landscape}. Importantly, this fixed point is \emph{unstable} for the non-driven system described by $\mathcal{H}_0(z,\varphi)$ in Eq.~\eqref{classical_ham}, hence leading to drastically different dynamics.


Altogether, the numerical studies presented in this Section~\ref{section_numerics} validate the effective description announced in Eq.~\eqref{NLS_effective_generalized} [see also Sections~\ref{sect_effective} and \ref{sect_classical_eff}], and hence, confirm the creation of effective interactions and nonlinearities through the driving sequence.

\section{Designing lattice systems with controllable drive-induced interactions}\label{section_lattice}




We described in Section~\ref{section_quantum} how activating coupling processes between two modes (or sites), according to a well-designed pulse sequence, generates effective pair-hopping processes [Eq.~\eqref{eff_bosonic}]. We now analyze the possibility of extending this scheme to lattice systems, both in the classical (mean-field) limit and in the regime of strongly-correlated quantum matter. 

\begin{figure}[t!]
\includegraphics[width = \linewidth]{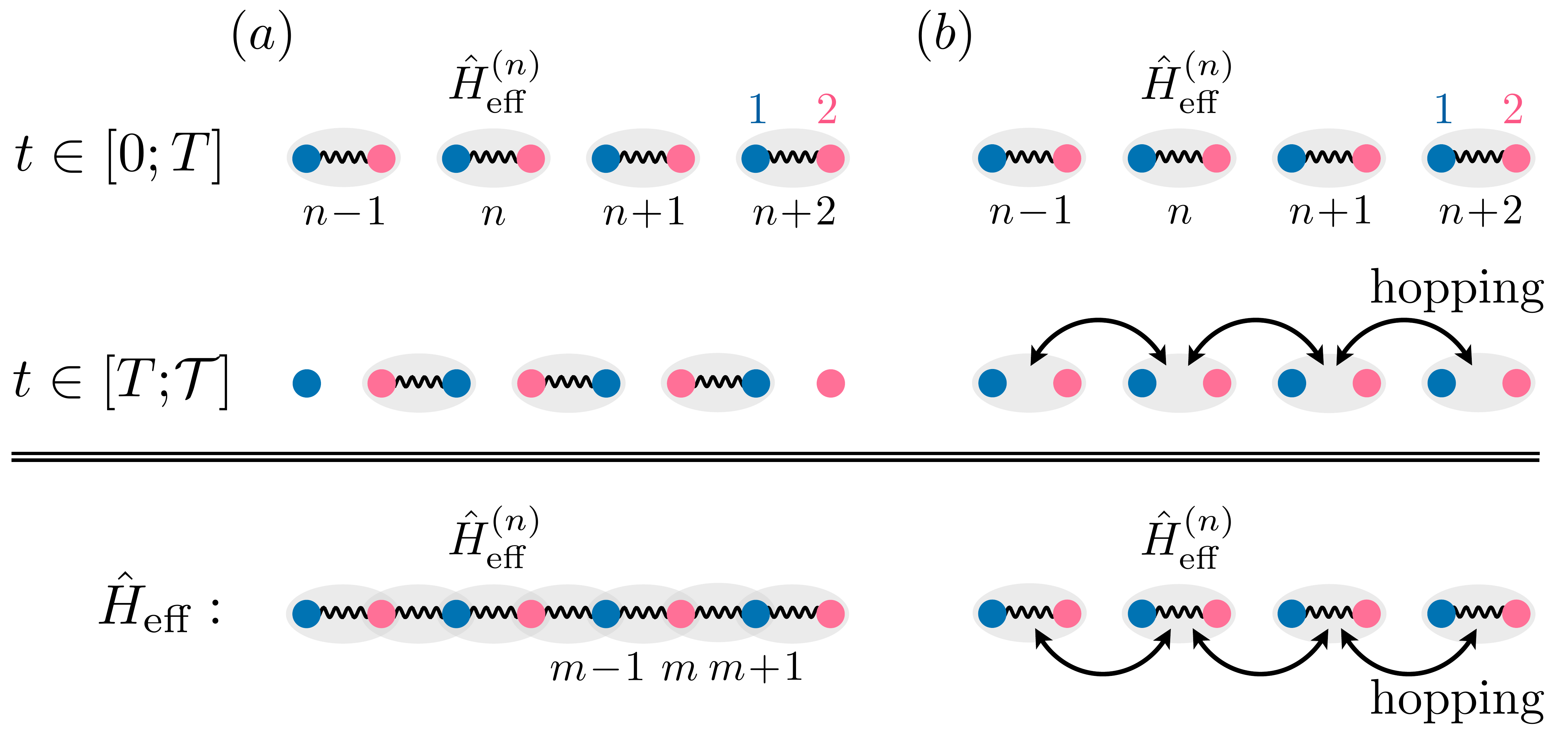}
\caption{Designing effective interactions in lattice systems using two types of sequences. (a) We apply the pulse sequence in Eq.~\eqref{U_sequence} within each individual dimer over a duration $T$, and then apply that same sequence to the complementary dimerization. The total sequence, of period $\mathcal{T}\!=\!2T$, realizes the extended Bose-Hubbard model in Eq.~\eqref{eff_bosonic_lattice}, which is characterized by drive-induced pair-hopping processes. (b) In the second type of sequence, one preserves the dimerized structure and activates hopping between the dimers during the second step. The resulting class of models exhibit orbital ordering, in direct analogy with $p$-band systems. 
}
\label{fig_lattice_effective}
\end{figure}

Here, we set the focus on two types of sequences, illustrated in Fig.~\ref{fig_lattice_effective}, which lead to different classes of lattice models. In the first scenario, one applies the pulse sequence in Eq.~\eqref{U_sequence} in a dimerized manner in view of generating uniform pair-hopping processes over the entire lattice; see Fig.~\ref{fig_lattice_effective}(a). In the second sequence, one first applies the pulse sequence on individual dimers and then activates hopping processes between them; see Fig.~\ref{fig_lattice_effective}(b). Interestingly, this second approach leads to a class of models that share similarities with p-band systems~\cite{li2016physics}, without recurring to higher bands; this aspect will be explored in Section~\ref{section_GS}.

We note that subjecting a lattice to a time-periodic modulation generically leads to various types of correlated tunneling processes and higher-order interaction processes~\cite{ma2011photon,di2014quantum,bukov2015universal,eckardt2015high,anisimovas2015role,meinert2016floquet,pieplow2018generation,zahn2022formation}. In our approach, tunable pair-hopping processes are generated at the level of individual dimers, hence allowing for highly controllable exotic lattice models. Possible experimental implementations will be discussed in Section~\ref{section_exp}.

\subsection{Generating uniform pair-hopping processes on a lattice}\label{section_uniform_pair}

Let us consider a dimerized lattice of $N_s$ sites, which we label as $(n; s)$ with
\begin{equation}
n=1, \dots , N_d, \quad s=1,2.\label{eq_dimers}
\end{equation}
Here, $N_d$ denotes the number of dimers and $s$ labels the two modes (or orbitals) within each dimer; see Fig.~\ref{fig_lattice_effective}(a). 

Each dimer $n$ is assumed to be described by a (static) Hamiltonian $\hat H_0^{(n)}$ of the form given in Eq.~\eqref{eq_parent_static}. Physical realizations include arrays of two-mode optical cavities~\cite{hafezi2011robust,anderson2016engineering}, two-component Bose gases in an optical lattice~\cite{trotzky2008time}, or quantum gases in tunable (dimerized) superlattices~\cite{tarruell2012creating,walter2022breakdown}.

We introduce the angular momentum operators [Eq.~\eqref{Schwinger}] associated with each dimer
\begin{align}
&\hat J_x^{(n)} = \frac{1}{2} \left (\hat a_{n,1}^{\dagger} \hat a_{n,2} + \hat a_{n,2}^{\dagger} \hat a_{n,1} \right ) ,  \notag \\
&\hat J_y^{(n)} = \frac{1}{2i} \left (\hat a_{n,2}^{\dagger} \hat a_{n,1} - \hat a_{n,1}^{\dagger} \hat a_{n,2} \right ), \notag \\
&\hat J_z^{(n)} = \frac{1}{2} \left (\hat a_{n,2}^{\dagger} \hat a_{n,2} - \hat a_{n,1}^{\dagger} \hat a_{n,1} \right ) ,\notag \\
&\hat N^{(n)} = \hat a_{n,1}^{\dagger} \hat a_{n,1} + \hat a_{n,2}^{\dagger} \hat a_{n,2},\label{Schwinger_lattice}
\end{align}
where $\hat a_{n,s}^{\dagger}$ creates a boson at the lattice site $(n; s)$. We then write the total (undriven) Hamiltonian as [Eq.~\eqref{Josephson}]
\begin{align}
\hat H_0&=\sum_n \hat H_0^{(n)} \label{H_zero_lattice}\\
&=\sum_n \left \{ \chi \left [\hat J_z^{(n)}\right ]^2 - \Omega_0\,  \hat J_x^{(n)} + \frac{\eta}{4} \left [\hat N^{(n)}\right ]^2 - \frac{1}{2}\hat N^{(n)} \right \}
,\notag 
\end{align}
where $\chi\!=\!1\!-\!\beta$ and $\eta\!=\!1\!+\!\beta$. In the following, individual dimers will be coupled such that the number of particles $\hat N^{(n)}$ will no longer be conserved at the level of each dimer. As a consequence, the interaction terms $\sim (\hat N^{(n)})^2$ in Eq.~\eqref{H_zero_lattice} cannot be ignored.


We now introduce the pulse sequence, which we split into two main steps:
\begin{itemize}
\item Step 1: We apply the pulse sequence in Eq.~\eqref{U_sequence} within each individual dimer over a duration $T$.
\item Step 2: We consider the complementary dimerization, $$\dots  \quad (n-1;2)-(n,1) \quad (n;2)-(n+1,1) \quad \dots $$ and we apply the pulse sequence in Eq.~\eqref{U_sequence} within those new dimers over a duration $T$.
\end{itemize}
The complete sequence of period $\mathcal{T}\!=\! 2T$ is illustrated in Fig.~\ref{fig_lattice_effective}(a).

Following the method of Section~\ref{section_quantum}, we readily obtain the effective Hamiltonian describing the evolution over Step 1:
\begin{align}
\hat H_{\text{eff}}^{\text{Step 1}}&=\sum_n \hat H_{\text{eff}}^{(n)} \\
&=\sum_n \bigg \{ \chi \left ( \alpha \left [\hat J_z^{(n)}\right ]^2 + (1- \alpha) \left [\hat J_y^{(n)}\right ]^2 \right ) \notag \\
&\hspace{0.4cm} - \Omega_0\,  \hat J_x^{(n)} + \frac{\eta}{4} \left [\hat N^{(n)}\right ]^2 - \frac{1}{2}\hat N^{(n)} \bigg \}.\label{H_zero_lattice_Step_1}
\end{align}
A similar expression can be derived for the complementary dimerization considered during Step 2.

The total effective Hamiltonian is then obtained through the time-evolution operator over a period $\mathcal{T}$ of the full sequence, 
\begin{equation}
 e^{-i \mathcal{T} \hat H_{\text{eff}}}\equiv \hat U (\mathcal{T};0) =e^{-i T \hat H_{\text{eff}}^{\text{Step 2}}}e^{-i T \hat H_{\text{eff}}^{\text{Step 1}}},\label{effective_definition}
\end{equation}
which can be estimated using the Trotter approximation
\begin{equation}
\hat H_{\text{eff}}\approx \frac{1}{2} \left (\hat H_{\text{eff}}^{\text{Step 1}} + \hat H_{\text{eff}}^{\text{Step 2}} \right ).\label{Trotter_steps}
\end{equation}

The pulse sequence strongly couples the original dimers in Eq.~\eqref{eq_dimers}, hence, it is relevant to relabel the sites using a single index $m\!=\!1, \dots 2 N_d$. We obtain the effective Hamiltonian $\hat H_{\text{eff}}$ in Eq.~\eqref{Trotter_steps} in terms of the bosonic operators $\hat a_{m}^{(\dagger)}$ as
\begin{align}
\hat H_{\text{eff}}&= \frac{U_1}{2} \sum_m  \hat a_m^{\dagger} \hat a_m^{\dagger} \hat a_m  \hat a_m \label{eff_bosonic_lattice} \\
&+U_2 \sum_m \left ( \hat a_{m+1}^{\dagger} \hat a_{m}^{\dagger} \hat a_{m+1}  \hat a_{m}+\hat a_{m-1}^{\dagger} \hat a_{m}^{\dagger} \hat a_{m-1}  \hat a_{m} \right )\notag \\
&+\frac{U_3}{2}\sum_m \left ( \hat a_{m+1}^{\dagger} \hat a_{m+1}^{\dagger} \hat a_{m}  \hat a_{m} + \hat a_{m-1}^{\dagger} \hat a_{m-1}^{\dagger} \hat a_{m}  \hat a_{m} \right )\notag \\
& -\frac{\Omega_0}{2} \sum_m \left (\hat a_{m+1}^{\dagger} \hat a_{m} + \hat a_{m-1}^{\dagger} \hat a_m  \right) + \mathcal{O} (T) , \notag
\end{align}
where the interaction parameters are given by
\begin{align}
&U_1= \left (1- \alpha (\beta-1) + \beta \right)/2, \notag \\
&U_2= \left (1+\alpha (\beta -1) \right)/2 , \notag \\
&U_3= (\alpha-1) (1-\beta)/4 . \label{g_def_lattice}
\end{align}
The effective Hamiltonian in Eq.~\eqref{eff_bosonic_lattice} contains three types of tunable interaction terms:~Hubbard (on-site) interactions of strength $U_1$, nearest-neighbor interactions of strength $U_2$ and pair-hopping processes of strength $U_3$. We point out that all interaction terms are uniformly defined over the entire lattice. Such an extended Bose-Hubbard model is known to exhibit a rich phase diagram~\cite{luhmann2016twisted}, which displays time-reversal-symmetry-broken superfluid phases, pair superfluid and supersolid phases, and unconventional Mott insulators. The driven setting described in this Section thus offers a realistic platform for the fine exploration of these intriguing phases of quantum matter.

We note that the two-step sequence presented in this Section can be generalized in multiple ways. For instance, one could modulate the strength of the bare interactions between the various steps of the sequence, and possibly exploit this feature within additional Trotter steps. Such schemes would allow for independent control over all interaction processes in the effective Hamiltonian~\eqref{eff_bosonic_lattice}.

\subsection{Lattice system with drive-induced four-wave mixing}

In the classical limit, $\hat a_{m} \rightarrow \psi_{m}$, the driven lattice system described above [Eq.~\eqref{eff_bosonic_lattice} and Fig.~\ref{fig_lattice_effective}(a)] is effectively described by the coupled NLSE
\begin{align}
&i \frac{\partial \psi_m}{\partial t} = U_1 \vert \psi_m \vert^2 \psi_m + U_2 \left (\vert \psi_{m+1} \vert^2 + \vert \psi_{m-1} \vert^2  \right ) \psi_m \notag \\
&+U_3 \psi_m^* \left(\psi_{m+1}^2 + \psi_{m-1}^2 \right ) - \frac{\Omega_0}{2} \left (\psi_{m+1} + \psi_{m-1} \right ) , \label{eff_NLSE_lattice}
\end{align}
where the strength of the various nonlinearities $U_{1,2,3}$ (self-phase modulation, cross-phase modulation and four-wave mixing) are provided in Eq.~\eqref{g_def_lattice}. 

This driven nonlinear setting is well-suited to explore the impact of exotic nonlinearities on discrete solitons. In particular, preliminary studies suggest that the drive-induced four-wave mixing in Eq.~\eqref{eff_NLSE_lattice} can stabilize inter-site solitons~\cite{marius}, which are generically unstable for on-site nonlinearities~\cite{kevrekidis2009discrete}.

\subsection{Dimerized lattice with effective pair hopping}\label{orbital_order_section}

In this Section, we consider a slightly different driving sequence, which preserves the dimerized structure of Eqs.~\eqref{eq_dimers}-\eqref{H_zero_lattice}. As in the previous Section~\ref{section_uniform_pair}, the pulse sequence is split into two main steps:
\begin{itemize}
\item Step 1: We apply the pulse sequence in Eq.~\eqref{U_sequence} within each individual dimer over a duration $T$.
\item Step 2: We activate single-particle hopping processes between neighboring dimers (to be specified), over a duration $T$.
\end{itemize}
The complete sequence of period $\mathcal{T}\!=\! 2T$ is illustrated in Fig.~\ref{fig_lattice_effective}(b).

A broad class of models can be designed using this driving sequence. For the sake of concreteness, we focus our study on a specific model obtained by setting the parameters $\alpha\!=\!\Omega_0\!=\!0$ at Step 1, such that the effective Hamiltonian describing the time-evolution over Step 1 reduces to 
\begin{align}
\hat H_{\text{eff}}^{\text{Step 1}}&=\sum_n \hat H_{\text{eff}}^{(n)} \notag \\
&=U \sum_n \left ( \left [\hat N^{(n)}\right ]^2 - \xi \left [\hat J_y^{(n)}\right ]^2 \right ) - \frac{1}{2}\sum_n\hat N^{(n)}, \label{H_eff_lattice_Step_1}
\end{align}
where $U\!=\!(\beta+1)/4$ and $\xi\!=\!4(\beta-1)/(\beta+1)$. Moreover, we consider the single-particle processes activated in Step 2 to be of the form
\begin{align}
\hat H^{\text{Step 2}}=& - 2 \Omega \sum_n \left (\hat a^{\dagger}_{n+1,1}\hat a_{n,1} + \hat a^{\dagger}_{n+1,2}\hat a_{n,2} + \text{h.c.} \right ) \notag \\
&- 2 \Omega_{12} \sum_n \left (\hat a^{\dagger}_{n+1,1}\hat a_{n,2} + \text{h.c.} \right ),
\end{align}
as we illustrate in Fig.~\ref{figVI_1}(a); see also Section~\ref{section_exp} on possible realizations.

Altogether, the total effective Hamiltonian is obtained through the time-evolution operator over a period $\mathcal{T}$ of the full sequence [Eq.~\eqref{effective_definition}], and it reads
\begin{align}
\hat H_{\text{eff}}&=\frac{U}{2} \sum_n \left ( \left [\hat N^{(n)}\right ]^2 - \xi \left [\hat J_y^{(n)}\right ]^2\right )  \label{Heff1}\\
&- \Omega \sum_n \left (\hat a^{\dagger}_{n+1,1}\hat a_{n,1} + \hat a^{\dagger}_{n+1,2}\hat a_{n,2} + \text{h.c.} \right ) \notag \\
&- \Omega_{12} \sum_n \left (\hat a^{\dagger}_{n+1,1}\hat a_{n,2} + \text{h.c.} \right ) \notag \\
&-\left[\frac{U}{2}\left(1-\frac{\xi}{4}\right)+\mu\right]\sum_{n}\hat{N}^{(n)}+ \mathcal{O} (T) , \notag
\end{align}
where we introduced the chemical potential $\mu$ in the last line for later convenience. This effective dimerized lattice model is illustrated in Fig.~\ref{figVI_1}(a).



The Hamiltonian in Eq.~\eqref{Heff1} features interaction terms of the form $-\left [\hat J_y^{(n)}\right ]^2$, which reminds the models describing interacting bosons in $p$-bands~\cite{hebert2013exotic,li2016physics}; see also Ref.~\cite{di2021chiral}. Indeed, repulsive bosons in $p_{x,y}$ orbitals experience a characteristic orbital-type coupling of the form $-\hat{L}_z^2$, where $\hat{L}_z\!=\!i(\hat{p}^{\dagger}_{x}\hat{p}_{y}-\hat{p}^{\dagger}_{y}\hat{p}_{x})$ is the angular-momentum operator. The operator $\hat J_y^{(n)}\!=\!(i/2) (\hat a_{n,1}^{\dagger} \hat a_{n,2} - \hat a_{n,2}^{\dagger} \hat a_{n,1} )$ entering the first line of Eq.~\eqref{Heff1} can thus be interpreted as a local angular momentum, with the two modes $\hat a_{1,2}$ playing the role of $p_{x,y}$ orbitals. 


It is the aim of the next Section~\ref{section_GS} to explore the consequences of these unconventional interactions and orbital structure on the ground-state properties of the dimerized lattice in Eq.~\eqref{Heff1}.

\section{Bosonic phases in a dimerized lattice with effective pair hopping}\label{section_GS}


\subsection{Orbital order and emergent magnetic fluxes}

\mbox{When setting $\xi\!>\!0$, the peculiar interaction term $-\xi\left [\hat J_y^{(n)}\right ]^2$} in Eq.~\eqref{Heff1} favors an orbital-ordered ground state, which exhibits finite ``angular momentum" at the level of each dimer:~\mbox{$\vert J_y^{(n)} \vert\!\ne\!0$} is maximized in the ground state, hence leading to a spontaneous breaking of time-reversal symmetry (TRS). Indeed, the angular-momentum states $\vert b_{\pm}^{(n)} \rangle$, which diagonalize the $\hat{J}_{y}^{(n)}$ operator, 
\begin{equation}
\hat{J}_{y}^{(n)} \vert b_{\pm}^{(n)} \rangle = (\pm) \vert b_{\pm}^{(n)} \rangle, \qquad \vert b_{\pm}^{(n)} \rangle = \hat{b}^{\dagger}_{n,\pm} \vert \emptyset \rangle,
\end{equation}
have a complex structure given by
\begin{align}
& \hat{b}^{\dagger}_{n,\sigma} = \frac{1}{\sqrt{2}} \left (\hat{a}^{\dagger}_{n,1} + i\sigma \hat{a}^{\dagger}_{n,2}\right ),  \qquad \sigma\!=\!\pm ,
 \label{Jbasis}\\
 &\mathsf{T}\, \hat{b}^{\dagger}_{n,\sigma} \, \mathsf{T}^{-1}\!=\!\hat{b}^{\dagger}_{n,\,\overline{\sigma}}, \qquad \hspace{1.5cm}\overline{\sigma}\!=\!-\sigma,\notag
\end{align}
where $\mathsf{T}$ is the TRS operator. The spontaneous breaking of TRS leads to rich phases and chiral dynamics in the ground state, as we describe below. Henceforth, we set $\xi\!>\!0$ except otherwise stated.

It is instructive to write the effective Hamiltonian~\eqref{Heff1} in the angular-momentum-state basis ($b_{\pm}$),
\begin{eqnarray}
\hat{H}_{\textrm{eff}} &=& \frac{1}{2}\sum_{n\sigma}\Bigg[U_{\xi} \, \hat{n}_{n,\sigma}(\hat{n}_{n,\sigma}-1) +W_{\xi} \, \hat{n}_{n,\sigma}\hat{n}_{n,\,\overline{\sigma}}\Bigg]
\notag \\
&-&\sum_{n\sigma} \left(t_{\sigma} \, \hat{b}^{\dagger}_{n+1,\sigma}\hat{b}_{n,\sigma}+\text{h.c.}\right )
\notag \\
&-&\sum_{n\sigma}\left(t_{\sigma \overline{\sigma}} \, \hat{b}^{\dagger}_{n+1,\sigma}\hat{b}_{n,\,\overline{\sigma}} + \text{h.c.}\right)-\mu\sum_{n\sigma}\hat{n}_{n,\sigma}, \qquad \label{Heffbb}
\end{eqnarray}
where $\hat{n}_{n,\sigma}\!=\!\hat{b}^{\dagger}_{n,\sigma}\hat{b}_{n,\sigma}$. The Hamiltonian in Eq.~\eqref{Heffbb} features intra-species and inter-species interactions of strength
\begin{equation}
U_{\xi}=U(1-\xi/4), \qquad  W_{\xi}=U(1+\xi/4) ,
\end{equation}
as well as \emph{complex} tunneling matrix elements given by
\begin{align}
&t_{\sigma}=\left(\Omega+\frac{\Omega_{12}}{2}e^{i\sigma\pi/2}\right)=\vert t_{\sigma} \vert e^{i \sigma \Phi}, \label{complex_hopping} \\
&t_{\sigma \overline{\sigma}}=\frac{\Omega_{12}}{2}e^{-i\sigma\frac{\pi}{2}}.
\end{align}

\begin{figure}[t]
\center
\includegraphics[width=0.9\columnwidth]{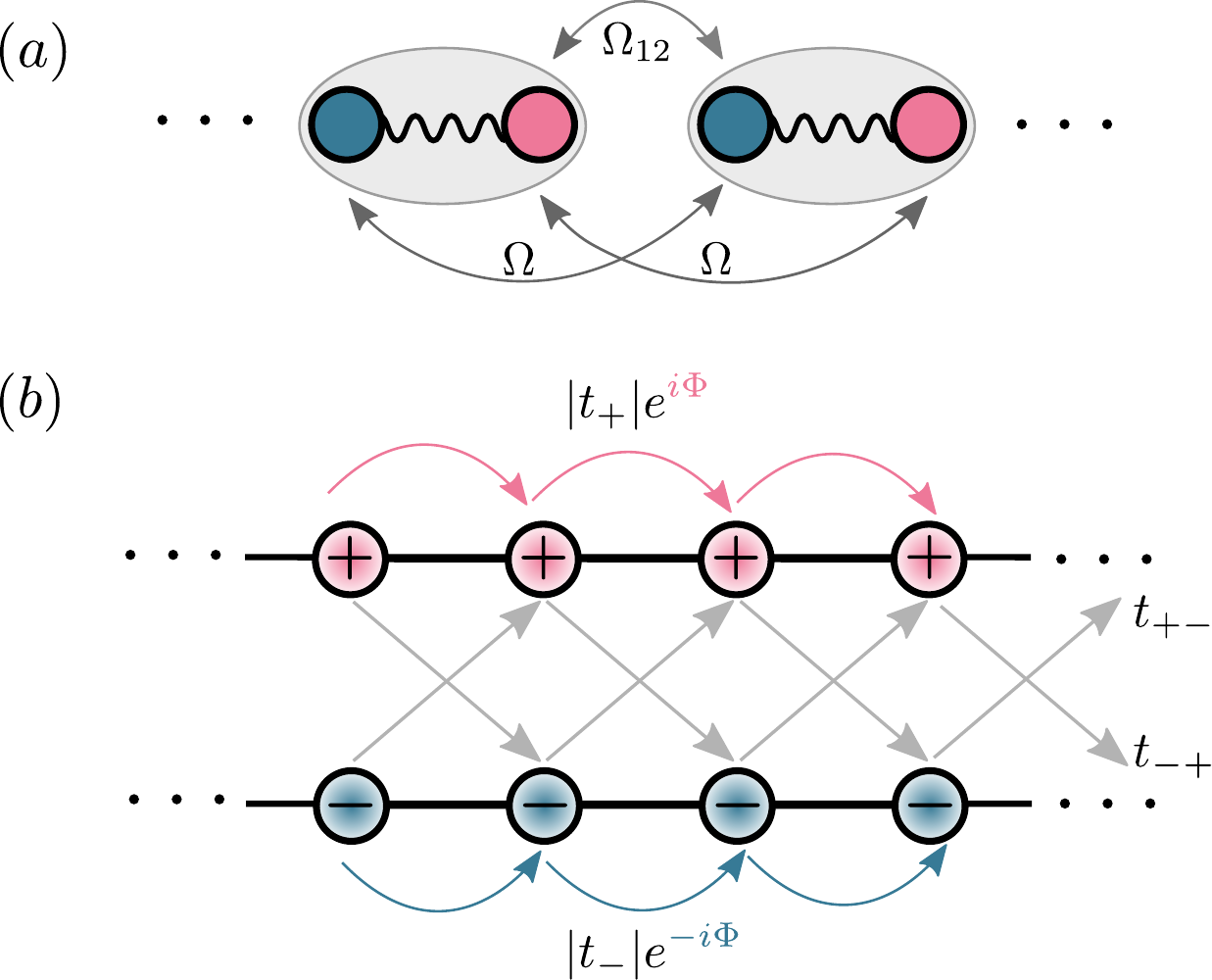}
\caption{(a) Illustration of the dimerized lattice model in Eq.~\eqref{Heff1}, with inter-dimer couplings $\Omega$ and $\Omega_{12}$. (b) Sketch of the fictitious Creutz-Hubbard ladder in Eq.~\eqref{Heffbb}, as obtained when considering the angular-momentum-state representation ($b_{\pm}$). The arrows depict hopping processes, and their color reflect the complex phase acquired upon tunneling ($+\Phi$ in red, $-\Phi$ in blue). When projected onto a single leg (i.e.~when spontaneous orbital ordering occurs), the lattice exhibits an emergent flux $\Phi\!=\!\pm \text{atan}(\Omega_{12}/2 \Omega)$, giving rise to persistent currents.}
\label{figVI_1}
\end{figure}

In this picture, the problem can be interpreted as a fictitious Creutz-Hubbard ladder~\cite{tovmasyan2013geometry,takayoshi2013phase,junemann2017exploring}, where each leg is entirely composed of either $b_{+}$ or $b_{-}$ orbitals; see the sketch in Fig.~\ref{figVI_1}(b). For \mbox{$\xi\!>\!0$}, the inter-species interaction $W_{\xi}$ always dominates over the intra-species interaction $U_{\xi}$, hence favoring the stabilization of an orbital ordering in the system. 

To capture this orbital order, we introduce the local orbital polarization, which is defined at each dimer as
\begin{equation}
m^{(n)}_0 \equiv -\frac{2}{\rho_n}\langle \hat{J}_y^{(n)}\rangle_{0} =\frac{1}{\rho_n} \left ( \langle \hat{n}_{n,+}\rangle_{0}-\langle\hat{n}_{n,-}\rangle_{0} \right ).\label{eq_polarization}
\end{equation}
Here, $\rho_n\!=\!\langle \hat{N}^{(n)}\rangle_0$ denotes the local density of bosons, and we henceforth use the notation $\langle \hdots \rangle_0$ to express the mean value of an operator in the ground state.

The effective Hamiltonian $\hat{H}_{\textrm{eff}}$ [Eqs.~\eqref{Heff1},\eqref{Heffbb}] displays two types of hopping terms. As we will demonstrate later in this Section, the kinetic terms $\sim\Omega$ stabilize a uniform ``ferromagnetic" ordering in the chain of dimers, while the terms $\sim\Omega_{12}$ (which are absent in p-band models~\cite{li2016physics}) are responsible for the emergence of an effective magnetic flux and chiral currents. This can be intuitively grasped from the complex tunneling matrix elements in Eq.~\eqref{complex_hopping}, and which are illustrated in Fig.~\ref{figVI_1}(b):~Each leg of the fictitious ladder ($\sigma\!=\!\pm$) is associated with a magnetic flux, 
\begin{equation}
\Phi_{\sigma}= \sigma \times \Phi , \qquad \Phi = \text{atan}(\Omega_{12}/2 \Omega),\label{the_flux}
\end{equation}
 such that a macroscopic occupation of a single leg (through spontaneous orbital ordering) leads to the emergence of a chiral persistent current:~a clear signature of TRS breaking in the system. This simple picture illustrates how the emergent chirality of the system is directly determined by the interaction-induced orbital polarization $m_0$ in Eq.~\eqref{eq_polarization}.

These peculiar properties will now be explored in detail, both in the mean-field limit (relevant for nonlinear optics and weakly-interacting bosonic gases) and in the quantum (strongly-correlated) regime. We will also present a practical quench protocol, which dynamically reveals the presence of orbital polarization in this unconventional lattice system.

\subsection{Mean-field regime: orbital polarization and chiral currents}

We start by analyzing the mean-field (classical) regime of the Creutz-Hubbard ladder in Eq.~\eqref{Heffbb}, which is obtained by performing the substitution
\begin{equation}
  \hat{b}_{n,\sigma}\rightarrow \langle \hat{b}_{n,\sigma}\rangle \equiv \psi_{n,\sigma}. 
\end{equation}
The corresponding NLSE (expressed in the $b_{\pm}$ basis) reads
\begin{align}
i &\frac{\partial \psi_{n,\sigma}}{\partial t} \!\!=\!\! \Bigg[U_{\xi}|\psi_{n,\sigma}|^2 +W_{\xi}|\psi_{n,\,\overline{\sigma}}|^2-\left(\frac{U_{\xi}}{2}+\mu\right)\Bigg]\psi_{n,\sigma} \notag \\
&-t_{\sigma} \, \psi_{n-1,\sigma} - t_{\sigma \overline{\sigma}} \, \psi_{n-1,\,\overline{\sigma}}-t_{\sigma}^{*} \, \psi_{n+1,\sigma} - t_{\sigma \overline{\sigma}}^{*} \, \psi_{n+1,\,\overline{\sigma}}.\notag \\
\label{DNLS}
\end{align}

We aim at determining the ground-state properties of this system, setting the focus on the emergence of orbital order in the regime $\xi\!>\!0$. Following a self-consistent mean-field approach, we obtain analytical predictions for the orbital polarization and the chiral persistent current in terms of the system parameters. We hereby summarize our findings, and refer the reader to Appendix~\ref{appendix_MF} for a detailed analysis. 

\subsubsection{The orbital polarization}

First of all, we find that the ground-state orbital polarization $m_0$ is directly related to the relative phase $\varphi$ between the two components of the condensate within each dimer [see Eq.~\eqref{psi_z_phi}],
\begin{equation}
\varphi = \textrm{atan}\left(\frac{m_0}{\sqrt{1-m_0^2}}\right).   
 \label{theta}
\end{equation}
Furthermore, we find that these local ground-state properties ($m_0$, $\varphi$) are uniform over the entire dimerized lattice. Hence, a ground state with finite polarization $m_0\!\neq\!0$ defines a `chiral' superfluid phase (CSF), which is characterized by a uniform twisting of the phase $\varphi$ over the dimerized lattice. We note that similar twisted superfluid phases have been identified in other classes of models supporting pair-tunneling processes~\cite{luhmann2016twisted}. 

When the coupling $\Omega_{12}\!=\!0$, condensation occurs at quasi-momentum $k_0\!=\!0$ and the system exhibits two degenerate ground states with opposite orbital polarizations $m_{0}\!=\!\pm 1$; according to Eq.~\eqref{theta}, this corresponds to a relative phase $\varphi\!=\!\pm\pi/2$ within each dimer. In this way, the ground state maximizes the ``angular momentum" $\vert J_y^{(n)}\vert$ at the level of each dimer and it spontaneously develops a ``ferromagnetic" ordering throughout the dimerized lattice:~orbital order emerges through the spontaneous breaking of TRS [Eq.~\eqref{Jbasis}].

For a small finite coupling $\Omega_{12}$, the ground-state polarization is found to decrease as
\begin{equation}
 m_0 \approx \pm\left(1 \mp \frac{1}{2}\left(\frac{\Omega_{12}}{\Omega_c}\right)^2\right), \qquad \text{for} \,\,\,  \Omega_{12}\ll \Omega_c,\label{pol_small}
\end{equation} 
where we introduced the critical value
\begin{equation}
\Omega_c = \frac{\rho}{2}(W_{\xi}-U_{\xi}) = U\rho\xi/4 .   \label{Omega_c}
\end{equation}
Here, $\rho\!=\!\sum_{\sigma}|\psi_{n,\sigma}|^2\!=\!N/N_d$ denotes the particle density, with $N$ the total number of bosons and $N_d$ the number of dimers, and we considered periodic boundary conditions. Furthermore, condensation is found to occur at finite quasi-momentum, which for $\Omega_{12} \ll \Omega_c$ reads
\begin{equation}
k_0 \approx \left ( \Omega_{12}/2\Omega \right ) \textrm{sgn}(m_0) \approx \Phi \, \textrm{sgn}(m_0),\label{finite_k}
\end{equation} 
hence reflecting the emergence of an effective flux $\Phi$ [Eq.~\eqref{the_flux}] dictated by orbital order.

For small dimerized lattices, such that \mbox{$N_d\!<\!2\pi\Omega/|\Omega_{12}|$}, condensation occurs at $k_0\!=\!0$ for any value of the coupling $\Omega_{12}$. In this case, the ground-state orbital polarization can be obtained analytically as
\begin{equation}
 m_0 = 
\begin{cases}
\label{pol_0}
\pm \sqrt{1-\left(\frac{\Omega_{12}}{\Omega_{c}}\right)^2}  & \Omega_{12}<\Omega_{c}\\
0 & \textrm{otherwise}.
\end{cases}
\end{equation}

\begin{figure}[t]
\center
\includegraphics[width=0.9\columnwidth]{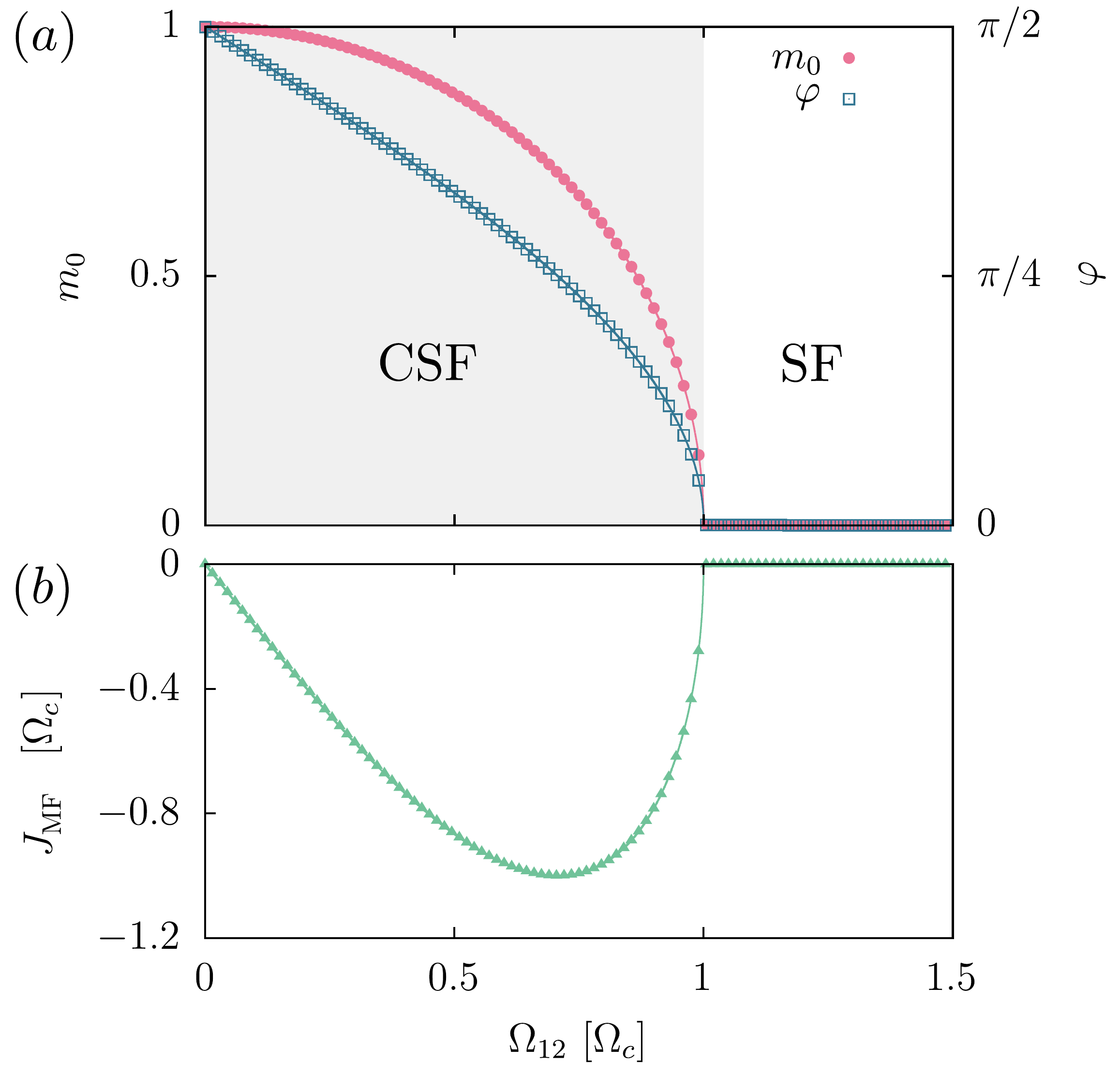}
\caption{(a) Ground-state orbital polarization $m_0$ and relative phase $\varphi$ as a function of the hopping amplitude $\Omega_{12}$, measured in units of the critical value $\Omega_c = U\xi\rho/4$. The shaded region depicts the chiral superfluid (CSF) phase with $\varphi\neq 0$, while the non-shaded region represents a more conventional superfluid (SF) phase. Note that the transition from $\varphi\!\ne\!0$ to $\varphi\!=\!0$ is analogous to the transition from Phase III to Phase I in Fig.~\ref{Fig:PhaseDiagram}; see Section~\ref{section_pendulum}. (b) Mean-field current as a function of $\Omega_{12}$. In both panels, the points where obtained through imaginary-time evolution of the NLSE in Eq.~\eqref{DNLS}, evaluating quantities in the ground state with $m_0\!>\!0$. The solid lines represent the analytical mean-field predictions given by Eqs.~\eqref{theta},~\eqref{pol_0} and~\eqref{simple_current}. The system contains $N_d\!=\!30$ dimers at filling $\rho\!=\!2$, and the interaction parameters are set to $U\!=\!0.2 \, \Omega$ and $\xi\!=\!4/3$.}
\label{figVI_2}
\end{figure}

We have validated these predictions by numerically solving the coupled NLSE~\eqref{DNLS} and performing imaginary-time evolution to reach the ground state. In order to favor one of the two degenerate TRS-broken ground states, we used an initial seed privileging the $+$ orbitals. Figure~\ref{figVI_2}(a) shows the obtained orbital polarization $m_0$ and relative phase $\varphi$ as a function of $\Omega_{12}/\Omega$. The analytical curve with $m_0\!>\!0$ [Eq.~\eqref{pol_0}] is plotted with a solid red line, presenting an excellent agreement with the numerical results (red dots). We note that the angle $\varphi$ evolves continuously from $\pi/2$ to zero, as described by Eq.~\eqref{theta}. 

We emphasize that the sharp transition displayed in Fig.~\ref{figVI_2}(a), from $m_0\!\neq\!0$ to $m_0\!=\!0$, is due to the condensation at $k_0\!=\!0$, which is imposed by the small system size ($N_d\!=\!30$ dimers). For sufficiently large lattices, we find that condensation occurs at a finite quasi-momentum $k_0$ [even beyond the limit of validity of Eq.~\eqref{finite_k}], leading to a smoother behavior of the orbital polarization; see the inset in Fig.~\ref{figVI_5}. This surprising behavior, which points to a crossover rather than a genuine phase transition, can be traced back to the peculiar form of the underlying mean-field functional; see Appendix~\ref{appendix_MF}.

Finally, it is worth noticing that the transition displayed in Fig.~\ref{figVI_2}(a), by which the relative phase changes from $\varphi\!\ne\!0$ to $\varphi\!=\!0$, is analogous to the transition from Phase III to Phase I discussed in Section~\ref{section_pendulum} for a single dimer; see Fig.~\ref{Fig:PhaseDiagram}. In the present case, the fixed points $\textrm{FP}_{*}$  are described by Eq.~\eqref{theta}, the discrete $S_2$ symmetry corresponds to TRS, and the role of the dimensionless coupling $\tilde\Omega_0$ is played by the ratio $\Omega_{12}/\Omega_c$.

\subsubsection{Chiral persistent currents}

The interplay of local orbital polarization and hopping processes gives rise to a chiral ground-state current on a ring geometry. In the mean-field regime, this chiral persistent current can be expressed in terms of the condensate's momentum $k_0$ and orbital polarization $m_0$ according to
\begin{align}
\notag
J_{\textrm{MF}}(k_0) = \rho\Bigg[&\sin(k_0)\left(2\Omega + \Omega_{12}\sqrt{1-m_0^2}\right)\\
&- 2\Omega_{12}\cos(k_0)m_0 \Bigg].     
\label{JMF}
\end{align}
For small system sizes, condensation occurs at zero momentum and we find a simple relation for the chiral current
\begin{equation}
J_{\textrm{MF}}(k_0\!=\!0)\!=\!-2\rho\Omega_{12}m_0.\label{simple_current}
\end{equation}
We have validated this analytical prediction for the chiral current by numerically solving the coupled NLSE~\eqref{DNLS}, as we show in Fig.~\ref{figVI_2}(b).

\subsubsection{The quantum regime:~Numerical validation beyond mean field}

\begin{figure}[b]
\center
\includegraphics[width=\columnwidth]{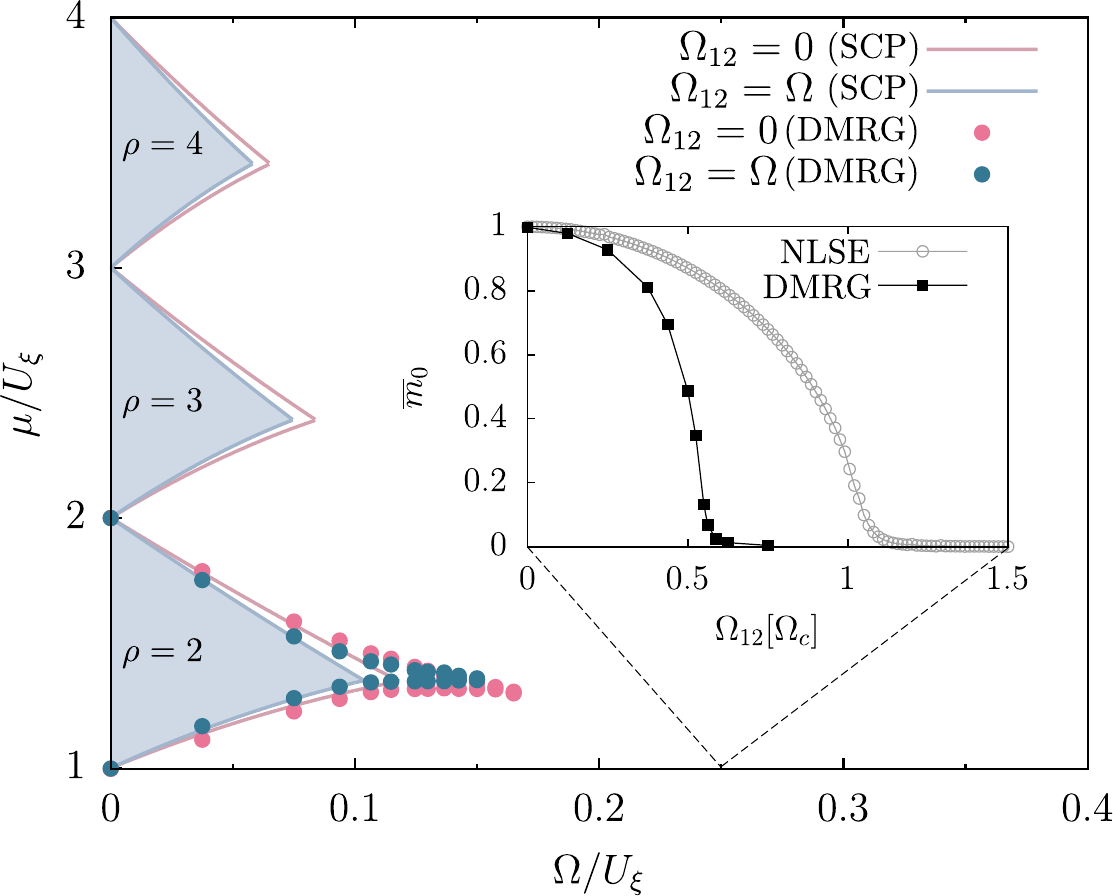}
\caption{Phase diagram of the fictitious Creutz-Hubbard ladder in Eq.~\eqref{Heffbb} as a function of the chemical potential and the ratio $\Omega/U_{\xi}$. Blue shaded areas represent the Mott insulating phases, obtained within a strong-coupling perturbation (SCP) theory for $\Omega_{12}\!=\!\Omega$ and $\rho\geq 2$. Red solid lines show the Mott-superfluid boundaries for $\Omega_{12}\!=\!0$. Filled points were obtained using a DMRG algorithm for $\Omega_{12}\!=\!0$ (red) and $\Omega_{12}\!=\!\Omega$ (blue), for a filling fraction $\rho\!=\!2$. The inset shows the evolution of the averaged ground-state orbital polarization $\overline{m}_0$ as a function of $\Omega_{12}$, for $\Omega/U_{\xi}\!=\!0.25$ and  filling $\rho\!=\!2$, in a chain with open boundary conditions ($N_d\!=\!100$ dimers); the DMRG result is compared to that obtained from the NLSE in Eq.~\eqref{DNLS}.}
\label{figVI_5}
\end{figure}

In order to validate the existence of the transition predicted by mean-field theory, we solved the full quantum many-body Hamiltonian in Eq.~\eqref{Heffbb} using density-matrix renormalization group (DMRG) methods~\cite{white1992density}. In practice, we select one of the TRS-broken ground states (with $+$ polarization) by adding a small polarizing field of strength $0.001\,\Omega$, and we keep up to 512 DMRG states to ensure a truncation error $\leq 10^{-6}$. Here, we consider a lattice containing $N_d\!=\!100$ dimers, with open boundary conditions, and we calculate the average ground-state orbital polarization,
\begin{equation}
 \overline{m}_0 = \frac{1}{\rho N_d} \sum_{n=1}^{N_d}\langle \hat{n}_{n,+}-\hat{n}_{n,-} \rangle_{0}, 
\end{equation}
for various values of the ratio $\Omega_{12}/\Omega_c$. This calculation is performed deep in the quantum regime (but still within the chiral superfluid phase), by setting $\Omega/U_{\xi}\!=\!0.25$ and a filling $\rho\!=\!2$. The resulting curve $\overline{m}_0(\Omega_{12})$ is depicted in the inset of Fig.~\eqref{figVI_5}, together with the mean-field prediction. Interestingly, the transition from the chiral superfluid ($\overline{m}_0\!\ne\!0$) to the conventional superfluid ($\overline{m}_0\!=\!0$) is still observed deep into the quantum regime. We note that the transition is qualitatively similar in that regime, although the transition point is slightly below the mean-field prediction ($\Omega_{12}\!=\!\Omega_c$). \\

\subsection{A quench protocol to measure the orbital polarization}\label{quench_section}

We have seen that the ground state of the system spontaneously breaks TRS by developing a finite orbital polarization $m_0$, whose sign reflects the privileged orbital order. This order parameter can be measured through a simple quench protocol, as we now explain. 

We assume that the system is initialized in the ground state. At $t\!=\!0$, all the dimers are suddenly decoupled from each other, so that the post-quench Hamiltonian ($t>0$) is of the form
\begin{equation}
 \hat{H}_Q\!=\!\frac{U_{\xi}}{2}\sum_{n\sigma}\hat{n}_{n,\sigma}(\hat{n}_{n,\sigma}-1) + W_{\xi}\sum_n\hat{n}_{n,+}\hat{n}_{n,-}-\!\mu \sum_{n\sigma}\hat{n}_{n,\sigma}.   
\end{equation}
The Heisenberg equations of motion for the bosonic creation and annihilation operators can be simply written as
\begin{eqnarray}
\frac{d\hat{b}_{n,\sigma}}{dt} &=& -\frac{i}{\hbar}[\hat{b}_{n,\sigma}(t),\hat{H}_Q]\\
\notag
&=& -\frac{i}{\hbar}\big(U_{\xi}\hat{n}_{n,\sigma} + W_{\xi}\hat{n}_{n,\overline{\sigma}}-\mu\big)\hat{b}_{n,\sigma}(t).
\end{eqnarray}
We point out that the operators $\hat{n}_{n,\sigma}(t)\!=\!\hat{n}_{n,\sigma}$ do not depend on time due to the fact that these quantities are conserved by the Hamiltonian $\hat{H}_Q$. Consequently, one can readily integrate these equations and find
\begin{eqnarray}
\hat{b}_{n,\sigma}(t) &=& e^{-\frac{i}{\hbar}\big(U_{\xi}\hat{n}_{n,\sigma} + W_{\xi}\hat{n}_{n,\,\overline{\sigma}}-\mu\big) t}\hat{b}_{n,\sigma}(t=0).
\end{eqnarray}
Taking the classical (mean-field) limit, and considering a ring geometry with a homogeneous density distribution, this translates into
\begin{eqnarray}
\psi_{n,\sigma}(t) &=& e^{-\frac{i}{\hbar}\big(U\rho - \sigma\Omega_c m_0 - \mu \big) t}\psi_{n,\sigma}(t=0),    
\label{evol_psi}
\end{eqnarray}
where $m_0$ is the ground-state orbital polarization (uniformly defined throughout the system). 

After the quench, the number of particles in the original orbitals $a_{1,2}$, defined at each dimer, evolves according to
\begin{eqnarray}
\notag
|\psi_{n,1}(t)|^2\!\!&=&\!\!\frac{1}{2}\left(\rho + \psi_{n,+}^{*}(t)\psi_{n,-}(t) +\psi_{n,-}^{*}(t)\psi_{n,+}(t)\right),\\
\notag
|\psi_{n,2}(t)|^2\!\!&=&\!\!\frac{1}{2}\left(\rho - \psi_{n,+}^{*}(t)\psi_{n,-}(t) -\psi_{n,-}^{*}(t)\psi_{n,+}(t)\right).\\
\label{nt}
\end{eqnarray}
Inserting Eq.~\eqref{evol_psi} into Eq.~\eqref{nt}, and using Eq.~\eqref{param} to parameterize the ground state at $t\!=\!0$, we find that the time evolution of the particle number at each dimer is described by
\begin{eqnarray}
\label{n1ev}
   |\psi_{n,1}(t)|^2 &=&  \frac{\rho}{2}\left[1+\sqrt{1-m_0^2}\sin\left(\frac{2\Omega_c m_0 t}{\hbar}\right)\right] , \\
\notag
   |\psi_{n,2}(t)|^2 &=&  \frac{\rho}{2}\left[1-\sqrt{1-m_0^2}\sin\left(\frac{2\Omega_c m_0 t}{\hbar}\right)\right].
\end{eqnarray}
As a consequence, the amplitude and frequency of these oscillations give direct information on the orbital polarization of the ground state, $m_0$, which is also straightforwardly related to the twisted superfluid angle $\varphi$ through Eq.~\eqref{theta}. 

\begin{figure}[t]
\center
\includegraphics[width=0.9\columnwidth]{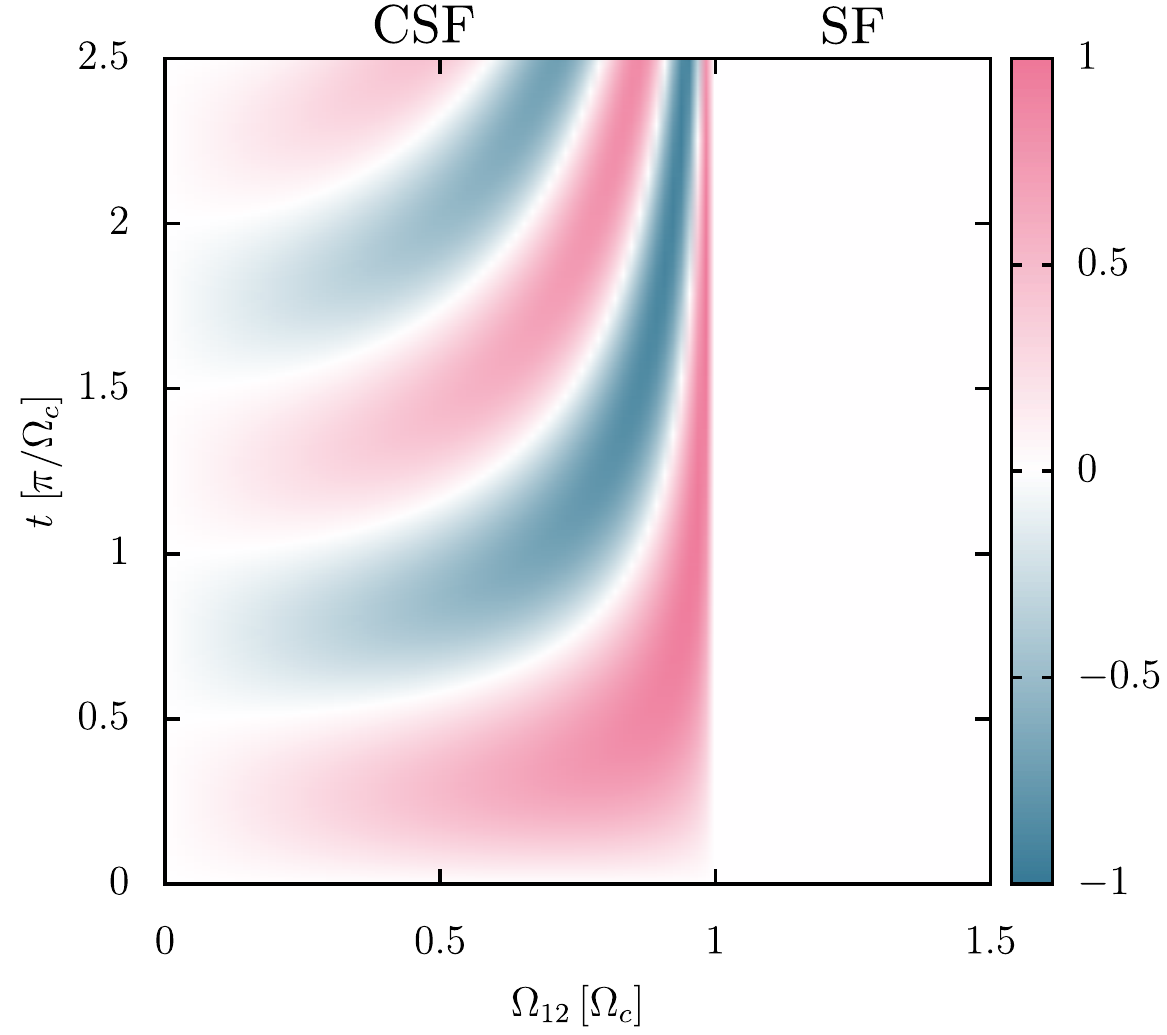}
\caption{Time evolution of the relative population at each dimer $z_n(t)$, as a function of the hopping amplitude $\Omega_{12}$, upon performing the quench protocol described in the main text. For each value of the hopping parameter, the initially evolved ground state corresponds to the one obtained in Fig.~\ref{figVI_2}, i.e.~a state with $m_0 \geq 0$.}
\label{figVI_4}
\end{figure}

Figure~\ref{figVI_4} shows the population imbalance measured at each dimer, 
\begin{equation}
z_n(t) = \frac{|\psi_{n,1}(t)|^2-|\psi_{n,2}(t)|^2}{\rho},    
\end{equation}
after numerically performing the quench protocol for different values of $\Omega_{12}$. For each value of the hopping amplitude, the initial ground state corresponds to that used in Fig.~\ref{figVI_2}, i.e.~an ordered state with $m_0 \geq 0$. We find that the dynamics obtained from these numerical simulations is perfectly described by the analytical prediction in Eq.~\eqref{n1ev}. In the conventional superfluid phase, where the system is completely depolarized, the particle number at each dimer remains unaltered.

\subsection{Strong-coupling regime \\ and the transition to the chiral Mott phase}

In the limit of strong interactions, $U \gg \Omega,\Omega_{12}$, and for a commensurate filling factor, $\rho\!=\!N/N_d\!>\!1$, the bosonic system described by Eq.~\eqref{Heffbb} is found to form a ``chiral" Mott insulating phase, characterized by an orbital ordering. As in the mean-field regime, this orbital order relies on having the interaction parameter $\xi\!>\!0$. We hereby set $\rho\!>\!1$, and treat the unit-filling case $\rho\!=\!1$ in the next Section~\ref{section_rho_one}.

When the hopping parameters are strictly zero, and when setting $\xi\!>\!0$, the particles either occupy the $b_+$ or the $b_-$ orbitals, so as to maximize the local angular momentum $\vert \hat{J}_{y}^{(n)}\vert$ at the level of each dimer. For $\rho\!=\!N/N_d\!>\!1$ bosons in each dimer, these Fock states are described by
\begin{equation}
    |\sigma_n\rangle = \frac{(\hat{b}^{\dagger}_{n,\sigma})^{\rho}}{\sqrt{\rho!}}|0\rangle,
    \label{spinstates}
\end{equation}
where $\hat{b}_{n,\sigma}$ is defined in Eq.~\eqref{Jbasis}.  In the absence of kinetic terms in the Hamiltonian~\eqref{Heffbb}, there is a macroscopic degeneracy of $2^{N_d}$ possible ground-state configurations $|\{\sigma_n\}\rangle$, which may be written as product states
\begin{equation}
 |\{\sigma_n\}\rangle = \prod_{n=1}^{N_d}|\sigma_n\rangle.\label{spinstates_bis}
\end{equation} 

The tunneling terms in the Hamiltonian~\eqref{Heffbb} do not couple these states at first order, but they do lift their degeneracy in second-order perturbation theory. Following the perturbative approach detailed in Appendix~\ref{appendix_Mott}, we obtain an effective Ising spin model 
\begin{equation}
\hat{H}^{\text{eff}}_{\text{Ising}} =K_{yy} \sum_n \hat{J}_y^{(n)} \hat{J}_y^{(n+1)}, 
 \label{exchange}
\end{equation}
with the exchange coupling
\begin{align}
K_{yy} = -\frac{4\Omega^2[W_{\xi}+\rho(W_{\xi}-U_{\xi})]}{U_{\xi}[U_{\xi} + \rho(W_{\xi}-U_{\xi})]}<0, \label{Ising_coupling}
\end{align}
where we assumed repulsive intraspecies interactions $U_{\xi}>0$ in Eq.~\eqref{Heffbb}, i.e.~$U\!>\!0$ and $\xi\!<\!4$ in the original Hamiltonian~\eqref{Heff1}. We point out that the effective Hamiltonian in Eq.~\eqref{exchange} only acts on the projected subspace spanned by the states in Eq.~\eqref{spinstates_bis}.

Importantly, the exchange coupling $K_{yy}\!<\!0$ in Eq.~\eqref{Ising_coupling} favors a uniform ``ferromagnetic" angular-momentum ordering. Moreover, the exchange coupling is found to be independent of the hopping parameter $\Omega_{12}$, at this order of perturbation theory. This analysis suggests that the orbital order identified in the superfluid phase (mean-field regime) should be preserved in the strongly-interacting regime. In analogy with $p$-band systems, we refer to this ordered (TRS-broken) Mott insulator as a ``chiral" Mott phase. We remark that $K_{yy}\!>\!0$ in $p$-bands~\cite{li2016physics}, such that the Mott phase is instead associated with a staggered ordering in that context.

The perturbative expansion described above can be further exploited to elucidate the boundaries between the chiral Mott and superfluid phases; see Appendix~\ref{appendix_Mott}. Indeed, applying a strong-coupling perturbation (SCP) theory up to second order in the hopping amplitudes, we obtained approximate boundaries for particle-type ($\mu_{+}$) and hole-type ($\mu_{-}$) excitations in the Mott phase
\begin{align}
&\frac{\mu_+}{U_{\xi}}\!\!=\!\!\rho - 2(\rho+1)\frac{|t_{\sigma}|}{U_{\xi}}+ \frac{|t_{\sigma}|^2}{U_{\xi}^2}\rho^2 - \frac{4(\rho+1)|t_{\sigma\overline{\sigma}}|^2\cos^2(\Phi)}{\rho U_{\xi }(W_{\xi}\!-\!U_{\xi})}\notag\\
&\hspace{0.67cm} -\!\!\frac{|t_{\sigma\overline{\sigma}}|^2}{U_{\xi}}\!\!\left(\frac{2\rho}{\rho(W_{\xi}\!-\!U_{\xi})\!+\!W_{\xi}\!+\!U_{\xi}}-\frac{4\rho}{\rho(W_{\xi}\!-\!U_{\xi})\!+\!U_{\xi}}\right), \notag\\ \notag\\
\vspace{0.8em}
& \frac{\mu_-}{U_{\xi}}\!\!=\!\!\rho\!-\!1\!+\!2\rho\frac{|t_{\sigma}|}{U_{\xi}}\!-\! \frac{|t_{\sigma}|^2}{U_{\xi}^2}(\rho+1)^2\!-\! \frac{2\rho|t_{\sigma\overline{\sigma}}|^2(1\!+\!2\sin^2(\Phi))}{U_{\xi}[\rho (W_{\xi}\!-\!U_{\xi})\!+\!U_{\xi}]} \notag\\
&\hspace{0.45cm} +\frac{|t_{\sigma\overline{\sigma}}|^2}{U_{\xi}}\!\!\left(\frac{2\rho}{\rho(W_{\xi}\!-\!U_{\xi})\!+\!U_{\xi}\!-\!W_{\xi}}+\frac{2(\rho-1)}{\rho(W_{\xi}\!-\!U_{\xi})\!+\!2U_{\xi}}\right).\notag \\
 \label{Mott_lobe}
\end{align}
The difference $\mu_{+}\!-\!\mu_{-}$ determines the charge gap in the Mott regime~\cite{Freericks1994}. Interestingly, due to the peculiar orbital structure of the model, the expressions in Eq.~\eqref{Mott_lobe} explicitly depend on the effective flux $\Phi$ that spontaneously emerge in the fictitious ladder for finite $\Omega_{12}$; see Eq.~\eqref{the_flux} and Refs.~\cite{Niemeyer1999,Oktel2015}. 

In the limit $\Omega_{12}\!=\!0$, the low-energy physics is entirely determined by one of the $\pm$ orbitals (the system exactly projects onto a single leg of the fictitious ladder in Fig.~\ref{figVI_1}), and we hence recover the Mott lobes of the more conventional 1D Bose-Hubbard model~\cite{Freericks1994}. The Mott-SF phase diagram resulting from Eq.~\eqref{Mott_lobe} is depicted in Fig.~\ref{figVI_5}, in terms of the chemical potential $\mu$ and ratio $\Omega/U_{\xi}$, for two values of $\Omega_{12}$: $\Omega_{12}\!=\!0$ and $\Omega_{12}\!=\!\Omega$ (i.e.~for a finite effective flux $\Phi\!\ne\!0$). 

We compare these analytical predictions with the phase diagram extracted from DMRG calculations (filled points in Fig.~\ref{figVI_5}), for the two cases $\Omega_{12}\!=\!0$ and $\Omega_{12}\!=\!\Omega$, and a filling factor $\rho\!=\!2$. We note that the SCP-theory prediction is in very good agreement with the more accurate numerical method for $\Omega \ll U_{\xi}$. Moreover, the DMRG calculations confirm the analytical prediction according to which the orbital order is unaltered by a finite coupling $\Omega_{12}$, deep in the chiral-Mott phase; see Fig.~\ref{figX} in Appendix~\ref{appendix_Mott}. As the system approaches the superfluid phase, the presence of the hopping term $\sim\Omega_{12}$ is found to destabilize the angular-momentum ordering, as we previously obtained in our mean-field analysis; see Fig.~\ref{figX} in Appendix~\ref{appendix_Mott}.

\subsection{Strongly-correlated phases at unit filling $\rho\!=\!1$}\label{section_rho_one}
For unit filling, $\rho\!=\!1$, the Ising-spin description in Eq.~\eqref{exchange} is insufficient:~it lacks spin-flip terms involving the spin-states in Eq.~\eqref{spinstates_bis}, which are now present and scale as the square of the tunneling amplitudes. In order to derive a proper low-energy theory at strong coupling, within second-order perturbation theory, we perform a canonical transformation and project the transformed Hamiltonian onto the subspace of unit occupation at each dimer; see Appendix~\ref{appendix_spinhalf} for details. Following this procedure, we map our original bosonic model to an effective spin-1/2 theory described by the effective spin Hamiltonian
\begin{align}
 \hat{H}^{\textrm{eff}}_{1/2}=& \sum_{n}\sum_{\nu=x,y,z} K_{\nu\nu}  \hat{J}_{\nu}^{(n)}\hat{J}_{\nu}^{(n+1)}+h_x\sum_n \hat{J}_{x}^{(n)} \notag\\
&-\bm{D}\cdot\sum_{n}\left(\hat{\bm{J}}^{(n)}\times \hat{\bm{J}}^{(n+1)}\right).\label{effective_spin_complete}
\end{align}
This model includes exchange couplings along all three spin directions, a magnetic field $h_x$ along the $x$-direction, as well as Dzyaloshinskii-Moriya interactions~\cite{dzyaloshinsky1958thermodynamic,moriya1960new,moriya1960anisotropic,dmitrienko2014measuring,liu2018chiral} characterized by the vector $\bm{D}$. The corresponding couplings are given by
\begin{align}
&K_{xx}=-\frac{4\Omega^2}{W_{\xi}}, \hspace{2.3cm} h_x= -4\Omega\Omega_{12}\left(\frac{1}{W_{\xi}}+\frac{1}{U_{\xi}}\right),\notag\\
&K_{yy}=-\frac{4\Omega^2}{W_{\xi}}\left(\frac{2 W_{\xi}}{U_{\xi}}-1\right), \hspace{0.3cm} \bm{D}= \left (0, \frac{4\Omega \Omega_{12}}{W_{\xi}}, 0 \right ) . \notag \\
&K_{zz}= -\frac{4\Omega^2}{W_{\xi}}\left(1-\frac{\Omega_{12}^2}{2\Omega^2}\right), \label{couplings_spinhalf}
\end{align}

Interestingly, the Dzyaloshinskii-Moriya interaction is found to originate from an interplay between the  inter-species interactions $W_{\xi}$ in Eq.~\eqref{Heffbb} and the single-particle coupling $\Omega_{12}$, which effectively produces a Rashba-type spin-orbit coupling~\cite{moriya1960new,moriya1960anisotropic}. We also note that the coupling constant $K_{yy}$ entering the Ising model in Eq.~\eqref{Ising_coupling} reduces to that in Eq.~\eqref{couplings_spinhalf} in the limit $\rho\!=\!1$.

In the limit $\Omega_{12}\!=\!0$, we obtain an XYX quantum spin-1/2 Heisenberg model
\begin{align}
\hat{H}_{\text{XYX}} &= \frac{K}{2}\sum_n \left(\hat{J}^{(n)}_{+}\hat{J}^{(n+1)}_{-}+\hat{J}^{(n)}_{-}\hat{J}^{(n+1)}_{+}\right) \label{Heisenberg}\\
&+ \, \Delta K \sum_n \hat{J}_y^{(n)}\hat{J}_y^{(n+1)},   \notag 
\end{align}
where we defined the operators $\hat{J}^{(n)}_{\pm}\!=\!\hat{J}^{(n)}_x \pm i\hat{J}^{(n)}_z$, and where we introduced the ferromagnetic coupling
\begin{equation}
K\!\equiv\!K_{xx}\!=\!K_{zz}\!=\!-4\Omega^2/W_{\xi},
\end{equation}
and the anisotropy parameter
 \begin{equation}
  \Delta =2(W_{\xi}/U_{\xi})-1.   
 \end{equation}
We note that a similar effective Hamiltonian was obtained for p-band bosons~\cite{Pinheiro2013}. In the present context, the anisotropy parameter can satisfy $\Delta >1$ upon setting $W_{\xi}\!>\!U_{\xi}$. In this case, our system privileges ferromagnetic order along the $y$-axis, hence forming a chiral Mott insulating phase with one boson per dimer. We expect that a small finite value of $\Omega_{12}$ will slightly depolarize this chiral Mott phase.

Last but not least, we note that similar Heisenberg models can be mapped onto an interacting Kitaev chain~\cite{Pinheiro2013,kitaev2001unpaired}, which suggests an interesting route towards Floquet-engineered topological superconductors~\cite{qi2011topological}.


\section{Experimental implementations \\ and concluding remarks}\label{section_exp}


\subsection{Optical cavities and photonic lattices}


This work introduces a method to engineer and tune nonlinearities in optical devices, using a designed pulse sequence that couples the optical modes in a fast and periodic manner. These repeated mixing operations simply correspond to the pulsed activation of a linear coupling between two optical modes, and they can thus be implemented in a broad range of two-mode nonlinear systems, ranging from optical resonators~\cite{cao2017experimental,cao2020reconfigurable,hill2020effects,garbin2020asymmetric} and waveguide arrays~\cite{szameit2010discrete,szameit2009inhibition} to circuit-QED platforms~\cite{roushan2017chiral}. 

In a two-mode optical cavity~\cite{cao2017experimental,hill2020effects,garbin2020asymmetric}, the pulsed operations could correspond to a coupling between the two polarization eigenmodes of the cavity, which can be directly realized by means of quarter-wave plates~\cite{kockaert2006fast,kozyreff2006fast}; see the sketch in Fig.~\ref{fig_sketch}(a). In optical-waveguide arrays~\cite{szameit2010discrete}, the two modes ($1$ and $2$) would describe light propagating in two adjacent waveguides. In this case, the pulsed linear couplings in Eqs.~\eqref{NLS_time}-\eqref{pulse} can be realized by abruptly changing the spatial separation between the two waveguides; see Fig.~\ref{fig_sketch}(b) for a sketch and Refs.~\cite{mukherjee2017experimental,maczewsky2017observation,mukherjee2020observation,mukherjee2021observation} for experimental realizations using ultrafast-laser-inscribed waveguides. Such optical-waveguide settings could benefit from the state-recycling technique of Refs.~\cite{mukherjee2018state,duncan2020synthetic}, where light is re-injected into the waveguides (and possibly modified) at every roundtrip; see also Refs.~\cite{kraych2019nonlinear,kraych2019statistical,kraych2020instabilites} regarding setups based on recirculating fiber loops. 

While we considered a generic setting that includes both self-phase and cross-phase modulations in the absence of the periodic drive [Eq.~\eqref{NLS}], we found that effective nonlinearities emerge even when a single type of bare nonlinearity is present. Importantly, we demonstrated that the strength (and sign) of effective nonlinearities can be tuned by simply adjusting the pulse sequence; see Eqs.~\eqref{NLS_effective_generalized}-\eqref{g_def} and Eq.~\eqref{g_def_lattice}.  We also emphasize that the parameter $\beta$ [i.e.~the relative strength and sign of the bare self-phase and cross-phase modulations in Eq.~\eqref{NLS}] can vary across a large number of experimental configurations~\cite{cambournac2002symmetry,delque2007polarization,hill2020effects}.

To detect the emergence of drive-induced nonlinearities, we proposed to study changes in the phase space's topology~\cite{zibold2010classical}, which can be explored by monitoring the dynamics of the relative intensity $z(t)$ and phase $\varphi(t)$ of the two optical modes. According to our numerical studies, these properties could already be revealed over ``time" scales of the order of $5-10T$, where $T$ denotes the period of the driving sequence. This is particularly appealing for waveguide settings~\cite{szameit2010discrete}, where the ``evolution time" associated with the propagation distance -- and hence the number of driving periods -- is limited. In this context, it would be interesting to combine such driving schemes with a state-recycling protocol~\cite{mukherjee2018state}.

While we considered a simple pulse sequence, characterized by the alternance of linear mixing operations and ``free" evolution [Fig.~\ref{fig_sequence}(a)], we note that more complicated protocols and configurations could be envisaged. For instance, different types of mixing processes could be activated within each period of the drive, including nonlinear processes.

The lattice models explored in Sections~\ref{section_lattice}-\ref{section_GS} could be implemented in nonlinear optics, by engineering appropriate couplings between photonic dimers. In optical-cavity implementations, each dimer would be represented by a two-mode cavity; one would then couple many of such dimers using mode-dependent couplings~\cite{hafezi2011robust,tikan2021emergent,komagata2021dissipative,tusnin2021dissipative,tikan2022protected}, hence realizing the models illustrated in Fig.~\ref{fig_lattice_effective}. These lattice models could also be realized in arrays of ultrafast-laser-inscribed waveguides~\cite{szameit2010discrete}, where the couplings between individual waveguides can be adjusted with high precision~\cite{mukherjee2017experimental,maczewsky2017observation,mukherjee2018state}. In this context, it would be exciting to study the interplay of drive-induced nonlinearities, solitons and topological band structures; see for instance Ref.~\cite{ivanov2021four}, where edge solitons were studied in the presence of four-wave mixing.

It would also be intriguing to explore the applicability of our scheme in the context of superconducting microwave cavities~\cite{chakram2021seamless}, where optical nonlinearities originate from the coupling to transmon ancillas. Indeed, it was recently shown that such optical nonlinearities can be modified by applying an off-resonant drive on the transmon ancillas~\cite{zhang2022drive}. Moreover, in circuit-QED platforms, the linear coupling between neighboring qubits can be modulated in a time-periodic manner~\cite{roushan2017chiral}; applying our pulse protocol to such settings could be used to modify the nonlinearity of the qubits, and hence, the interaction between microwave photons. In general, we anticipate that drive-induced nonlinearities, such as the effective four-wave mixing studied in this work, could be useful for nonlinear optics applications~\cite{agrawal2001applications,agrawal2012nonlinear}.

We remark that the present work relies on a non-dissipative theoretical framework. Our scheme could nevertheless be applied to driven-dissipative optical devices~\cite{carusotto2013quantum}, such as fiber ring cavities or microresonators described by the Lugiato-Lefever equation~\cite{lugiato1987spatial,haelterman1992dissipative,leo2010temporal,coen2013modeling,garbin2020asymmetric,coen2023nonlinear}, upon treating dissipation within the Floquet analysis ~\cite{higashikawa2018floquet,schnell2020there,schnell2021high}.

Finally, we note that modifying, possibly enhancing, optical nonlinearities represents a central theme throughout the realm of photonics~\cite{carusotto2013quantum,ozawa2019topological}. In general, effective photon-photon interactions are obtained by coupling a light field to a mediator, e.g.~an atom or a mechanical mode. In such settings, the effective interactions (e.g.~four-wave mixing) can be further controlled by an additional parametric drive, which either acts on the optical cavity~\cite{lemonde2013nonlinear} or on the mediator mode~\cite{lemonde2016enhanced}. In contrast, our Floquet driving scheme involves a periodic modulation of the linear coupling between optical modes [Eq.~\eqref{NLS_time}], a general method that does not rely on the nature (nor on the origin) of the bare nonlinearity. This method  allows for highly tunable drive-induced interactions, whose strength $\sim\!\chi$ remains comparable to the bare interactions; see Eqs.~\eqref{eff_final} and \eqref{eff_bosonic}. Moreover, we emphasize that this simple approach relies on a single condition:~the time-scale separation $T\!\ll\!1/\chi$. Generically, driving a non-linear setting periodically in time  leads to parametric instabilities; however, these instabilities are suppressed in the high-frequency limit $T\!\rightarrow\!0$; see Refs.~\cite{lellouch2017parametric,boulier2019parametric,wintersperger2020parametric,Carli2023}. We note that this stable regime is compatible with the aforementioned time-scale separation $T\!\ll\!1/\chi$.

\subsection{Ultracold atomic gases}

The bosonic Josephson junction (BJJ) Hamiltonian in Eq.~\eqref{Josephson} can be experimentally realized by manipulating ultracold gases of bosonic atoms~\cite{smerzi1997quantum,albiez2005direct,schumm2005bose,zibold2010classical,gross2010nonlinear,pigneur2018relaxation,recati2022coherently}. 
In the following paragraphs, we discuss possible implementations of the driving pulse sequence in Eqs.~\eqref{U_sequence}-\eqref{mix_operator}, for systems of cold atoms that either employ their internal or external degrees of freedom. We also propose ways to probe the effects associated with drive-induced nonlinearities through various observables.

\subsubsection{Two-mode systems using atomic internal states}

When using two internal states of an atom (e.g.~$^{87}$Rb) as a pseudo-spin, the interaction term $\sim \chi \hat J_z^2$ entering the BJJ Hamiltonian in Eq.~\eqref{Josephson} directly reflects atomic collisions in the two internal states; see Fig.\,\ref{fig_sketch}(c). In this context, the linear coupling $\sim\Omega_0 \hat J_x$ can be generated with high control, using coherent coupling with oscillatory (microwave) magnetic fields~\cite{zibold2010classical}. The mixing operator in Eq.~\eqref{mix_operator} is implemented using the same microwave drive, with a Rabi frequency $\Omega_\tau$ chosen such that $\Omega_\tau \tau\!=\!\pi/2$, where $\tau$ is the pulse duration [Fig.~\eqref{fig_sequence}]. The Rabi frequency $\Omega_\tau$ can be made much larger than other frequency scales in the system, such that the Floquet pulses and the internal dynamics have well separated time scales.
The strength of the non-linearity $\chi$ is typically limited by the atomic properties, however it can be tuned with the help of a Feshbach resonance~\cite{chin2010feshbach}.

The readout of the relevant observables (i.e.~the relative population $z$ and relative phase $\varphi$ in the two internal states) is routinely performed using state-selective imaging of the atomic densities. To extract the relative phase, the imaging is combined with a $\pi/2$-rotation around the $y-$direction in order to map the phase on measurable atomic densities.

\subsubsection{Atoms in double-well potentials}
Two-mode atomic systems can also be implemented using external (spatial) degrees of freedom, namely, by loading the atoms into optical~\cite{albiez2005direct} or magnetic~\cite{hofferberth_radio-frequency_2006} double-well potentials; see Fig.\,\ref{fig_sketch}(d). 

Here, the Hubbard interaction term in Eq.~\eqref{eq_parent_static} [which is equivalent to $\sim \chi \hat J_z^2$ in the BJJ Hamiltonian in Eq.~\eqref{Josephson}] is directly generated by the on-site atomic interactions. The tunneling between the wells provides the coherent coupling term and can be tuned by either changing the separation of the wells or the height of the barrier.
To implement a pulse of the driving sequence [Eqs.~\eqref{U_sequence}-\eqref{mix_operator}], the tunneling term has to be made dominant during the pulse duration. We note that a different Floquet scheme has been recently applied in a double-well experiment to control the amplitude and phase of the tunneling matrix elements~\cite{Ji_floquet2022}.

In the double-well system, the population imbalance $z$ can be directly evaluated by measuring the number of atoms in the two wells. Moreover, the relative phase $\varphi$ can be accessed by interference measurements~\cite{albiez2005direct,pigneur2018relaxation}.

The Hamiltonian in Eq.~\eqref{eq_parent_static}, and the derivation that leads to the effective Hamiltonian in Eq.~\eqref{eff_bosonic}, assumes that each well contains a single orbital:~this is equivalent to the single-mode approximation in spinor condensates~\cite{raghavan1999coherent,qu2020probing}. This scheme thus requires very limited excitations within each well over the driving pulse sequence. This can be achieved by using sequences that are slow compared to trapping frequencies; we note that the high degree of experimental control over designed potentials allows for the implementation of optimal-control schemes to optimize the performance~\cite{van2016optimal,borselli_two-particle_2021}



\subsubsection{Arrays of dimers and engineered lattice models}

The lattice models introduced in Section~\ref{section_lattice}, and represented in Fig.~\eqref{fig_lattice_effective}, could be designed by assembling an array of dimers, e.g.~using optical tweezer setups~\cite{SparTweezer}. Alternatively, one could trap two internal states of an atom at each site of an optical lattice (a ``dimer") and then activate state-dependent hopping over the lattice using laser-assisted tunneling methods~\cite{gerbier2010gauge,goldman2013realizing,wu2016realization,lin2020novel}. This scheme would allow for a fine control over the inter-dimer couplings (i.e.~the parameters $\Omega$, $\Omega_{12}$), but also, on the inter-particle interactions (i.e.~the parameter $\beta$).

\subsubsection{Probing pair-hopping processes and orbital order}

The phase space associated with the effective classical Hamiltonian in Eq.~\eqref{classical_eff_imbalanced}, which was analyzed in Section~\ref{section_pendulum}, can be finely studied using atomic Bose gases~\cite{raghavan1999coherent,zibold2010classical,zibold2012classical}. This can be readily performed by measuring the mean values of the relative population $z(t)$ and phase $\varphi(t)$ for different times and initial conditions. This would allow for the characterization of the effective Hamiltonian on a ``classical" level.

Ultracold atomic systems offer the possibility to access genuine quantum properties, such as coherent spin squeezing~\cite{esteve2008squeezing,gross2010nonlinear,julia2012dynamic,strobel2014fisher,evrard2019enhanced,kunkel_simultaneous_2019}. In particular, generalized measurements can be used to evaluate non-commuting observables (such as the imbalance $z$ and relative phase $\varphi$) within the same experimental realization~\cite{kunkel_simultaneous_2019}. Besides, the Husimi distribution [Fig.~\eqref{fig_husimi}] can be reconstructed from projective measurements~\cite{evrard2019enhanced}. 

\begin{figure}[t]
\center
\includegraphics[width=0.9\columnwidth]{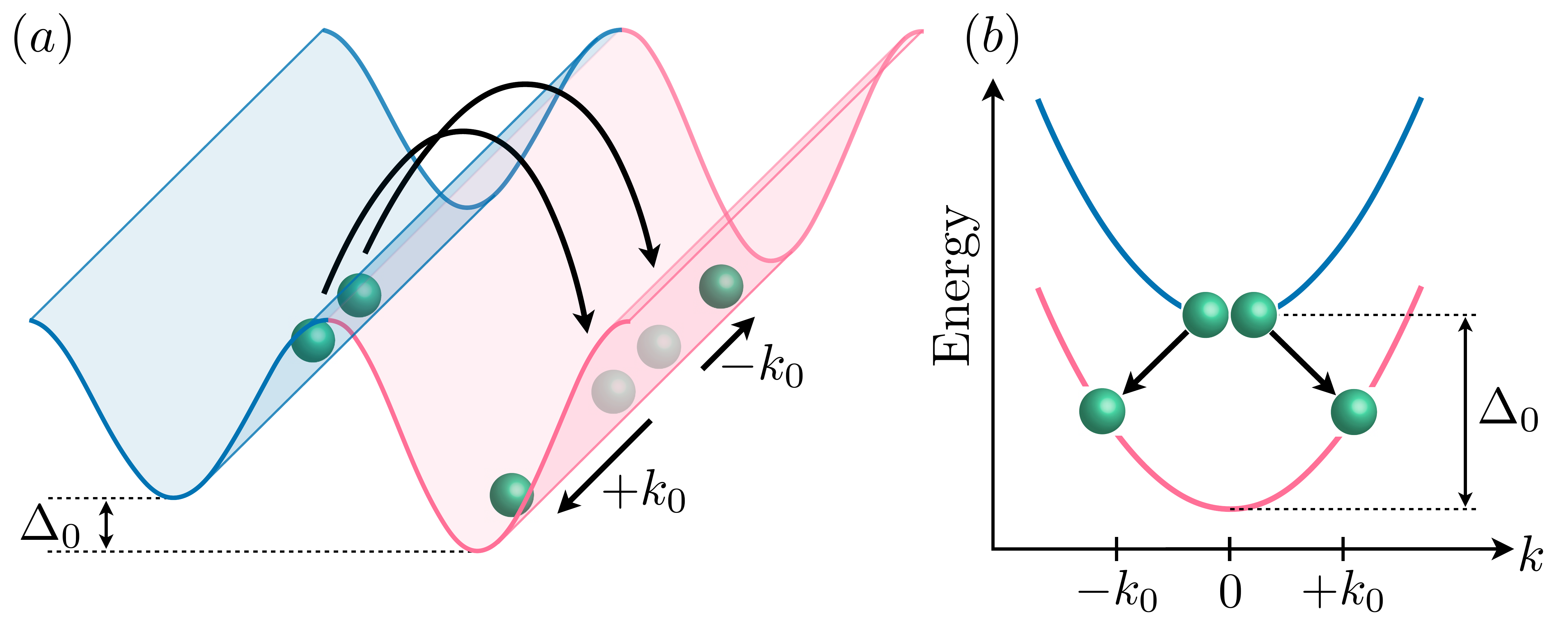}
\caption{Detecting pair-hopping processes in a driven double well potential. (a) An additional spatial degree of freedom, combined with an energy offset $\Delta_0$ between the two wells, allows for specific pair-hopping processes, which result in pairs of atoms with opposite momentum $\pm k_0$. (b) Dispersion relation associated with the two wells, which are shifted in energy by an amount $\Delta_0$. A pair-hopping process converts two atoms in the `blue' well with momentum $k\!=\!0$, into a pair of atoms in the `red' well with momentum $\pm k_0$. These finite-momentum pairs can be detected after a long TOF, hence revealing the effective pair-hopping processes generated by the driving sequence.}
\label{fig_pair_measure}
\end{figure}

Drive-induced pair-hopping processes are a striking feature of the effective Hamiltonian in Eq.~\eqref{eff_bosonic}. To detect this effect, we propose to exploit an additional spatial degree of freedom (``tube" geometry), as we illustrate in Fig.~\ref{fig_pair_measure}(a). Specifically, we apply an energy offset $\Delta_0$ to one of the wells (colored in blue), and assume that atoms are initially prepared at momentum $k\!\approx\!0$. When activating the driving sequence, pair-hopping processes are effectively generated, and atoms would then be allowed to hop by pairs to the other well (colored in red), where they would acquire a finite momentum $\pm k_0$; see Figs.~\ref{fig_pair_measure}(a)-(b). The momentum correlation could then be revealed experimentally by letting the cloud expand for a long time-of-flight~\cite{borselli_two-particle_2021}:~the finite momentum leads to a separation of the atom pairs, such that counting the number of atoms at $\pm k_0$ would reveal the correlation in the reduced variance (compared to a binomial distribution) of the population imbalance.

Finally, we note that the quench protocol introduced in Section~\ref{quench_section} could be directly implemented in a quantum-gas experiment, in view of revealing the orbital order and TRS-broken nature of the chiral superfluids and Mott phases analyzed in Section~\ref{section_GS}. As illustrated in Fig.~\ref{figVI_4}, the finite orbital polarization in the ground state can be unambiguously detected by monitoring the time-evolving population imbalance $z_n(t)$, locally defined at the level of each dimer, upon performing the quench protocol.


\paragraph*{Acknowledgments} 
This work was initiated through discussions with J. Fatome and S. Coen, who are warmly acknowledged. The authors also thank M.~Bukov, I.~Carusotto, N.~R. Cooper, J. Dalibard, A.~Eckardt, N.~Englebert, M. J\"urgensen, Yun Li, B.~Mera,  F.~Petiziol, M. C. Rechtsman,  J. Schmiedmayer, A.~Schnell and H.~Strobel for various discussions. We are very grateful to S. Coen, N.~Englebert, J. Fatome, P.~Kockaert  and S. Mukherjee for their comments on an early version of this manuscript. 

N.~G and L.~P.~G. are supported by the FRS-FNRS (Belgium), the ERC Starting Grants TopoCold and LATIS and the EOS project CHEQS. O.~K.~D. acknowledges funding from the International Max Planck Research School for Quantum Science and Technology (IMPRS - QST). L.~B. acknowledges funding from Politecnico di Torino, starting package Grant No. 54 RSG21BL01. The work in Vienna was performed under the QUANTERA project MENTA (FWF: I-6006) and M.~P. has received funding from the European Union’s Horizon 2020 research and innovation program under the Marie Sklodowska-Curie grant agreement No 101032523. M.~D.~L. is supported by the Rita Levi Montalcini Program.


\begin{appendix}

\section{Useful formulas}\label{sect_app}

The Section~\ref{section_quantum} uses two families of operators:~the bosonic operators $\hat a_{1}^{(\dagger)}$ and $\hat a_{2}^{(\dagger)}$ associated with the two modes, and which satisfy the canonical bosonic commutation relations, $[\hat a_{s},\hat a_{s'}^{\dagger}]=\delta_{s,s'}$, where $s\!=\!1,2$; and the angular momentum (Schwinger) operators, defined as
\begin{align}
&\hat J_x = \frac{1}{2} \left (\hat a_1^{\dagger} \hat a_2 + \hat a_2^{\dagger} \hat a_1 \right ) , \quad \hat J_y = \frac{1}{2i} \left (\hat a_2^{\dagger} \hat a_1 - \hat a_1^{\dagger} \hat a_2 \right ), \notag \\
&\hat J_z = \frac{1}{2} \left (\hat a_2^{\dagger} \hat a_2 - \hat a_1^{\dagger} \hat a_1 \right ) , \quad \hat N = \hat a_1^{\dagger} \hat a_1 + \hat a_2^{\dagger} \hat a_2. \label{app_Schwinger}
\end{align}
These operators satisfy the spin commutation relations $[\hat J_{\mu}, \hat J_{\nu}]=i\varepsilon_{\mu \nu \lambda} \hat J_{\lambda}$, and the operator $\hat N$ counts the total number of bosons in the system (assumed to be constant). 

In view of expressing interaction processes with Schwinger operators, it is useful to note that
\begin{align}
&\hat J_z^2 = \frac{1}{4} \left ( \hat a_1^{\dagger} \hat a_1^{\dagger} \hat a_1  \hat a_1 + \hat a_2^{\dagger} \hat a_2^{\dagger} \hat a_2 \hat a_2  - 2 \hat a_1^{\dagger} \hat a_2^{\dagger} \hat a_1  \hat a_2 + \hat N \right ) ,\notag\\
&\hat J_y^2 = \frac{1}{4} \left (2 \hat a_1^{\dagger} \hat a_2^{\dagger} \hat a_1  \hat a_2 - \hat a_1^{\dagger} \hat a_1^{\dagger} \hat a_2  \hat a_2 - \hat a_2^{\dagger} \hat a_2^{\dagger} \hat a_1 \hat a_1   + \hat N \right ) ,\notag \\
&\hat N^2 = \left (\hat a_1^{\dagger} \hat a_1^{\dagger} \hat a_1  \hat a_1 + \hat a_2^{\dagger} \hat a_2^{\dagger} \hat a_2 \hat a_2  + 2 \hat a_1^{\dagger} \hat a_2^{\dagger} \hat a_1  \hat a_2 + \hat N \right ) .\label{useful}
\end{align}
Hence, both $\hat J_z^2$ and $\hat N^2$ contain intra-mode (Hubbard) and inter-mode (cross) interactions, while $\hat J_y^2$ contains a combination of inter-mode interactions and pair-hopping processes [Fig.~\ref{fig_process}]. We point out that $\hat J_z^2$ is related to $\hat J_y^2$ through a unitary transformation; see Eq.~\eqref{BCH}.

From Eq.~\eqref{useful}, we can express the intra-mode (Hubbard) interaction terms as
\begin{equation}
\frac{1}{2} \left ( \hat a_1^{\dagger} \hat a_1^{\dagger} \hat a_1  \hat a_1 + \hat a_2^{\dagger} \hat a_2^{\dagger} \hat a_2 \hat a_2 \right ) = \hat J_z^2 + \text{constant},
\end{equation}
where the irrelevant constant term reads $\hat N (\hat N -2)/4$. Similarly, the inter-mode (cross) interaction term reads
\begin{equation}
\hat a_1^{\dagger} \hat a_2^{\dagger} \hat a_1  \hat a_2 = - \hat J_z^2 + \text{constant},
\end{equation}
with the irrelevant constant term $\hat N^2/4$. These expressions were used to derive the Hamiltonian in Eq.~\eqref{Josephson} from Eq.~\eqref{eq_parent_static}.

Finally, it is useful to note that a combination of intra-mode (Hubbard) interactions and pair-hopping processes can be expressed as
\begin{align}
\hat J_z^2 + \hat J_y^2&= \frac{1}{4} \left ( \hat a_1^{\dagger} \hat a_1^{\dagger} \hat a_1  \hat a_1 + \hat a_2^{\dagger} \hat a_2^{\dagger} \hat a_2 \hat a_2 \right ) \notag \\
&- \frac{1}{4} \left (\hat a_1^{\dagger} \hat a_1^{\dagger} \hat a_2  \hat a_2 + \hat a_2^{\dagger} \hat a_2^{\dagger} \hat a_1 \hat a_1 \right ) + \text{constant}.
\end{align}

\section{Orbital order and phase transitions from a mean-field analysis}\label{appendix_MF}

In this Appendix, we provide a detailed mean-field analysis of the extended Bose-Hubbard model in Eq.~\eqref{Heffbb}.

Upon performing the mean-field substitution
\begin{equation}
  \hat{b}_{n,\sigma}\rightarrow \langle \hat{b}_{n,\sigma}\rangle \equiv \psi_{n,\sigma},
\end{equation}
we obtain the mean-field functional
\begin{eqnarray}
\notag
\mathcal{F}\!&=&\!\frac{1}{2}\sum_{n\sigma}\Bigg[U_{\xi}|\psi_{n,\sigma}|^2(|\psi_{n,\sigma}|^2-1)+ W_{\xi}|\psi_{n,\sigma}|^2|\psi_{n,\,\overline{\sigma}}|^2\Bigg]\\
\label{Fbb}
\!&-&\!\sum_{n\sigma} \Bigg[\left(\Omega+\frac{\Omega_{12}}{2}e^{i\sigma\pi/2}\right)\psi^{*}_{n+1,\sigma}\psi_{n,\sigma}+h.c.\Bigg]\\
\notag
\!&-&\!\frac{\Omega_{12}}{2}\sum_{n\sigma}\left(e^{-i\sigma\frac{\pi}{2}}\psi^{*}_{n+1,\sigma}\psi_{n,\,\overline{\sigma}}+h.c.\right)-\mu\sum_{n\sigma}|\psi_{n,\sigma}|^2 ,
\end{eqnarray}
from which we derive the coupled NLSE in Eq.~\eqref{DNLS}.

When imposing periodic boundary conditions (ring geometry), and in the limit of weak interactions $U\!\ll\!\Omega$, the mean-field ground state is expected to have a uniform density distribution over the entire chain. We can then propose Bloch states as stationary solutions of Eq.~\eqref{DNLS}, namely    
\begin{align}
&\psi_{n,\sigma}(t)=e^{-i(\varepsilon(k)-\mu) t/\hbar} \, e^{i k n} \, \phi_{k,\sigma}, \notag \\
&\phi_{k,\sigma}=\sqrt{\frac{\rho}{2}[1+\sigma m(k)]}e^{i\Theta_{k,\sigma}} .
 \label{param}
\end{align}
In this way, the local density $\rho_n\!=\!\rho\!=\!\sum_{\sigma}|\psi_{\sigma}|^2\!=\!N/N_d$ is constant, with $N$ the total number of bosons and $N_d$ the number of unit cells (dimers) in the ring. We note that the orbital polarization entering Eq.~\eqref{param} is given by
\begin{equation}
 m(k) = \frac{1}{\rho}(|\phi_{k,+}|^2-|\phi_{k,-}|^2). 
 \label{mk}
\end{equation}
Inserting the ansatz~\eqref{param} into the NLSE in Eq.~\eqref{DNLS}, we find that $\phi_{k}=(\phi_{k,+}, \phi_{k,-})^{T}$ should be an eigenstate of the Gross-Pitaevskii Hamiltonian written in Bloch representation,
\begin{eqnarray}
\hat{H}_{\textrm{GP}}(k)&=&\left[\frac{U_{\xi}}{2}(\rho-1)+\frac{W_{\xi}}{2}\rho-2\Omega\cos(k)\right]\hat 1\\
\notag
&-&\left[\Omega_c m(k)+\Omega_{12}\sin(k)\right] \hat \sigma_z-\Omega_{12}\cos(k) \hat \sigma_y,    
\end{eqnarray}
where $m(k)$ should be determined self-consistently [Eq.~\eqref{mk}], and where $\Omega_c$ is given by Eq.~\eqref{Omega_c}. Note that, for each value of $k$, this Hamiltonian exactly maps to the mean-field Hamiltonian of a transverse spin-$1/2$ Ising model with an additional longitudinal magnetic field. Indeed, $m(k)$ represents the self-consistent magnetization along the $z$-direction with $\Omega_c$ the magnitude of the ferromagnetic coupling constant. The magnitudes of the transverse field along the $y$-direction and the longitudinal field along the $z$-direction are respectively given by $\Omega_{12}\cos(k)$ and $\Omega_{12}\sin(k)$.

The solution with lowest eigenenergy determines the mean-field energy functional
\begin{eqnarray}
\notag
\varepsilon_{\textrm{MF}}(k) &=&  -\sqrt{\Omega_{12}^2+\Omega_c^2 m(k)^2 + 2 m(k) \Omega_{12}\Omega_c\sin(k)}\\
\label{EMF}
&-&2\Omega\cos(k)+\frac{U_{\xi}}{2}(\rho-1)+\frac{W_{\xi}}{2}\rho , 
\end{eqnarray}
which should be minimized. We will denote by $k_0$ the value of $k$ that achieves this minimization (to be specified below).
 
By inserting the eigenstate of this low-energy branch into Eq.~\eqref{mk}, we obtain that the orbital polarization $m(k)$ should satisfy the self-consistent condition
\begin{equation}
 m(k) = \frac{m(k)\Omega_c + \Omega_{12}\sin(k)}{\sqrt{\Omega_{12}^2+\Omega_c^2 m(k)^2 + 2 m(k) \Omega_{12}\Omega_c\sin(k)}}.
 \label{SC}
\end{equation}
Moreover, we find that the phases $\Theta_{k,\sigma}\!=\!\Theta_{\sigma}$ are independent of the wavevector, and that they are determined by
\begin{equation}
\Theta= \Theta_{+}-\Theta_{-} = - (\pi/2) \textrm{sgn}(\Omega_{12}).\label{Theta_condition}
\end{equation}

It is insightful to analyze how these conditions on the nature of the ground-state translate in the original basis of Eq.~\eqref{Heff1}. In this representation, the mean-fields are given by $\langle \hat{a}_{n,s}\rangle_0 \equiv \psi_{n,s}$, with $s\!=\!1,2$, and they read [Eqs.~\eqref{Jbasis} and \eqref{param}]
\begin{eqnarray}
\psi_{n,1}
&=& e ^{ik_0 n}\sqrt{\rho}\left(\frac{\sqrt{1+m_0}e^{i\Theta_{+}} +\sqrt{1-m_0}e^{i\Theta_{-}}}{2}\right) \notag\\
\psi_{n,2} &=& i e^{ik_0 n}\sqrt{\rho}\left(\!\frac{\sqrt{1+m_0}e^{i\Theta_{+}}-\sqrt{1-m_0}e^{i\Theta_{-}}}{2}\right).\notag\\
\label{psi_original_basis}   
\end{eqnarray}
Here, we explicitly evaluated the fields at $k\!=\!k_0$ and we introduced the notation $m(k_0)\!=\!m_0$; we also omitted the trivial dynamical phase.

The condition in Eq.~\eqref{Theta_condition} then simply corresponds to having the $a_{1,2}$ orbitals equally populated in the ground state, i.e.~$|\psi_{n,1}|^2\!=\!|\psi_{n,2}|^2\!=\!\rho/2$. Without loss of generality, we henceforth set $\Omega_{12}\!>\!0$, and we express the relative phase $\varphi$ between the components $\psi_{n,2}$ and $\psi_{n,1}$ according to
\begin{equation}
\frac{\psi_{n,2}}{\psi_{n,1}} = e^{i\varphi} =  \sqrt{1-m_0^2} + i m_0,
\label{theta_def}
\end{equation}
where we used Eqs.~\eqref{Theta_condition}-\eqref{psi_original_basis}.
We thus obtain a simple relation between the local relative phase (internal angle) and the ground-state orbital polarization given in Eq.~\eqref{theta}.

The interplay of local orbital polarization and hopping processes gives rise to a ground-state current on a ring geometry. This can be obtained by evaluating the current operator derived from Eq.~\eqref{Heffbb},
\begin{eqnarray}
\label{Jop}
\hat{J} &=& i \frac{\Omega}{\hbar N_d}\sum_{n\sigma} \left(\hat{b}^{\dagger}_{n+1,\sigma}\hat{b}_{n,\sigma}- \text{h.c.}\right)\\
\notag
&+ & \frac{\Omega_{12}}{2\hbar N_d}\sum_{n \sigma} \left(\sigma\hat{b}^{\dagger}_{n+1,\sigma}\hat{b}_{n,\,\overline{\sigma}}-\sigma\hat{b}^{\dagger}_{n+1,\sigma}\hat{b}_{n,\sigma}+ \text{h.c.}\right).
\end{eqnarray}
In the mean-field solution, the current flowing through the ring is given by Eq.~\eqref{JMF}, where $k_0$ and $m_0$ are still to be determined below.

Finding the minimum of Eq.~\eqref{EMF} by imposing the self-consistent condition given by Eq.~\eqref{SC} can be  cumbersome from an analytical point of view. Hence, it is useful to consider particular limits. When $\Omega_{12}\!=\!0$, the minimum energy is precisely reached at $k_0\!=\!0$, which leads to two possible degenerate polarizations $m_{0}\!=\!\pm 1$; this situation corresponds to a relative phase $\varphi\!=\!\pm \pi/2$. A finite coupling $\Omega_{12} \ll \Omega_c$ leads to a non-zero ground-state quasi-momentum $k_0 \approx (\Omega_{12}/2\Omega) \, \textrm{sgn}(m_0) \approx \Phi \, \textrm{sgn}(m_0)$ and to a small depolarization of the system [see Eq.~\eqref{pol_small}]. We recall that $\Phi$ is the effective flux generated by the complex tunneling in Eq.~\eqref{Heffbb}; see Eq.~\eqref{the_flux} and Fig.~\ref{figVI_1}. Importantly, condensation at a finite quasi-momentum activates an effective longitudinal field in the Ising picture---recall that this field scales as $\Omega_{12}\sin(k)$---leading to a smoothing of the transition to the unpolarized state and an eventual absence of critical behavior. As a technical note, we remark that reaching this finite $k_0$ requires sufficiently large lattices satisfying $N_d\!>\!2\pi\Omega/|\Omega_{12}|$. In any case, to lowest order in $\Omega_{12}$, the ground-state polarization decreases according to Eq.~\eqref{pol_small}.

The chiral current flowing through the ring is activated by $\Omega_{12}$; see Eq.~\eqref{Jop}. In the mean-field ground state, the leading-order contribution is given by
\begin{equation}
 J_{\textrm{MF}}(k_0) \approx -\textrm{sgn}(m_0) \rho \Omega_{12}.\label{MF_current}   
\end{equation}
Importantly, the sign of the persistent current in Eq.~\eqref{MF_current} depends on the orbital order that spontaneously emerges in the system. This emergent chirality is a striking signature of the spontaneous breaking of TRS; see also the main text.

\section{Orbital order in the strongly-correlated regime}\label{appendix_Mott}

In this Appendix, we derive the effective Ising spin model in Eq.~\eqref{exchange} and we obtain the chiral superfluid-to-Mott phase diagram in the strong-coupling regime [Fig.~\ref{figVI_5}].
In the absence of kinetic terms in the Hamiltonian~\eqref{Heffbb}, there is a macroscopic degeneracy of $2^{N_d}$ possible ground-state configurations $|\{\sigma_n\}\rangle$, which may be written as product states
\begin{equation}
 |\{\sigma_n\}\rangle = \prod_{n=1}^{N_d}|\sigma_n\rangle, 
 \label{states0}
\end{equation}
with $|\sigma_n\rangle$ the states having well defined angular momentum along the $y$-direction, namely $J^{(n)}_y=\sigma_n \rho/2$ with $\sigma_n = \pm 1$; see Eq.~\eqref{spinstates} in the main text. The corresponding ground-state energy reads 
\begin{equation}
E^{(0)}_N(|\{\sigma_n\}\rangle) = U_{\xi}N(\rho-1)/2 -\mu N.
\end{equation}

The tunneling terms in the Hamiltonian~\eqref{Heffbb} do not couple these states at first order, but they do lift their degeneracy in second-order perturbation theory. Indeed, the first non-trivial correction to the energy of these $N$-particle states is given by
\begin{equation}
\Delta E (| \{\sigma_n\}\rangle) = \sum_{l}\frac{|\langle l|\hat{H}_T| \{\sigma_n\}\rangle |^2}{E^{(0)}_N(|\{\sigma_n\}\rangle)-E^{(0)}_N(|l\rangle)},  
\end{equation}
where $\hat{H}_T$ contains all the tunneling terms of Eq.~\eqref{Heffbb}, and where $|l\rangle$ is an excited state. Since the Hamiltonian~\eqref{Heffbb} only couples first nearest neighbors, this expression can be further simplified as a sum of pair contributions,
\begin{equation}
\Delta E (| \{\sigma_n\}\rangle) = \sum_{n} \Delta E(|\sigma_n\rangle|\sigma_{n+1}\rangle).
\end{equation}
The energy corrections for each pair are readily obtained as
\begin{align}
&\Delta E (|+\rangle |+\rangle)=\Delta E(|-\rangle|-\rangle) \notag \\
&\hspace{1cm}=-\frac{2|t_{\sigma}|^2 \rho(\rho+1)}{U_{\xi}}-\frac{2|t_{\sigma\overline{\sigma}}|^2\rho}{[\rho(W_{\xi}-U_{\xi}) + U_{\xi}]}, \notag \\
&\Delta E (|+\rangle |-\rangle)=\Delta E (|+\rangle |-\rangle) \notag \\
&\hspace{1cm}=-\frac{ 2|t_{\sigma\overline{\sigma}}|^2\rho(\rho+1)}{U_{\xi}}-\frac{2|t_{\sigma}|^2\rho}{[\rho(W_{\xi}-U_{\xi}) + U_{\xi}]}.\notag
\end{align}

\begin{figure}[t]
\center
\includegraphics[width=0.8\columnwidth]{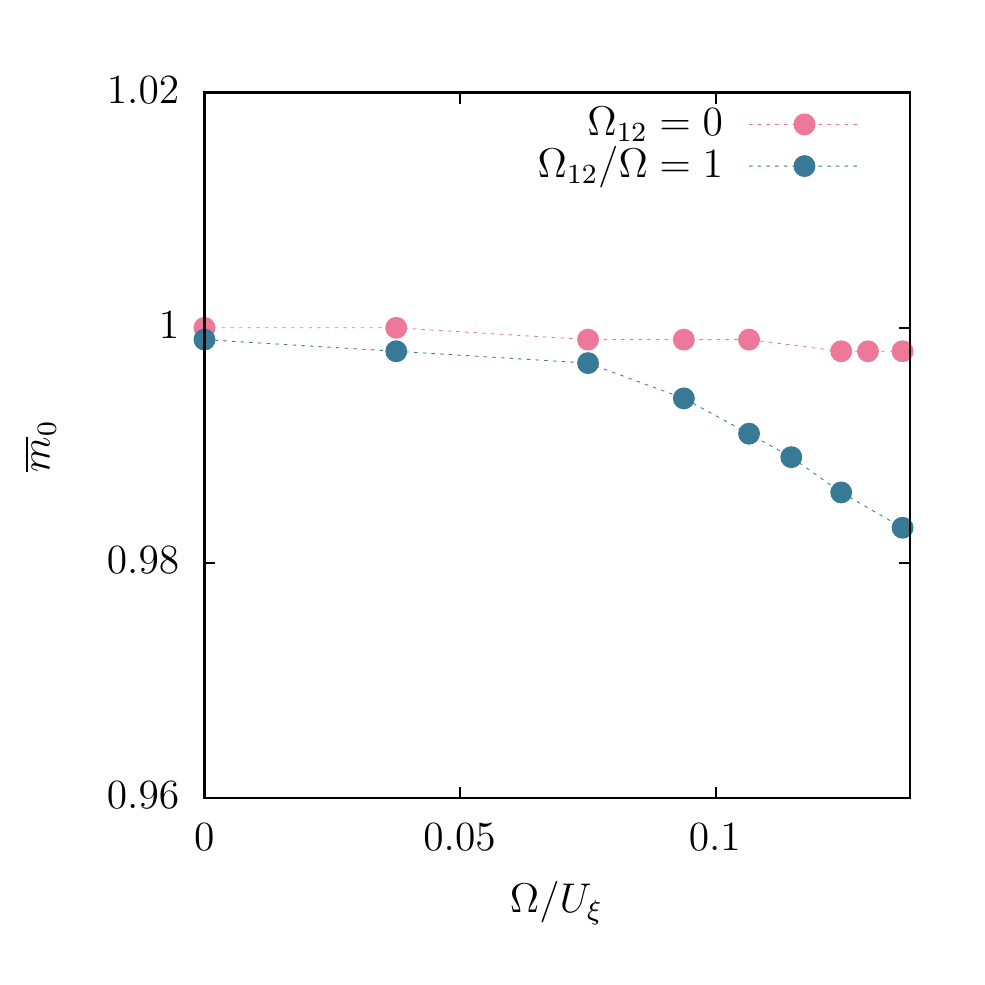}
\caption{Evolution of the averaged ground-state orbital polarization $\overline{m}_0$, within the Mott regime, for $\rho\!=\!2$ and $\Omega_{12}=0$ (red) and $\Omega_{12}/\Omega=1$ (blue). The points were obtained using a DMRG algorithm in a chain with open boundary conditions and $N_d=100$ dimers.}
\label{figX}
\end{figure}

We note that this approach is valid whenever $\xi\!<\!4$, which ensures repulsive intraspecies interactions $U_{\xi}>0$ in Eq.~\eqref{Heffbb}. The correction to the energy of the manifold of states given by Eq.~\eqref{states0}, up to second order, can hence be expressed as a constant shift (which is independent of the configuration $\{\sigma_n\}$) plus an orbital exchange interaction,
\begin{equation}
 \Delta E(|\{\sigma_n\}\rangle) = \sum_n E_0 + K_{yy} \sum_n J_y^{(n)} J^{(n+1)}_y, 
 \label{exchange_appendix}
\end{equation}
with $J_y^{(n)}=\sigma_n\rho/2$. The shift and exchange coupling are given by
\begin{eqnarray}
\notag
E_0 &=& -\frac{\left(|t_{\sigma}|^2+|t_{\sigma\overline{\sigma}}|^2\right)\rho[W_{\xi}\rho(\rho+1)-U_{\xi}(\rho^2-2)]}{ U_{\xi}[U_{\xi} + \rho(W_{\xi}-U_{\xi})]}\\
\notag
K_{yy} &=& -\frac{ 4\left(|t_{\sigma}|^2-|t_{\sigma\overline{\sigma}}|^2\right) [W_{\xi}+\rho(W_{\xi}-U_{\xi})]}{U_{\xi}[U_{\xi} + \rho(W_{\xi}-U_{\xi})]}<0.\\
\label{coupling_constant_appendix}
& &
\end{eqnarray}
These results lead to the effective Ising spin model displayed in Eq.~\eqref{exchange}. We remark that the exchange coupling at this order only depends on the tunneling $\Omega$ and that it favors a uniform `ferromagnetic' ordering. In a one-dimensional ring geometry, the approximated ground-state energy can then be expressed as
\begin{equation}
 E_\textrm{GS}(N) = E^{(0)}_N + N_d\left(E_0+K_{yy}\frac{\rho^2}{4}\right).   
\end{equation}
With the aim of comparing this analytical model with a more accurate numerical tool, we performed DMRG simulations to analyze the evolution of the ground-state orbital polarization within the chiral-Mott phase. The results are presented in Fig.~\ref{figX} for both $\Omega_{12}=0$ and $\Omega_{12}/\Omega=1$. The orbital order is practically unaltered by the existence of a finite $\Omega_{12}$, in agreement with what we expect
from the effective spin model:~the exchange
coupling constant $K_{yy}$ in Eq.~\eqref{coupling_constant_appendix} does not depend on $\Omega_{12}$. On
the other hand, as the system gets closer to the superfluid
phase, the presence of this hopping term destabilizes the
angular momentum ordering.

The perturbative expansion described above can be further used to elucidate the boundaries between the (chiral) Mott and superfluid phases, by additionally considering how the ground state energy changes by adding or removing a particle in the system:~the phase boundary precisely occurs when this particle-hole excitation gap vanishes. Since we are interested in the regime $W_{\xi}>U_{\xi}$, the relevant low-energy manifold to consider upon adding one extra particle is given by $N_d$ states of the form
\begin{equation}
|n_{N+1}\rangle = \frac{\left(\hat{b}^{\dagger}_{1,\sigma}\right)^{\rho}\hdots\left(\hat{b}^{\dagger}_{n,\sigma}\right)^{\rho+1}\hdots \left(\hat{b}^{\dagger}_{N_d,\sigma}\right)^{\rho}}{\sqrt{(\rho!)^{N_d-1}(\rho+1)!}}|0\rangle, 
\label{states_plusone}
\end{equation}
so that the state $|n_{N+1}\rangle$ has one more boson in the $n-$th dimer of a ferromagnetic state with $\sigma$-order. The unperturbed energy of these states is
\begin{equation}
 E_{N+1}^{(0)}(\{|n_{N+1}\rangle\})=U_{\xi}N(\rho-1)/2+U_{\xi}\rho-\mu(N+1). 
\end{equation}
The ground-state energy can then be approximately found via a canonical transformation procedure~\cite{Slichter1990}, in which an effective Hamiltonian $\hat{H}^{'}$ that takes into account up to second order processes in the tunneling amplitudes is defined within the manifold of the original set of $\{|n\rangle_{N+1}\}$-states. The matrix elements of $\hat{H}'$ are determined as
\begin{eqnarray}
\langle n^{}_{N+1}|\hat{H}^{'}|n'_{N+1}\rangle &=& E^{(0)}_{N+1} + \langle n^{}_{N+1}|\hat{H}_T|n'_{N+1}\rangle\\
\notag
&+&\sum_l \frac{\langle n^{}_{N+1}|\hat{H}_T|l^{}_{}\rangle \langle l^{}_{}|\hat{H}_T|n'_{N+1}\rangle}{E^{(0)}_{N+1}-E_{N+1}^{(0)}(|l\rangle)},
\end{eqnarray}
with $|l\rangle$ the excited states and $E_{N+1}^{(0)}(|l\rangle)$ their corresponding unperturbed energy. The $N_d\times N_d$ matrix has a tridiagonal form and can be analytically diagonalized, revealing the splitting of the degenerate states of Eq.~\eqref{states_plusone} into a band described by
\begin{eqnarray}
\notag
E'_{N+1}(k) &=& E^{(0)}_{N+1}+\Sigma_{N+1}-2|t_{\sigma}|(\rho+1)\cos(k-\Phi_{\sigma})\\
\notag
&-&\frac{2(\rho+1)|t_{\sigma\overline{\sigma}}|^2}{(W_{\xi}-U_{\xi})\rho}\cos(2 k)\\
&-&\frac{2\rho(\rho+1)|t_{\sigma}|^2}{U_{\xi}}\cos(2(k-\Phi_{\sigma})),
\label{EpNp1}
\end{eqnarray}
with 
\begin{eqnarray}
\notag
\Sigma_{N+1} &=& -\frac{|t_{\sigma}|^2\rho(\rho+2)}{U_{\xi}}-\frac{2\rho|t_{\sigma\overline{\sigma}}|^2}{\rho(W_{\xi}-U_{\xi})+W_{\xi}+U_{\xi}}\\
\notag
&-&\frac{2(\rho+1)|t_{\sigma\overline{\sigma}}|^2}{\rho(W_{\xi}-U_{\xi})}-\frac{2(N_d-2)\rho(\rho+1)|t_{\sigma}|^2}{U_{\xi}}\\
&-&\frac{2(N_d-2)\rho|t_{\sigma\overline{\sigma}}|^2}{\rho(W_{\xi}-U_{\xi})+U_{\xi}}.
\end{eqnarray}
In Eq.~\eqref{EpNp1}, we introduced the flux $\Phi_{\sigma}\!=\!\sigma \arctan(\Omega_{12}/2\Omega)$ and the quasi-momentum $k=2\pi j/N_d$, with $j=0,\hdots,N_d-1$. Interestingly, at this order, the extra boson feels the presence of the flux $\Phi_{\sigma}$ in the lattice via effective hopping processes at first and second nearest neighbors. 

In the thermodynamic limit, the minimum energy of this band will be precisely at $k=\Phi_{\sigma}$, so that the ground-state energy with one extra boson is obtained as
\begin{equation}
 E_{\textrm{GS}}(N+1) = E'_{N+1}(\Phi_{\sigma}).   
\end{equation}
We follow a similar procedure to consider the effect of the hopping terms in the low-energy manifold with one boson less. When removing one particle from the $\sigma$-ordered $N$-particle ground state, the relevant manifold of states where the perturbation theory should be applied will be given by
\begin{equation}
|n_{N-1}\rangle = \frac{\left(\hat{b}^{\dagger}_{1,\sigma}\right)^{\rho}\hdots\left(\hat{b}^{\dagger}_{n,\sigma}\right)^{\rho-1}\hdots \left(\hat{b}^{\dagger}_{N_d,\sigma}\right)^{\rho}}{\sqrt{(\rho!)^{N_d-1}(\rho-1)!}}|0\rangle,    
\end{equation}
which have a zero-th order energy of
\begin{equation}
 E_{N-1}^{(0)}(\{|n_{N-1}\rangle\})=U_{\xi}N(\rho-1)/2-U_{\xi}(\rho-1)-\mu(N-1).   
\end{equation}
By diagonalizing the corresponding canonically transformed Hamiltonian, we obtain the broadening of these states into a band described by
\begin{eqnarray}
\notag
E'_{N-1}(k) &=& E^{(0)}_{N-1}+\Sigma_{N-1}-2|t_{\sigma}|\rho\cos(k+\Phi_{\sigma})\\
\notag
&-& \frac{2\rho |t_{\sigma\overline{\sigma}}|^2}{(W_{\xi}-U_{\xi})\rho+U_{\xi}}\cos(2k)\\
&-&\frac{2\rho(\rho+1)|t_{\sigma}|^2}{U_{\xi}}\cos(2(k+\Phi_{\sigma})),
\end{eqnarray}
with
\begin{eqnarray}
\notag
 \Sigma_{N-1} &=& -\frac{|t_{\sigma}|^2(\rho^2-1)}{U_{\xi}}-\frac{2\rho |t_{\sigma\overline{\sigma}}|^2}{\rho(W_{\xi}-U_{\xi})+U_{\xi}-W_{\xi}}\\
 \notag
 &-&\frac{2(\rho-1)|t_{\sigma\overline{\sigma}}|^2}{\rho(W_{\xi}-U_{\xi})+2U_{\xi}}-\frac{2(N_d-2)\rho(\rho+1)|t_{\sigma}|^2}{U_{\xi}}\\
 &-&\frac{2(N_d-2)\rho|t_{\sigma\overline{\sigma}}|^2}{\rho(W_{\xi}-U_{\xi})+U_{\xi}}.
\end{eqnarray}
Note that the hole-like excitation feels the opposite flux ($-\Phi_{\sigma}$). We then find that, in the thermodynamic limit, the ground-state energy with one boson less is given by
\begin{equation}
E_{\textrm{GS}}(N-1) = E'_{N-1}(-\Phi_{\sigma}).    
\end{equation}
The phase boundary between the (chiral) Mott insulator and superfluid phases is determined by the conditions 
\begin{align}
&E_{\textrm{GS}}(N+1)\!=\!E_{\textrm{GS}}(N), \notag \\
&E_{\textrm{GS}}(N)\!=\!E_{\textrm{GS}}(N-1).\notag
\end{align}
Solving these equations separately for $\mu$, we find the boundaries for the particle sector ($\mu_{+}$) and hole sector ($\mu_{-}$), which are displayed in Eq.~\eqref{Mott_lobe}. The difference $\mu_{+}-\mu_{-}$ determines the charge gap in the Mott phase. The resulting Mott-SF phase diagram is depicted in Fig.~\ref{figVI_5}, in terms of the chemical potential $\mu/U_{\xi}$ and ratio $\Omega/U_{\xi}$.


\section{Strong-coupling expansion \\ and the effective Hamiltonian for filling $\rho=1$}
\label{appendix_spinhalf}
At unit filling, the spin-states in Eq.~\eqref{spinstates} are coupled to each other via spin-flip processes, which also scale as the square of the tunneling amplitudes. In order to describe the corresponding low-energy manifold with an effective theory, we must perform a canonical transformation procedure and project the resulting effective Hamiltonian into the subspace of unit filling. Since the Hamiltonian only couples nearest neighbors, we can focus on the effective theory for just two dimers ($n$ and $n+1$) and then sum over all the lattice links connecting them. For simplicity, we will work in the basis of the original orbitals $a_{n,s}$ in each dimer (with $s=1,2$). 

In the absence of hopping terms, there are four possible degenerate states with 2 particles in the nearest neighbor dimer configuration that satisfy the unit filling condition. Their corresponding energy is given by $E^{(0)}_2=-2\mu$ and they can be expressed as:
\begin{equation}
|s^{}_n s'_{n+1}\rangle = \hat{a}^{\dagger}_{n,s}\hat{a}^{\dagger}_{n+1,s'}|0\rangle,
\label{D1}
\end{equation}
where we introduced the notation $|s_n\rangle = \hat{a}^{\dagger}_{n,s}|0\rangle$ with $s=1,2$. 

The matrix elements of the canonically transformed Hamiltonian $\hat{H}'$ within this subspace are given by
\begin{eqnarray}
\langle s_n s'_{n+1}|\hat{H}'|s''_n s'''_{n+1}\rangle \!\!\!&=&\!\!\!E_2^{(0)}+\langle s_n s'_{n+1}|\hat{H}_T|s''_n s'''_{n+1}\rangle\\
\notag
\!\!\!&+&\!\!\!\sum_l \frac{\langle s_n s'_{n+1}|\hat{H}_T|l^{}_{}\rangle \langle l^{}_{}|\hat{H}_T|s''_n s'''_{n+1}\rangle}{E^{(0)}_{2}-E_2^{(0)}(|l\rangle)},       
\end{eqnarray}
with $|l\rangle$ the excited states and $E_2^{(0)}(|l\rangle)$ their unperturbed energy. In particular, the states in Eq.~\eqref{D1} are coupled through $\hat{H}_T$ to the following high-energy virtual states: 
\begin{align}
\notag
|1\rangle\!&=\!\frac{1}{\sqrt{2}}\!\left(\hat{a}^{\dagger 2}_{n,1} +\hat{a}^{\dagger 2}_{n,2}\right)\!\!|0\rangle & |2\rangle\!&=\!\frac{1}{\sqrt{2}}\!\left(\hat{a}^{\dagger 2}_{n+1,1} +\hat{a}^{\dagger 2}_{n+1,2}\right)\!\!|0\rangle\\
\notag
|3\rangle\!&=\!\frac{1}{\sqrt{2}}\!\left(\hat{a}^{\dagger 2}_{n,1} -\hat{a}^{\dagger 2}_{n,2}\right)\!\!|0\rangle & |4\rangle\!&=\! \frac{1}{\sqrt{2}}\!\left(\hat{a}^{\dagger 2}_{n+1,1} -\hat{a}^{\dagger 2}_{n+1,2}\right)\!\!|0\rangle\\
|5\rangle\!&=\! \hat{a}^{\dagger}_{n,1}\hat{a}^{\dagger}_{n,2}|0\rangle & |6\rangle \!&=\!\hat{a}^{\dagger}_{n+1,1}\hat{a}^{\dagger}_{n+1,2}|0\rangle,
\end{align}
where the first two states have energy
$W_{\xi}-2\mu$ and  the others $U_{\xi}-2\mu$. 


Neglecting irrelevant constant energy shifts, we can express the projected canonically transformed Hamiltonian in the following form
\begin{align}
\notag
\hat{H}'_{n,n+1}\!\!&=\!\!\!\!\sum_{\substack{s,s
 '\\ s'',s
'''}}\!\!\langle s_n s'_{n+1}|\hat{H}'|s''_n s'''_{n+1}\rangle 
\hat{a}^{\dagger}_{n,s}\hat{a}^{\dagger}_{n+1,s'}\hat{a}^{}_{n,s''}\hat{a}^{}_{n+1,s'''}\\
&=\sum_s\left(\gamma_{ss}\hat{n}_{n,s}\hat{n}_{n+1,s}+\gamma_{s\overline{s}}\hat{n}_{n,s}\hat{n}_{n+1,\overline{s}}\right) \label{very_long_eq}\\
\notag
&+\Gamma_1\sum_s \hat{a}^{\dagger}_{n,s}\hat{a}^{\dagger}_{n+1,s}\hat{a}^{}_{n,\overline{s}}\hat{a}^{}_{n+1,\overline{s}}\\
\notag
&+\Gamma_2\sum_s \hat{a}^{\dagger}_{n,s}\hat{a}^{\dagger}_{n+1,\overline{s}}\hat{a}^{}_{n,\overline{s}}\hat{a}^{}_{n+1,s}\\
\notag
\!\!&+\!\! 
\notag
(\hat{a}^{\dagger}_{n,1}\hat{a}^{}_{n,2}+\hat{a}^{\dagger}_{n,2}\hat{a}^{}_{n,1})(\Gamma_3 \hat{n}_{n+1,1} + \Gamma_4 \hat{n}_{n+1,2})\\
\notag
\!\!&+\!\!(\hat{a}^{\dagger}_{n+1,1}\hat{a}^{}_{n+1,2}+\hat{a}^{\dagger}_{n+1,2}\hat{a}^{}_{n+1,1})(\Gamma_5 \hat{n}_{n,1} + \Gamma_6 \hat{n}_{n,2}),
\end{align}
where $\overline{1}=2$ and $\overline{2}=1$. We note that the tunneling Hamiltonian effectively leads to the emergence of correlated hopping terms within each dimer (the last two terms in Eq.~\eqref{very_long_eq}), which are absent in p-band bosonic models; see Ref.~\cite{Pinheiro2013}. The corresponding coupling constants are explicitly given by
\begin{align}
 \gamma_{11} &= -2\Omega^2\left(\frac{1}{U_{\xi}}+\frac{1}{W_{\xi}}\right)-\frac{\Omega_{12}^2}{U_{\xi}} & \gamma_{12} &=-\frac{2\Omega^2}{U_{\xi}}\\
 \notag
 \gamma_{21} &=-2\Omega_{12}^2\left(\frac{1}{W_{\xi}}+\frac{1}{U_{\xi}}\right) -\frac{2\Omega^2}{U_{\xi}} & \gamma_{22}&=\gamma_{11}\\
 \notag
 \Gamma_1 &=2\Omega^2\left(\frac{1}{U_{\xi}}-\frac{1}{W_{\xi}}\right) & \Gamma_2 &= -\frac{2\Omega^2}{U_{\xi}}\\
 \notag
 \Gamma_3 &= -\Omega\Omega_{12}\left(\frac{2}{W_{\xi}}+\frac{1}{U_{\xi}}\right) & \Gamma_4 &=-\frac{\Omega\Omega_{12}}{U_{\xi}}\\
 \notag
 \Gamma_{5} &= \Gamma_{4} & \Gamma_{6}&=\Gamma_{3}.
\end{align}

By further employing the Schwinger angular momentum representation [see Eq.~\eqref{Schwinger_lattice} in the main text] and the constraint $\hat{N}^{(n)}=\hat{n}_{n,1}+\hat{n}_{n,2}=1$, the effective Hamiltonian at each link can be expressed as
\begin{eqnarray}
 \hat{H}'_{n,n+1}&=& \sum_{\nu=x,y,z}K_{\nu\nu}  \hat{J}_{\nu}^{(n)}\hat{J}_{\nu}^{(n+1)}\\
 \notag
 &+&\left(\frac{\gamma_{22}-\gamma_{11}}{2}\right)\left(\hat{J}_{z}^{(n)}+\hat{J}_{z}^{(n+1)}\right)\\
 \notag
 &+&\left(\frac{\gamma_{21}-\gamma_{12}}{2}\right)\left(\hat{J}_{z}^{(n)}-\hat{J}_{z}^{(n+1)}\right)\\
\notag
&-&D\left(\hat{J}_{z}^{(n)}\hat{J}_{x}^{(n+1)}-\hat{J}_{x}^{(n)}\hat{J}_z^{(n+1)}\right)\\
\notag
&+&(\Gamma_3 + \Gamma_4)\hat{J}_{x}^{(n)} +(\Gamma_5 + \Gamma_6)\hat{J}_{x}^{(n+1)},
\end{eqnarray}
with $K_{xx}=2(\Gamma_1+\Gamma_2)$, $K_{yy}=2(\Gamma_2-\Gamma_1)$, $K_{zz}=\gamma_{11}+\gamma_{22}-\gamma_{12}-\gamma_{21}$ and $D=2(\Gamma_4-\Gamma_3)$. The magnitudes of these couplings, expressed in terms of the original hopping and interaction parameters of the model, are provided in Eq.~\eqref{couplings_spinhalf} in the main text. 

The effective spin-$1/2$ Hamiltonian in the one-dimensional lattice is finally written as 
\begin{align}
\notag
\hat{H}^{\textrm{eff}}_{1/2} &= \sum_n \hat{H}'_{n,n+1}\\
\notag
&=  \sum_n\sum_{\nu=x,y,z}K_{\nu\nu}  \hat{J}_{\nu}^{(n)}\hat{J}_{\nu}^{(n+1)} + h_x\sum_n \hat{J}_x^{(n)}\\
\label{effective_spinhalf_appendix}
&\hspace{0.2cm}-\bm{D}\sum_n\left(\hat{\bm{J}}^{(n)}\times\hat{\bm{J}}^{(n+1)}\right),   
\end{align}
where $\bm{D}=(0,D,0)$ and $h_x = \Gamma_3 + \Gamma_4 + \Gamma_5 + \Gamma_6$.

\end{appendix}


\begin{thebibliography}{100}
\bibitem{footnote}
The nonlinear Schr\"odinger equation in Eq.~\eqref{NLS} assumes that the
  parameter $\gamma$ is the same for both modes. If this is not the case, then
  Eq.~\eqref{NLS} would display a term of the form $\sim (- \partial^2/\partial
  x^2) \hat \sigma_z$. Since the Pauli matrix $\hat \sigma_z$ does not commute
  with the mixing operator $\hat \sigma_x$ in Eq.~\eqref{pi_over_two}, this
  kinetic term will be modified in the effective NLSE
  [Eq.~\eqref{NLS_effective_generalized}].
\end{thebibliography}



\end{document}